\documentclass[
 aip,
 amsmath,amssymb,
reprint,%
]{revtex4-1}

\usepackage{graphicx}
\usepackage{dcolumn}
\usepackage{bm}
\usepackage{upgreek}

\usepackage[utf8]{inputenc}
\usepackage[T1]{fontenc}
\usepackage{mathptmx}
\usepackage{etoolbox}
\usepackage{comment}
\usepackage{algorithm, fancyvrb, amsmath}
\makeatletter
\def\@email#1#2{%
 \endgroup
 \patchcmd{\titleblock@produce}
  {\frontmatter@RRAPformat}
  {\frontmatter@RRAPformat{\produce@RRAP{*#1\href{mailto:#2}{#2}}}\frontmatter@RRAPformat}
  {}{}
}%
\makeatother
\begin{document}

\preprint{AIP/123-QED}

\newcommand{\EPFL}{Laboratory of Computational Chemistry and Biochemistry, École Polytechnique Fédérale de Lausanne, CH-1015 Lausanne, Switzerland}
\newcommand{\DTU}{DTU Chemistry, Technical University of Denmark (DTU), DK-2800 Kongens Lyngby, Denmark}
\newcommand{\FZJ}{Computational Biomedicine, Institute of Advanced Simulations IAS-5/Institute for Neuroscience and Medicine INM-9, Forschungszentrum J\"{u}lich GmbH, J\"{u}lich 52428, Germany}
\newcommand{\CINECA}{CINECA, 40033 Casalecchio di Reno, BO, Italy}
\newcommand{\UMichME}{Department of Mechanical Engineering, University of Michigan, Ann Arbor, Michigan 48109, USA}
\newcommand{\UMichMS}{Department of Materials Science and Engineering, University of Michigan, Ann Arbor, Michigan 48109, USA}
\newcommand{\UFe}{Dipartimento di Scienze Chimiche, Farmaceutiche ed Agrarie (DOCPAS), Università degli Studi di Ferrara (Unife), I-44121 Ferrara, Italy}
\newcommand{\RWTH}{Department of Physics, RWTH Aachen University, Aachen 52074, Germany}

\title{MiMiC: A High-Performance Framework for Multiscale Molecular Dynamics Simulations}

\author{Andrej Antalík}
\affiliation{\EPFL}

\author{Andrea Levy}
\affiliation{\EPFL}

\author{Sonata Kvedaravičiūtė}
\affiliation{\DTU}

\author{Sophia K. Johnson}
\affiliation{\EPFL}

\author{David Carrasco-Busturia}
\altaffiliation[Current address: ]{
Division of Theoretical Chemistry and Biology, School of Engineering Sciences in Chemistry, Biotechnology and Health, KTH Royal Institute of Technology, SE-100 44 Stockholm, Sweden
}
\affiliation{\DTU}

\author{Bharath Raghavan}
\affiliation{\FZJ}
\affiliation{\RWTH}

\author{François Mouvet}
\affiliation{\EPFL}

\author{Angela Acocella}
\affiliation{\CINECA}

\author{Sambit Das}
\affiliation{\UMichME}

\author{Vikram Gavini}
\affiliation{\UMichME}
\affiliation{\UMichMS}

\author{Davide Mandelli}
\affiliation{\FZJ}

\author{Emiliano Ippoliti}
\affiliation{\FZJ}

\author{Simone Meloni}
\affiliation{\UFe}

\author{Paolo Carloni}
\affiliation{\FZJ}
\affiliation{\RWTH}

\author{Ursula Rothlisberger}
\affiliation{\EPFL}

\author{Jógvan Magnus Haugaard \surname{Olsen}}
\email[Author to whom correspondence should be addressed: ]{jmho@kemi.dtu.dk}
\affiliation{\DTU}

\date{\today}

\begin{abstract}
MiMiC is a framework for performing multiscale simulations in which loosely coupled external programs describe individual subsystems at different resolutions and levels of theory.
To make it highly efficient and flexible, we adopt an interoperable approach based on a multiple-program multiple-data (MPMD) paradigm, serving as an intermediary responsible for fast data exchange and interactions between the subsystems.
The main goal of MiMiC is to avoid interfering with the underlying parallelization of the external programs, including the operability on hybrid architectures (e.g., CPU/GPU), and keep their setup and execution as close as possible to the original.
At the moment, MiMiC offers an efficient implementation of electrostatic embedding QM/MM that has demonstrated unprecedented parallel scaling in simulations of large biomolecules using CPMD and GROMACS as QM and MM engines, respectively.
However, as it is designed for high flexibility with general multiscale models in mind, it can be straightforwardly extended beyond QM/MM.
In this article, we illustrate the software design and the features of the framework, which make it a compelling choice for multiscale simulations in the upcoming era of exascale high-performance computing.
\end{abstract}

\maketitle

\section{\label{sec:introduction}Introduction}
Chemical and biological phenomena span large temporal and spatial scales.
An accurate representation of the physics across these diverse scales requires the development of multiscale simulation methods able to employ different resolutions and levels of theory within the scope of a single simulation.
An approach involving quantum mechanics (QM) is essential to describe the reorganization of electrons and nuclei, e.g., during reactive events involving bond breaking and formation, yet a molecular mechanics (MM) model characterizes most protein and nucleic-acid properties with sufficient accuracy.
Furthermore, applying even lower resolutions such as coarse-grained (CG)\cite{Klein_2012, Pereira_2022, Voth_2022, Souza_2023} models enables simulations of larger systems over longer time scales.
Probably the most well-known example of a multiscale model is QM/MM, where a small part of a system is treated at a QM level of theory, and the remainder is modeled using an MM force field.~\cite{warshel1976theoretical, Singh_QMMM, Karplus_QMMM, Senn_QMMM, Rothlisberger_review, Estrin_review, Mennucci_review}

Multiscale methods are usually implemented by extending a program with the functionality of another, e.g., in the case of QM/MM, a QM program with MM functionality, or vice versa.
These implementations range from fully monolithic, through linked libraries with ad hoc interfaces to stand-alone programs, up to general interfaces between independent programs.
Although most QM/MM implementations adopt the first two of these strategies, some software packages implement more flexible interfaces that allow coupling to, in principle, any other program without substantial effort.\cite{AMBER_QM_MM_MD_interface, NAMD_goes_quantum, TCPB_2023}
Apart from these already more general interfaces, an even more flexible approach is to implement multiscale methods through integrative frameworks.
These do not themselves implement any QM, MM, or other such functionality for calculating the properties of individual subsystems but instead rely completely on external programs.\cite{MSCALE, QMMMW, PUPIL_soft_integration, PUPIL_soft_integration2, Cuby, ASE, Alto2007, Weingart2018, LICHEM, LICHEM2, Janus, Molssi_driver, QM3, ChemShell_QM_MM, ChemShell_QM_MM_redevelopment, Lu2023, QMMM2023}

An important aspect of multiscale implementations involving several distinct software components is their coupling, which can be classified as tight or loose based on the degree of interdependency between them.
Loose coupling helps to avoid code duplication and greatly reduces maintenance costs, as modifications made to one component do not impact the rest, and replacing individual components becomes almost trivial.
Furthermore, new features implemented in a given component are instantly available with practically no additional effort.
Loose coupling to the external programs is fundamental to essentially all general multiscale frameworks.
However, obtaining high computational efficiency can pose a challenge due to potentially slow data exchange compared to the tight-coupling approach, where the relevant data is available in memory.
In particular, most frameworks rely on file-based communication, i.e., using input and output files, which can have a substantial impact on performance.\cite{Isborn_jctc, QUICK_AMBER}
To overcome these issues, some adopt a library-linking approach that requires converting the external programs into libraries, thus losing some of the flexibility, while others are realized through network-based communication.

\begin{figure}[t]
\centering
\includegraphics[width=0.475\textwidth]{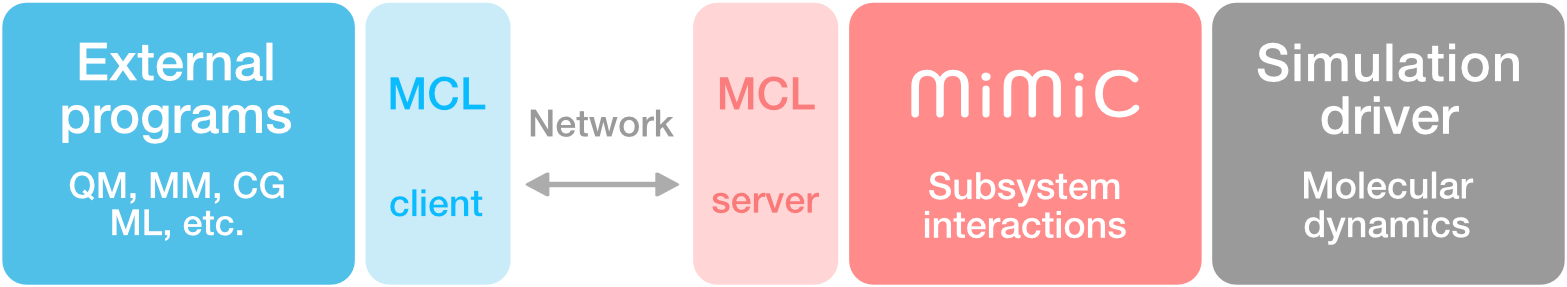}
\caption{Illustration of the strategy used by the MiMiC framework}
\label{fig:general_workflow}
\end{figure}

Recently, we introduced MiMiC\cite{olsen2019mimic}, a high-performance multiscale modeling framework that combines the best of both worlds, the flexibility of loose coupling, while achieving high computational efficiency\cite{mimic_jctc_hpc}.
It attains these advantages by employing a client-server approach together with a multiple-program multiple-data (MPMD) model through which it loosely couples external programs that concurrently calculate different subsystem contributions while MiMiC itself calculates interactions between them, see illustration in Fig.~\ref{fig:general_workflow}.
MiMiC has a dual-sided application programming interface (API).
One side communicates with a molecular dynamics (MD) driver that integrates the equations of motion, while the other interfaces to external programs.
To ensure efficient and straightforward interfacing with the latter, we developed the MiMiC communication library (MCL) that features a simple API and, moreover, provides a means to achieve efficient network-based communication while minimizing the impact on the underlying parallelization of the external programs.
This allows us to exploit the efficiency and features of specialized software packages, thus granting access to multiscale simulations that can take full advantage of state-of-the-art high-performance computing (HPC) architectures.

In this article, we give a detailed presentation of the software design of the MiMiC framework and showcase its features, performance, and recent applications involving large biomolecular systems.
Section \ref{sec:design} describes the internal organization of the code and the key aspects of communication, which might also prove useful in potential future interfaces.
Section \ref{sec:workflow} illustrates workflows of multiscale simulations with MiMiC, focusing on electrostatic and polarizable embedding QM/MM schemes and a multiple time step (MTS) approach for QM/MM MD.
Sections \ref{sec:features} and \ref{sec:upcoming_clients} provide an overview of current and upcoming features either implemented directly in MiMiC or accessible through external programs.
Section \ref{sec:applications} summarizes some notable studies in which MiMiC enabled the efficient and accurate modeling of complex biophysical and biochemical processes.
We conclude this paper by outlining the road ahead.

\section{\label{sec:design}Software Design}

The design of the MiMiC framework is guided by our primary objectives: achieving optimal flexibility and performance while minimizing implementation and maintenance efforts.
Flexibility is attained through a modular design where individual subsystems are assigned to external client programs that compute the relevant properties needed for a multiscale simulation.
In this MPMD setup, MiMiC serves as the intermediary that collects and distributes data among the concurrently running client programs and efficiently calculates contributions from the interactions between the subsystems.
To accommodate different types of subsystems, particles, and other quantities, we draw extensively on the object-oriented capabilities of modern Fortran and organize data structures into a hierarchy of derived types that can be easily extended.
Once a given type of subsystem interaction is implemented, it is then trivially transferable to any compatible program with no additional coding effort besides the implementation of the interface itself.
Apart from easy access to new features in the client programs and the MD driver, this approach avoids code duplication and substantially reduces the maintenance effort to keep the MiMiC ecosystem operational.

To fulfill our high-performance objective, we aim to minimize computational overheads associated with using multiple loosely coupled client programs.
To avoid their repeated startup and shutdown, we came up with an interoperable approach that does not interfere with the parallelization of these programs while keeping their setup and execution as close as possible to the original.
For this purpose, we developed MCL that facilitates data exchange between MiMiC and the clients through a clean interface, agnostic of the underlying communication mechanism, and which allows us to make use of the features and optimizations offered by the client programs.
MCL is written in C++ with simple C and Fortran APIs, thus ensuring interoperability with basically any programming language.
Currently, it supports two MPI-based modes of communication, but other mechanisms can be implemented essentially without any changes to the interfaces in the client programs.

\subsection{\label{sec:interface_design}Interfaces}

An essential feature of the loose-coupling paradigm is the design of a well-defined and general interface between individual programs, as opposed to several program-specific interfaces.
For practical purposes, the MiMiC framework is split into two separate libraries, namely, the main MiMiC library and MCL.
The first is linked to the simulation driver program, which also acts as the server that manages communication between the individual clients, keeps track of all simulation-related information, and computes subsystem interactions.
However, in order to become a client, an external program responsible for a particular subsystem requires only MCL, which facilitates data exchange with the server and can be introduced with only minimal changes in the original code.

\begin{figure*}
    \centering
    \includegraphics[width=0.6\linewidth]{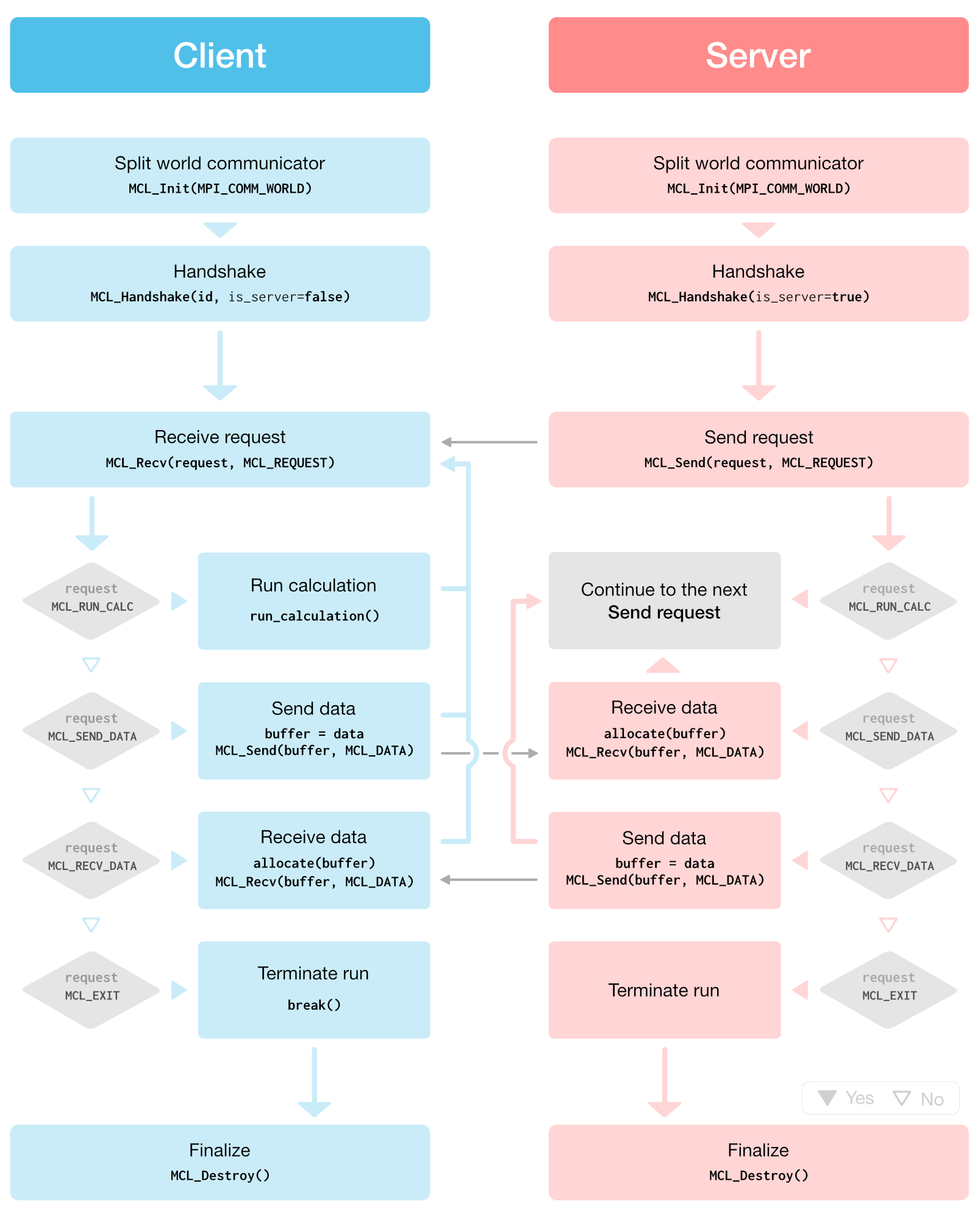}
    \caption{Schematic representation of the communication between the server and an external client program during a simulation.}
    \label{fig:sw_clt_srv}
\end{figure*}

\subsubsection{Client-Server Communication}

To support our minimal interference approach for clients, the MCL API consists of only a handful of procedures similar to those of MPI.
Here, we describe client-server communication of a generic run within the context of MPI MPMD, illustrated in Fig.~\ref{fig:sw_clt_srv}.

The simulation starts by launching all programs as an MPI MPMD job, meaning that MPI launches these programs simultaneously.
All of them now share a single \verb|MPI_COMM_WORLD| communicator that is intercepted by invoking \verb|MCL_Init()| right after the MPI initialization with \verb|MPI_Init()|.
This splits the communicator and returns a new local "world" communicator for each individual program.
All communication between the clients and the server is then channeled through intercommunicators.
The next step is to establish the client-server connections by calling the \verb|MCL_Handshake()| procedure in all running programs to specify the executable that will be the server and, at the same time, to identify individual clients by assigning them IDs.

At this stage, the server (MD driver) does not have any information about the subsystems, as these are specified in the inputs of the client programs.
Therefore, it first collects this information from the clients, including details relevant to the allocation of data structures in the MD driver (numbers of atoms and species/atom types), as well as details about the particular objects within the simulation needed for computing the subsystem interactions and propagating the system in time (masses, atomic numbers, etc.).
Once everything is set, the program enters the main MD loop, which, in terms of communication, predominantly consists of distributing coordinates to and collecting forces from the clients.

To enable minimal interference with the infrastructure of client programs while maintaining maximum flexibility, we coordinate actions between the clients and the server using a request-based approach.
In general, a client initializes as it normally would and then enters a construct that we refer to as the \emph{MiMiC loop}, which is merely a conditional loop in which two actions are performed in every iteration.
First, it receives a request via \verb|MCL_Recv()|, and then it executes the specified action given that it has already been implemented.
For an implementation example, see Algorithm \ref{alg:mimic_loop}.

The entire MiMiC loop basically reduces to the execution of one of four possible types of instructions.
Requests prepended by \verb|MCL_SEND| instruct a client to send the requested data.
This could involve allocating a buffer, filling it with requested data, and then sending it using \verb|MCL_Send()|.
The reciprocal action is a receive data request, which is prepended by \verb|MCL_RECV|.
The last type of practical request, denoted by the prefix \verb|MCL_RUN|, instructs a client to execute a calculation, e.g., to compute the energy, forces, or both.
Finally, the server can order clients to stop the program execution by issuing the \verb|MCL_EXIT| request.
Once received, the client program wraps up the run, deallocates memory, and ultimately terminates the program.
This typically occurs at the end of a simulation or when one of the programs raises an error.

\begin{algorithm}[H]
\caption{Pseudocode example of the MiMiC loop.}
\label{alg:mimic_loop}
\begin{Verbatim}[commandchars=\\\{\}]

{\bf while} ({\bf not} is_last_step) {\bf do}
    request = mimic_receive_request()
    {\bf if} (request == MCL_RUN_ENERGY_FORCES) {\bf then}
        energy, forces = calculate_energy_forces()
    {\bf else if} (request == MCL_SEND_FORCES) {\bf then}
        mimic_send_forces(forces)
    {\bf else if} (request == MCL_...) {\bf then}
            \vdots
    {\bf else if} (request == MCL_EXIT) {\bf then}
        is_last_step = true
    {\bf else}
        {\bf Abort}("Unrecognized MiMiC request!")
    {\bf end}
{\bf end}
\end{Verbatim}
\end{algorithm}

\subsubsection{MD Driver API}

So far, we have described only the client-side API, where MCL is used as a communication channel.
However, a program that acts as a server in terms of communication and as a driver in terms of an MD simulation is also needed.
Unlike the client-side interface, this program has to be linked against the MiMiC library, which then provides access to the MiMiC infrastructure via a simple API.
It consists of a collection of setter-, getter-, and runner-type procedures that can be easily incorporated into a typical MD loop.
At the moment, only a Fortran API is available, but we intend to extend it to other languages.

Let us demonstrate this with a short example of the procedure calls during a single step of an MD simulation.
Once the MD driver updates the particle positions, they are passed to MiMiC using the \verb|set_coordinates()| function.
Then, \verb|run_calculation()| is called at the point in the code where the MD driver typically performs a force calculation.
By invoking this function, MiMiC carries out all the necessary steps to provide the MD driver with the forces on all particles, i.e., it distributes coordinates to individual clients via MCL, issues commands to start energy and force calculations, collects relevant quantities, and computes interactions between the subsystems.
Finally, forces are collected and updated in the MD driver via \verb|get_forces()|, and the simulation proceeds with the rest of the MD step.

\subsection{\label{sec:data_structure_design}Data Structures}

\begin{figure*}[ht]
    \centering
    \includegraphics[width=0.81\linewidth]{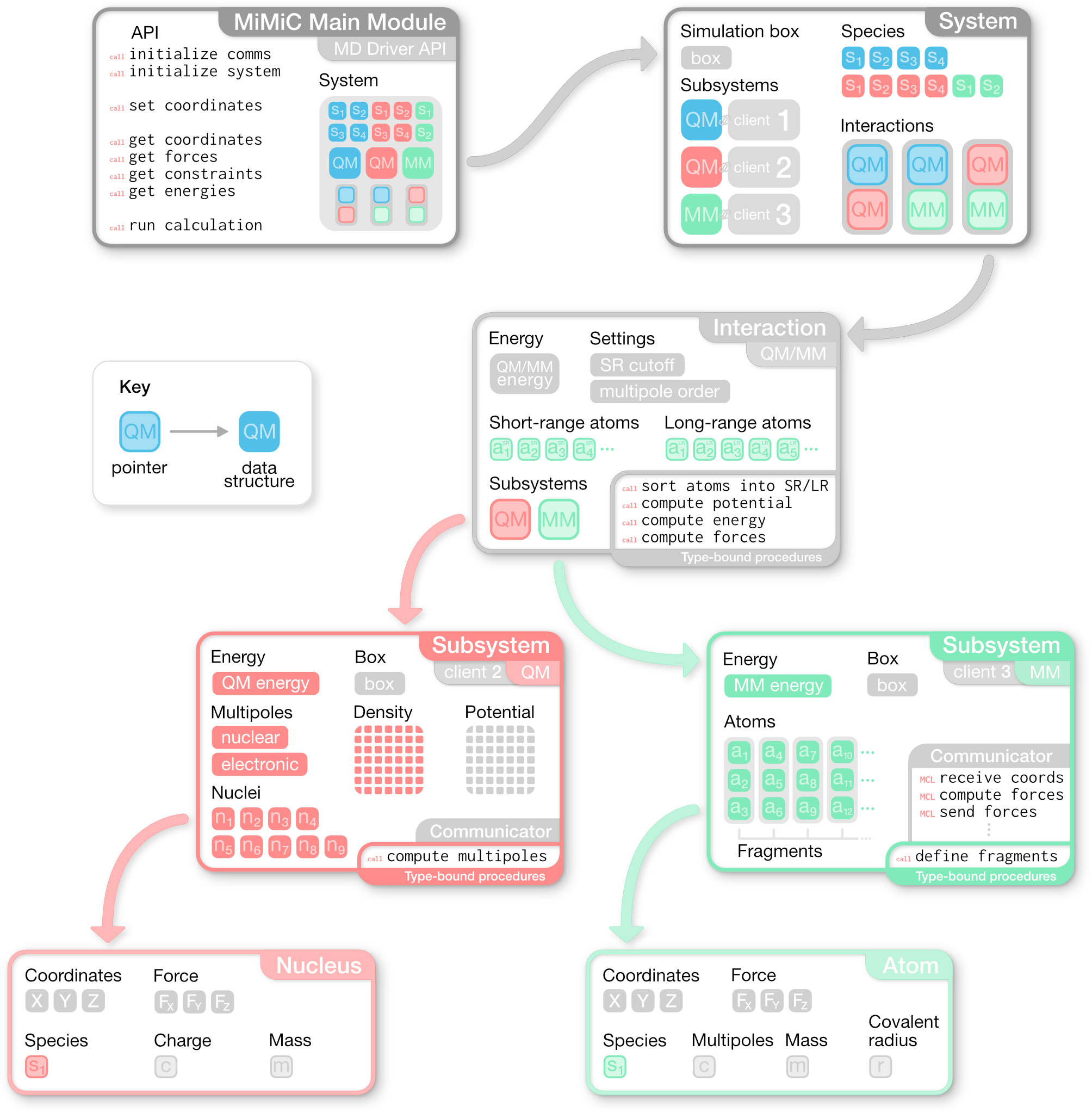}
    \caption{Illustration of the internal structure of the MiMiC main module as a hierarchy of classes.
    Note that this structure is by no means exhaustive, and not all of these features are implemented.}
    \label{fig:sw_data_structs}
\end{figure*}

The MiMiC main module comprises two parts, a public API for MD drivers and a private instance of the \emph{system} class (which here is a Fortran module with a derived type) that contains all the information about the simulation, such as the simulation box parameters, \emph{species} objects with particle properties, and, most importantly, \emph{subsystem} and \emph{interaction} objects.
All system data are thus stored in a hierarchy of objects illustrated in Fig.~\ref{fig:sw_data_structs}.
This data structure ladder plays a crucial role in the modularity of MiMiC, particularly \emph{subsystem} and \emph{interaction} classes, which are the focus of this section.

The fundamental objects in MiMiC are the instances of the \emph{subsystem} class, as each of them is associated with an external program that calculates its properties.
In its essence, the base \emph{subsystem} class contains the energy of a given subsystem, its box parameters (if present), and a set of particles, which can be clustered in fragments.
Because of its association with a client program, it also includes several methods (called type-bound procedures in Fortran) that are used for data transfer and communication via MCL, and to which we collectively refer to as a communicator.
In the initial phase of a simulation, they are used to collect data about the subsystem, like the number of particles and their species, and then, during the MD loop, to distribute particle coordinates to individual clients, instruct them to execute a calculation, and, once it is finished, collect the final energies, forces, and possibly other quantities.
So far, we have implemented two derived \emph{subsystem} classes, namely an MM subsystem and a QM subsystem.
The latter can be straightforwardly used for programs that have a real-space grid representation of the electron density and external potentials, such as those based on plane waves (PWs).
For other basis sets, such as Gaussian-type orbitals (GTOs), it may be necessary to implement additional functionality, such as mapping the electron density onto a real-space grid or including the external potential to the QM Hamiltonian, e.g., from MM point charges.

The \emph{interaction} class groups objects required for the calculation of subsystem interactions.
Here, we use an electrostatic embedding QM/MM scheme as an example to explain the essence of this class.
Among its attributes, we include pointers to the involved subsystems, two arrays referencing the short-range (sr) and long-range (lr) MM atoms, and settings related to the atom sorting into sr/lr subdomains (for the theoretical basis, see Sec.~\ref{sec:elec_emb}).
Finally, its methods constitute an interface for computing different quantities representing the interaction, such as the potential of MM atoms acting on the QM subsystem, QM/MM energy, or QM/MM forces.

Possible extensions of the functionality, for example, to perform MM/CG simulations, can be achieved by a few additions.
First, it is necessary to implement a new \emph{subsystem} class to accommodate a CG client program and all its characteristics.
One can, for instance, introduce a new class for CG beads by extending the base \emph{particle} class, which would then represent the particles in a new CG \emph{subsystem}.
Moreover, a new \emph{interaction} class needs to be implemented to calculate the interactions between the CG beads and MM atoms.
Further extensions to enable, e.g., a three-layer multiscale model, proceed in a similar way.

\subsection{\label{sec:parallelization}Parallelization Strategy}

MiMiC computes various quantities related to the interactions between subsystems, and it is, therefore, assigned its own computational resources.
To avoid introducing a bottleneck, we use parallelization strategies that are flexible enough to run efficiently on diverse hardware architectures, while at the same time being highly scalable.
Therefore, for computationally intensive routines, most data is stored in flat, contiguous arrays, ensuring that the number of nested loops is kept to a minimum.
This allows to potentially tailor parallelization to specific needs that might emerge in the future, such as the addition of new subsystems and interaction types.

In terms of employed parallelization schemes, MiMiC currently features a hybrid MPI-based distributed-memory and OpenMP-based shared-memory approach, which is particularly important in the calculation of electrostatic QM/MM interactions, as described in Sec.~\ref{sec:workflow}.
Within this scheme, the most computationally intensive operation is the numerical integration over the grid storing either an electron density or an external potential, which has to be performed for each MM atom interacting with the QM subsystem.
In a typical application, the number of atoms $N_\mathrm{atom}$ is on the order of $10^4$ to $10^5$, while the number of grid points $N_\mathrm{grid}$ is on the order of $10^6$ to $10^7$.
However, the number of atoms treated explicitly can be reduced to the order of $10^3$ by employing the sr/lr scheme (see Sec.~\ref{sec:elec_emb}), thus resulting in a total operation count proportional to $N_\mathrm{atom} \times N_\mathrm{grid}$, i.e., between $10^{9}$ and $10^{10}$.
In the current implementation, it then reduces to two nested loops: the outer loop over the MM atoms is distributed over MPI processes and both loops are further parallelized employing OpenMP-based shared-memory approaches.

\subsection{\label{sec:mimicpy}Preprocessing Toolkit}

The loose-coupling, minimally invasive strategy adopted by MiMiC requires preparing inputs for multiple external programs that are run concurrently.
Setting up such simulations can sometimes be challenging, especially since complex biomolecular systems are partitioned into two or more subsystems that need to be set up independently.
For this purpose, we provide MiMiCPy\cite{raghavan2023mimicpy}, a Python-based modular preprocessing toolkit for MiMiC that supports various formats of topology, coordinates, and input files and is easily extendable to new ones.

This set of tools facilitates the preparation of input files for MiMiC-based simulations by determining the mappings of the particles that are treated simultaneously by different programs, like QM atoms in a QM/MM simulation, where the MM client is responsible for calculating bonded and van der Waals QM/MM interactions.
MiMiCPy can be used from the command line, but also like a Python library and, notably, we implemented user-friendly plugins for commonly used visualization programs for biomolecular simulations like VMD\cite{humphrey1996vmd} and PyMOL\cite{schrodinger2015pymol}.
This feature is especially helpful for selecting complex QM regions that require careful visual inspection.

MiMiCPy supports three major file types widely used in computational chemistry, namely, topology, coordinates, and input files.
Internally, each of these has a corresponding module that further contains multiple parsers specific for every supported file format.
As parsers are only accessible from within their enclosing modules, each module offers a dedicated wrapper that serves as a unified interface for data exchange.
This means that adding support for a new file format only involves adding a parser class that interfaces with the appropriate wrapper, with no further changes to the core program. This allows for easy extensibility and quick support for new programs added to MiMiC.

\subsection{\label{sec:development}Software Development Practices}

The MiMiC framework is free and open-source software distributed under the GNU Lesser General Public License (LGPL) version 3.0 or later.
The license was chosen to support the free and open-source software movement without imposing severe restrictions on the client programs since proprietary software is still common in the molecular simulation community.
Thus, MCL may be used or integrated into essentially any software without being required to release its source code.
The MiMiC source code is hosted on GitLab\cite{mimic-projects} with releases deposited on Zenodo\cite{olsen_2022_7304688, bolnykh_2023_7497400} using versioning according to the semantic versioning specification\cite{semver}.
We strive to maintain high code quality through code review and continuous integration (CI) pipelines, including unit and integration tests.

Community contributions are highly appreciated, whether through bug reporting, feature suggestions, or direct code contributions.
To facilitate community engagement, we offer user support through a dedicated GitLab issue tracker (\url{https://gitlab.com/mimic-project/user-support}), apart from the project-specific GitLab issue trackers, and a discussion group hosted on Google Groups (\url{https://groups.google.com/g/mimic-project}).
General information about MiMiC, including installation instructions and tutorials, is available on our website \url{https://mimic-project.org}.

\begin{figure*}
\includegraphics[width=\textwidth]{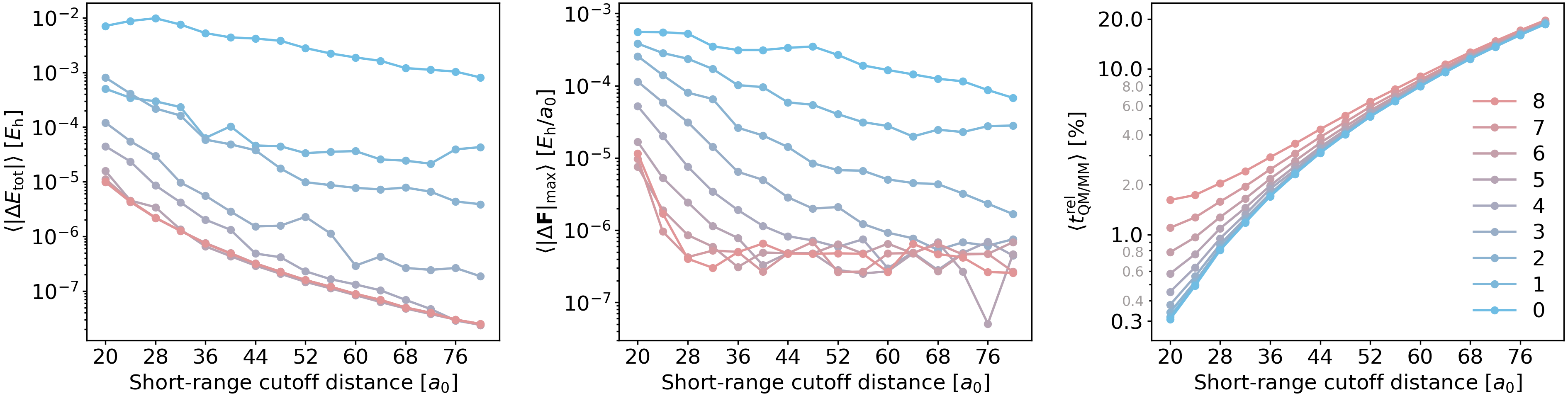}
\caption{Demonstration of the efficiency and tunability of MiMiC's electrostatic embedding scheme for a system consisting of an $n$-butanol molecule (QM subsystem) solvated in acetone (MM subsystem), with a total of 139\:495 atoms.
The plots show the convergence of the total energy (left) and the forces (center), as well as the relative wall times for the calculation of the QM/MM contributions (right) with respect to the short-range cutoff distance and the order of the multipole expansion (line colors).
The points correspond to mean absolute errors in energies, mean of the maximum absolute errors in forces, and mean relative wall times based on single-point calculations on ten snapshots using the full short-range coupling as reference.
Data taken from Ref.~\citenum{olsen2019mimic}.}
\label{fig:sr_lr}
\end{figure*}
\section{\label{sec:workflow}Workflow Examples}
The overall workflow for multiscale simulations varies depending on the system under study and the specific multiscale method employed. 
For multiscale QM/MM MD simulations, we typically start with a system that has been equilibrated at the classical MM level.
This MM equilibration phase involves the usual steps for classical MD simulations, such as energy minimization and equilibration to target conditions (e.g., ambient temperature) in the selected thermodynamic ensemble.
Subsequently, input files for all programs involved in the simulation are prepared.
This includes providing MiMiC with the information needed for client communication, the definition of subsystems, and the computation of subsystem interactions.
The MiMiCPy preprocessing toolkit, described in Sec.~\ref{sec:mimicpy}, is designed to facilitate this part of the workflow\cite{raghavan2023mimicpy}.
Following the preparation, the multiscale simulation is performed.
Like in any classical or ab inito MD workflow, this includes benchmarking to ensure optimal use of computational resources, equilibration, and production simulations.
The generated trajectory data can be post-processed to evaluate various properties of interest.
Although we currently rely on the output files produced by the external programs that can be analyzed using standard tools for MD analysis, we plan to introduce our own output to facilitate and homogenize the post-processing phase.

MiMiC-based multiscale simulations can proceed in several different ways depending on the method and its implementation.
In this section, we describe in detail three such workflows: electrostatic embedding QM/MM in Sec.~\ref{sec:elec_emb}, polarizable embedding QM/MM in Sec.~\ref{sec:pol_emb}, and MTS QM/MM in Sec.~\ref{sec:mts_md}.

\subsection{\label{sec:elec_emb}Electrostatic Embedding QM/MM MD} 
\subsubsection*{Theoretical Basis}
In electrostatic embedding QM/MM, the fixed point charges (or multipoles) of the MM subsystem generate an external field that polarizes the electron density of the QM subsystem.
The QM equations are thus solved in the presence of this field, but since the MM point charges are fixed, the MM subsystem is not explicitly polarized.
A possible way of implementing QM/MM is to employ an additive scheme in which the total time-independent Hamiltonian of the system,
\begin{equation}
    H_\mathrm{tot} = H_\mathrm{QM} + H_\mathrm{MM} + H_\mathrm{QM/MM} \ ,
\end{equation}
yields the energies of the QM and MM subsystems and the interactions between them.

In QM/MM MD, all simulated particles, i.e., QM nuclei and MM atoms, are propagated with classical equations of motion, which implies invoking the Born--Oppenheimer approximation for the QM subsystem.
Consequently, the nuclei move on a potential energy surface corresponding to a particular solution of the electronic QM equations for each nuclear configuration.
In the context of Kohn-Sham density functional theory (KS-DFT), the energy contributions arising from the QM subsystem include the kinetic energy of the electrons and the potential energy arising from classical Coulombic electron--electron, nuclear--electron, and nuclear--nuclear interactions, as well as the exchange--correlation energy.
On the other hand, the MM subsystem energy is determined by evaluating a force field, whose contributions are customarily divided into bonded (stretching, bending, and torsional energies) and non-bonded terms (van der Waals and electrostatic energies).

The QM/MM interaction Hamiltonian,
\begin{equation}
\label{eq:Hqmmm}
H_{\mathrm{QM/MM}} = V^{\mathrm{bonded}}_{\mathrm{QM/MM}} + V^{\mathrm{vdW}}_{\mathrm{QM/MM}}  + V^{\mathrm{es}}_{\mathrm{QM/MM}} \ ,
\end{equation}
includes all interactions between the two subsystems, i.e., bonded and non-bonded van der Waals (vdW) and electrostatic (es) interactions.
The MiMiC framework includes an implementation of the Hamiltonian electrostatic coupling scheme by \citeauthor{laio2002hamiltonian}\cite{laio2002hamiltonian}, where the electrostatic QM/MM interactions are split into short-range (sr) and long-range (lr) contributions,
\begin{equation}
\label{eq:V_es}
V^{\mathrm{es}}_{\mathrm{QM/MM}} = V^{\mathrm{es, sr}}_{\mathrm{QM/MM}} + V^{\mathrm{es, lr}}_{\mathrm{QM/MM}} \ .
\end{equation}

Short-range interactions are calculated using an exact expression
\begin{eqnarray}
\label{eq:v_es_sr}
    V^{\mathrm{es, sr}}_{\mathrm{QM/MM}} = \sum_{i=1}^{N_{\mathrm{MM}}^{\mathrm{sr}}} q_{i}\left( \int \mathrm{d} \mathbf{r}\,   
    T_{\mathrm{mod}}^{(0)}(\mathbf{R}_{\mathrm{MM},i}, \mathbf{r}) \rho(\mathbf{r}) \right. \nonumber \\
    + \left. \sum_{j=1}^{N_{\mathrm{QM}}}  T_{\mathrm{mod}}^{(0)}(\mathbf{R}_{\mathrm{MM},i}, \mathbf{R}_{\mathrm{QM},j})Z_{j} \right) \ ,
\end{eqnarray}
where $q_{i}$ is the point charge of the $i$-th MM atom, $Z_{j}$ is the nuclear (or core) charge of the $j$-th QM atom, $\rho$ is the electron density of the QM subsystem, and 
\begin{equation}
\label{eq:Tmod_0}
T_{\mathrm{mod}}^{(0)}(\mathbf{R}_a, \mathbf{R}_b) =  \frac{r_{\mathrm{cov},\, a }^4 - {\left| \mathbf{R}_b - \mathbf{R}_a\right|}^4}{r_{\mathrm{cov},\, a}^5 - {\left| \mathbf{R}_b - \mathbf{R}_a\right|}^5}
\end{equation}
is the modified zeroth-order interaction tensor, where $r_{\mathrm{cov},\,a}$ corresponds to the covalent radius of atom $a$.
This form of the interaction tensor effectively prevents electron spill-out, an artifact due to the lack of electronic Pauli repulsion between the QM and MM subsystems and the resulting over-polarization of electron density by nearby MM point charges\cite{laio2002hamiltonian}.

\begin{figure*}
\includegraphics[width=0.87\textwidth]{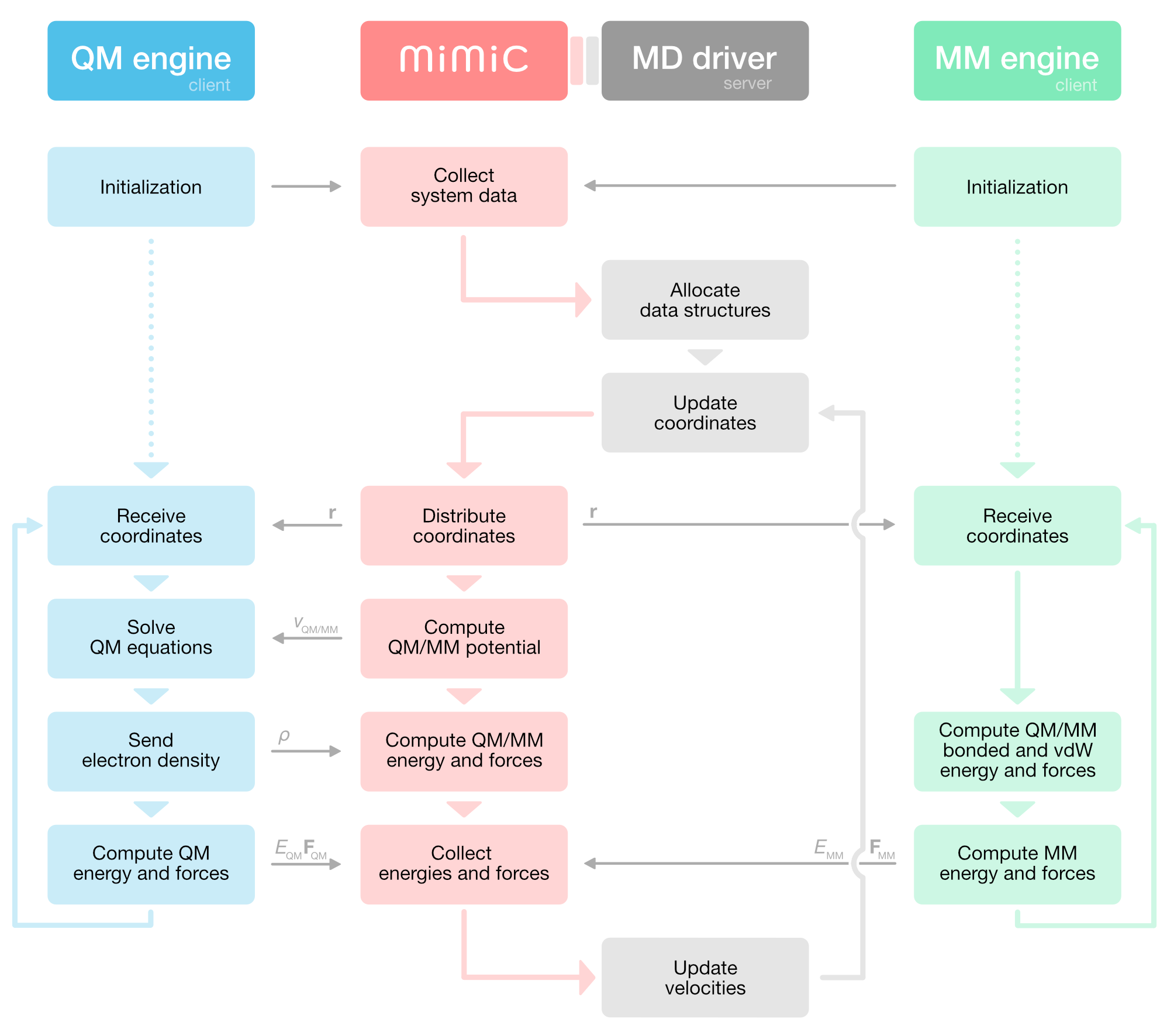}
\caption{Schematic workflow for electrostatic embedding QM/MM MD using MiMiC. Thick lines represent the workflow, while thin horizontal lines represent communication via MCL.}
\label{fig:workflow_elec_emb}
\end{figure*}

Our implementation of the long-range interactions generalizes the original formulation\cite{laio2002hamiltonian} to include open-ended multipole expansions of the QM electrostatic potential\cite{olsen2019mimic}.
For this, the QM multipoles are calculated as
\begin{equation}
\label{eq:multipole_lr}
 M_{\mathrm{QM}}^{[\alpha]} = \int \mathrm{d} \mathbf{r} \,  \rho(\mathbf{r}){\left(\mathbf{r}-\bar{\mathbf{R}}_{\mathrm{QM}}\right)}^\alpha + \sum_{i=1}^{N_{\mathrm{QM}}}Z_{i} {\left(\mathbf{R}_{\mathrm{QM},i}-\bar{\mathbf{R}}_{\mathrm{QM}}\right)}^\alpha \ ,
\end{equation}
where $\bar{\mathbf{R}}_{\mathrm{QM}}$ is the origin of the expansion, usually the centroid of the QM subsystem, and $\alpha = (\alpha_x, \alpha_y, \alpha_z)$ is a multi-index\cite{Raymond1991} indicating the Cartesian component.
In this notation, the sum of components and factorial are defined as 
$\left|\alpha\right|=\alpha_x+\alpha_y+\alpha_z$ and $\alpha!=\alpha_x!\cdot\alpha_y!\cdot\alpha_z!$, respectively.
For more details, we refer to Ref. \citenum{olsen2019mimic}.
The electrostatic energy for the long-range interactions can then be expressed as
\begin{equation}
\label{eq:v_es_lr}
    V^{\mathrm{es, lr}}_{\mathrm{QM/MM}} = \sum_{i=1}^{N_{\mathrm{MM}}^{\mathrm{lr}}}\sum_{\left|\alpha\right|=0}^{A_{\mathrm{QM}}}\frac{(-1)^{\left|\alpha\right|}}{\alpha!} q_i  T^{[\alpha]}(\mathbf{R}_{\mathrm{MM,}i}, \bar{\mathbf{R}}_{\mathrm{QM}}) M_{\mathrm{QM}}^{[\alpha]} \ ,
\end{equation}
where $A_{\mathrm{QM}}$ is the order of the expansion and $T^{[\alpha]}$ is a non-modified interaction tensor, defined through
\begin{equation}
T^{[\alpha]}(\mathbf{R}_a, \mathbf{R}_b) = 
\frac{\partial^{\left|\alpha\right|}}
{\partial R_{b,\,x}^{\alpha_x}
\partial R_{b,\,y}^{\alpha_y}
\partial R_{b,\,z}^{\alpha_z}}
\frac{1}{\left|\mathbf{R}_a - \mathbf{R}_b\right|} \ .
\end{equation}
This approach drastically reduces the overall computational cost by increasing the accuracy of the long-range contributions, allowing the selection of smaller short-range regions whose interactions are substantially more expensive to compute.
In particular, with a proper selection of the short-range cutoff, which dictates how many MM atoms are treated in short- and long-range domains, and the order for the multipole expansion, it is possible to reach the same accuracy as if all the electrostatic interactions were treated exactly, i.e. according to Eq.~\ref{eq:v_es_sr}, yet, at a fraction of the cost, as shown in Fig.~\ref{fig:sr_lr}.

A detailed theoretical treatment of the electrostatic embedding QM/MM scheme implemented using MiMiC can be found in Ref.~\citenum{olsen2019mimic}. It includes explicit expressions for the polarizing potential acting on the QM subsystem due to MM atoms, i.e., the functional derivative of the QM/MM interaction energy with respect to the electron density, and for the forces acting on QM and MM atoms, i.e., the derivatives of the QM/MM interaction energy with respect to the atomic positions.

\subsubsection*{Simulation Workflow}
Electrostatic embedding QM/MM is implemented using MiMiC by interfacing two external programs, a QM and an MM client, to an MD driver. The general simulation workflow is illustrated in Fig.~\ref{fig:workflow_elec_emb}.
After the initialization, consisting of data collection and allocation, the simulation enters the main MD loop.
At each MD step, MiMiC distributes atomic coordinates among the clients, which then concurrently compute energy and forces that are collected by MiMiC and passed to the MD driver to update velocities and subsequently calculate a new set of coordinates.

The input to the MM client includes both QM and MM atoms, but all QM--QM and electrostatic QM--MM interactions must be excluded, as they are treated at the QM and QM/MM levels and thus calculated by a QM client and MiMiC, respectively.
The MM client thus computes the potential energy of the MM subsystem and the remaining QM/MM interaction terms (Eq.~\ref{eq:Hqmmm}), which are described purely at the MM level.
The latter corresponds to van der Waals interactions between QM--MM atoms as well as all bonded terms that cross the QM/MM boundary and involve at least one MM atom.
To saturate broken bonds on the QM side, boundary MM atoms bonded to a boundary QM atom are augmented with monovalent pseudopotentials.

The QM client solves the QM equations in the presence of the external MM potential computed by MiMiC, sends the converged electron density to MiMiC, and then proceeds to compute the QM forces while MiMiC computes the corresponding electrostatic QM/MM contributions.
Finally, MiMiC collects energies and forces from all client programs and forwards them to the MD driver, which then updates velocities and positions starting a new MD step.

\subsection{\label{sec:pol_emb}Polarizable Embedding QM/MM MD}
\subsubsection*{Theoretical Basis}
In electrostatic embedding, the QM system is polarized by the electric field generated by the fixed point charges of the MM subsystem.
As such, this approach does not account for the explicit polarization of the MM subsystem, which is important in many cases and even crucial in some, e.g., in processes involving absorption or emission of photons and electron or proton transfer\cite{Bondanza2020-lp}.
Polarizable embedding QM/MM schemes address this issue by using an MM force field that includes a polarizable component.
For instance, in the frequently used induced point dipole model, the additional dynamic degrees of freedom are modeled using electric polarizabilities that describe an electronic charge distribution distorted by an electric field.
Here, we briefly summarize the theoretical basis for polarizable embedding QM/MM MD using the polarizable AMOEBA\cite{Ponder2010-nr} force field to model the MM subsystem (QM/AMOEBA).
For further details, we refer to Refs.~\citenum{locohybridqmmmmd, Nottoli2020-ty}.

The electrostatic interactions in the AMOEBA force field are modeled using permanent multipoles, up to and including quadrupoles, and induced dipoles, from which electrostatic and polarization QM/MM interaction energies arise.
The former is computed similarly as in electrostatic embedding QM/MM using modified interaction tensors (Eq.~\ref{eq:Tmod_0}).
The polarization energy is expressed in terms of induced dipoles and electric fields, so it is necessary to first determine the induced dipoles through coupled polarization equations.
A characteristic aspect of AMOEBA is its non-variational nature, which stems from the fact that the electric field from the permanent multipoles used to determine the induced dipoles differs from the one used to calculate the polarization energy.
The former is referred to as the \emph{direct field} and the latter as the \emph{polarization field}.
The difference appears due to the use of different scaling factors and exclusion rules for each of these fields.
Consequently, it is not possible to derive the polarization equations by simply requiring the stationarity of the energy functional with respect to the polarizable degrees of freedom.
To avoid the complexity of non-variational optimization, a variational formulation has been adopted, resulting in two sets of polarization equations and, therefore, two sets of induced dipoles\cite{locohybridqmmmmd, Nottoli2020-ty}.

For the polarization QM/MM interaction energy in the QM/AMOEBA model, we follow the formulation by \citeauthor{locohybridqmmmmd}\cite{locohybridqmmmmd} in which it is expressed as
\begin{eqnarray}
\label{eq:V_ind}
V^{\mathrm{pol}}_{\mathrm{QM/MM}}
= 
\frac{1}{2} \bm{\mu}_\mathrm{d}^\mathrm{T} \mathbf{A}\boldsymbol{\mu}_\mathrm{p}
- \frac{1}{2} \left( \boldsymbol{\mu}_\mathrm{p}^\mathrm{T} \mathbf{E}_\mathrm{d} + \boldsymbol{\mu}_\mathrm{d}^\mathrm{T} \mathbf{E}_\mathrm{p} \right) \nonumber \\
- \frac{1}{2} (\boldsymbol{\mu}_{\mathrm{p}}+\boldsymbol{\mu}_{\mathrm{d}})^{\mathrm{T}} \mathbf{E}_\mathrm{QM} \ ,
\end{eqnarray}
where $\boldsymbol{\mu}_\mathrm{d}$ and $\boldsymbol{\mu}_\mathrm{p}$ are the AMOEBA induced dipoles due to the direct field, $\mathbf{E}_\mathrm{d}$, and the polarization field, $\mathbf{E}_\mathrm{p}$, respectively, as well as the field from the electrons and nuclei in the QM subsystem, $\mathbf{E}_\mathrm{QM}$, and $\mathbf{A}$ is the classical response matrix, which contains inverse polarizabilities on the diagonal blocks and Thole damped interaction tensors\cite{Thole1981-po, Ponder2010-nr} on the off-diagonal blocks.
This contribution thus includes the energies from the polarization of the MM subsystem and the mutual polarization interactions between the QM and MM subsystems.

On top of providing the MM energy and forces as in the electrostatic embedding workflow, the MM client now also has to solve two sets of polarization equations
\begin{subequations}
    \label{eq:dips}
    \begin{align}
        \mathbf{A}\boldsymbol{\mu}_\mathrm{d} &= \mathbf{E}_\mathrm{d} + \mathbf{E}_\mathrm{QM} \ ,\label{eq:d_dips} \\
        \mathbf{A}\boldsymbol{\mu}_\mathrm{p} &= \mathbf{E}_\mathrm{p} + \mathbf{E}_\mathrm{QM} \ , \label{eq:p_dips}
    \end{align}
\end{subequations}
to obtain the induced dipoles.
In polarizable embedding QM/MM models, the induced dipoles depend on the QM electric field as seen in Eqs.~\ref{eq:dips} and thus have to be solved iteratively until self-consistency for electron density and induced dipoles is reached.
This ensures fully self-consistent polarization between the QM and MM subsystems.

\begin{figure*}[t]
\includegraphics[width=0.615\textwidth]{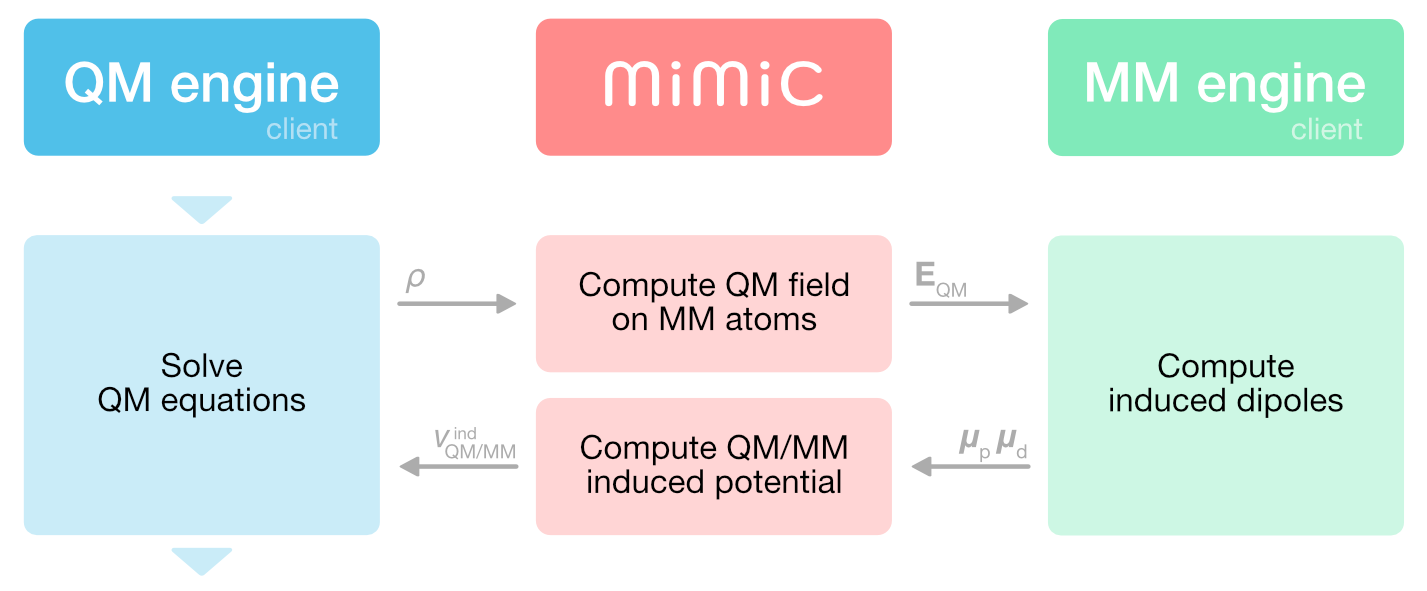}
\caption{Schematic workflow of the iterative cycle used in polarizable embedding QM/MM through MiMiC. Thin horizontal lines represent communication via MCL.}
\label{fig:workflow_pol_emb}
\end{figure*}
\subsubsection*{Simulation Workflow}
Enabling fully polarizable QM/MM MD simulations using MiMiC requires modifications of the algorithm used to solve the QM equations.
Achieving full self-consistency entails iteratively adjusting the induced dipoles in the MM subsystem and the electron density in the QM subsystem until both are converged with respect to the field generated by the other subsystem.
This iterative process is characteristic of most polarizable embedding QM/MM models.
The general workflow of a polarizable embedding QM/MM MD simulation using MiMiC is quite similar to the one for electrostatic embedding QM/MM shown in Fig.~\ref{fig:workflow_elec_emb}.
It requires extending the \emph{Solve QM equations} step with the iterative cycle illustrated in Fig.~\ref{fig:workflow_pol_emb}.

The extended workflow can be summarized as follows.
MiMiC receives an electron density from the QM client and uses it to compute the QM electric field at all MM atoms.
The resulting field is sent to the MM client, which solves the polarization equations (Eqs.~\ref{eq:dips}).
The induced dipoles are then passed to MiMiC, which computes the corresponding induced energy and potential and relays them to the QM client, which proceeds with the next iteration.
This procedure is repeated until the convergence criteria for the electron density and induced dipoles are met, thus solving the QM equations fully self-consistently with the polarizable MM subsystem.

The presented workflow is unique for the AMOEBA force field, but other polarizable models can be straightforwardly enabled with minimal or no changes in MiMiC.

\subsection{\label{sec:mts_md}Multiple Time Step QM/MM MD}
\subsubsection*{Theoretical Basis}
MTS algorithms reduce the cost of MD simulations by separating forces into \emph{fast} and \emph{slow} components and integrating the latter less frequently, i.e., with a larger time step.
Assuming a time-scale separation between these components, it is possible to split the classical Liouville operator of an $N$ particle system as
\begin{align}
    iL
    & =
    \sum_{j=1}^{N} \left(
    \dot{\mathbf{q}}_j \cdot \nabla_{\mathbf{q}_j}
    +
    \mathbf{F}^{\mathrm{fast}}_j \cdot \nabla_{\mathbf{p}_j}
    + 
    \mathbf{F}^{\mathrm{slow}}_j \cdot \nabla_{\mathbf{p}_j}
    \right) \nonumber \\
    & = 
    i L_\mathrm{q} + i L_\mathrm{p}^{\mathrm{fast}} + i L_\mathrm{p}^{\mathrm{slow}} \ .
\end{align}
As introduced by \citeauthor{tuckerman1992reversible}\cite{tuckerman1992reversible}, this allows the Trotter factorization of the classical time propagator 
\begin{equation}
\label{eq:MTS_propagator}
    e^{iL\Delta t} =
    e^{iL_\mathrm{p}^{\mathrm{slow}} \left(\frac{\Delta t}{2}\right)} \left[
    e^{iL_\mathrm{p}^{\mathrm{fast}} \left(\frac{\Delta t}{2n}\right)} \,
    e^{iL_\mathrm{q} \left(\frac{\Delta t}{n}\right)} \,
    e^{iL_\mathrm{p}^{\mathrm{fast}} \left(\frac{\Delta t}{2n}\right)}
    \right]^n
    e^{iL_\mathrm{p}^{\mathrm{slow}} \left(\frac{\Delta t}{2}\right)} \ ,
\end{equation}
where the terms of higher than second order are neglected.
The operators in square brackets correspond to $n$ inner propagation steps for the fast components with a time step $\Delta t/n$, and the ones outside to an outer step for the slow components, thus completing a full time step $\Delta t$.

The MTS approach can also be employed in Born--Oppenheimer MD, for which it can be reformulated in terms of higher and lower levels of theory owing to the observation that the differences between forces calculated with lower- or higher-level methods vary smoothly and slowly over time\cite{steele2013communication, liberatore2018versatile, mouvet2022recent}.
This is because quantum chemical methods usually only differ in the treatment of electron correlation (and exchange in DFT).
In this QM$^{\text{low}}$--QM$^{\text{high}}$ MTS approach, the forces calculated by a low-level method (QM$^{\text{low}}$) are used as fast components, while the slow components are represented by a correction from a higher-level method (QM$^{\text{high}}$)
\begin{subequations}
\begin{align}
        \mathbf{F}^{\mathrm{fast}}\; &= \:\mathbf{F}^{\mathrm{low}} \ , \label{eq:F_fast}\\
        \mathbf{F}^{\mathrm{slow}}   &= \: \mathbf{F}^{\mathrm{high}} - \,\mathbf{F}^{\mathrm{low}} \ . \label{eq:F_slow}
\end{align}
\end{subequations}
Therefore, as QM$^{\text{high}}$ forces need to be calculated only so often, adopting this scheme effectively reduces the overall computational cost of high-level methods by at least a factor of five\cite{liberatore2018versatile}.

In the context of electrostatic embedding QM/MM, the scheme introduced above can be implemented by choosing
\begin{subequations}
\label{eq:mts_F_qmmm}
\begin{align}
        \label{eq:Flow_qmmm}
        \mathbf{F}^{\mathrm{low}} \:&=\: \mathbf{F}^{\mathrm{low}}_{\mathrm{QM}} + \mathbf{F}^{\mathrm{low}}_{\mathrm{QM/MM}} + \mathbf{F}^{\mathrm{bonded+vdW}}_{\mathrm{QM/MM}} +
        \mathbf{F}_{\mathrm{MM}} \ , \\
        \label{eq:Fhigh_qmmm}
        \mathbf{F}^{\mathrm{high}} &=\: \mathbf{F}^{\mathrm{high}}_{\mathrm{QM}} + \mathbf{F}^{\mathrm{high}}_{\mathrm{QM/MM}} + \mathbf{F}^{\mathrm{bonded+vdW}}_{\mathrm{QM/MM}} +
        \mathbf{F}_{\mathrm{MM}} \ ,
\end{align}
\end{subequations}
where $\mathbf{F}^{\mathrm{low}}_{\mathrm{QM}}$ and $\mathbf{F}^{\mathrm{low}}_{\mathrm{QM/MM}}$ are the low-level forces arising from the QM and QM/MM terms, respectively, just as  $\mathbf{F}^{\mathrm{high}}_{\mathrm{QM}}$ and $\mathbf{F}^{\mathrm{high}}_{\mathrm{QM/MM}}$ represent their high-level counterparts.
Note that in the calculation of Eq. \ref{eq:F_slow}, within an electrostatic embedding QM/MM scheme, the high- and low-level bonded and vdW QM/MM forces as well as the MM forces cancel out.
The latter implies that the MM atoms always evolve with the inner time step $\Delta t / n$.

\begin{figure*}[ht]
\includegraphics[width=0.87\textwidth]{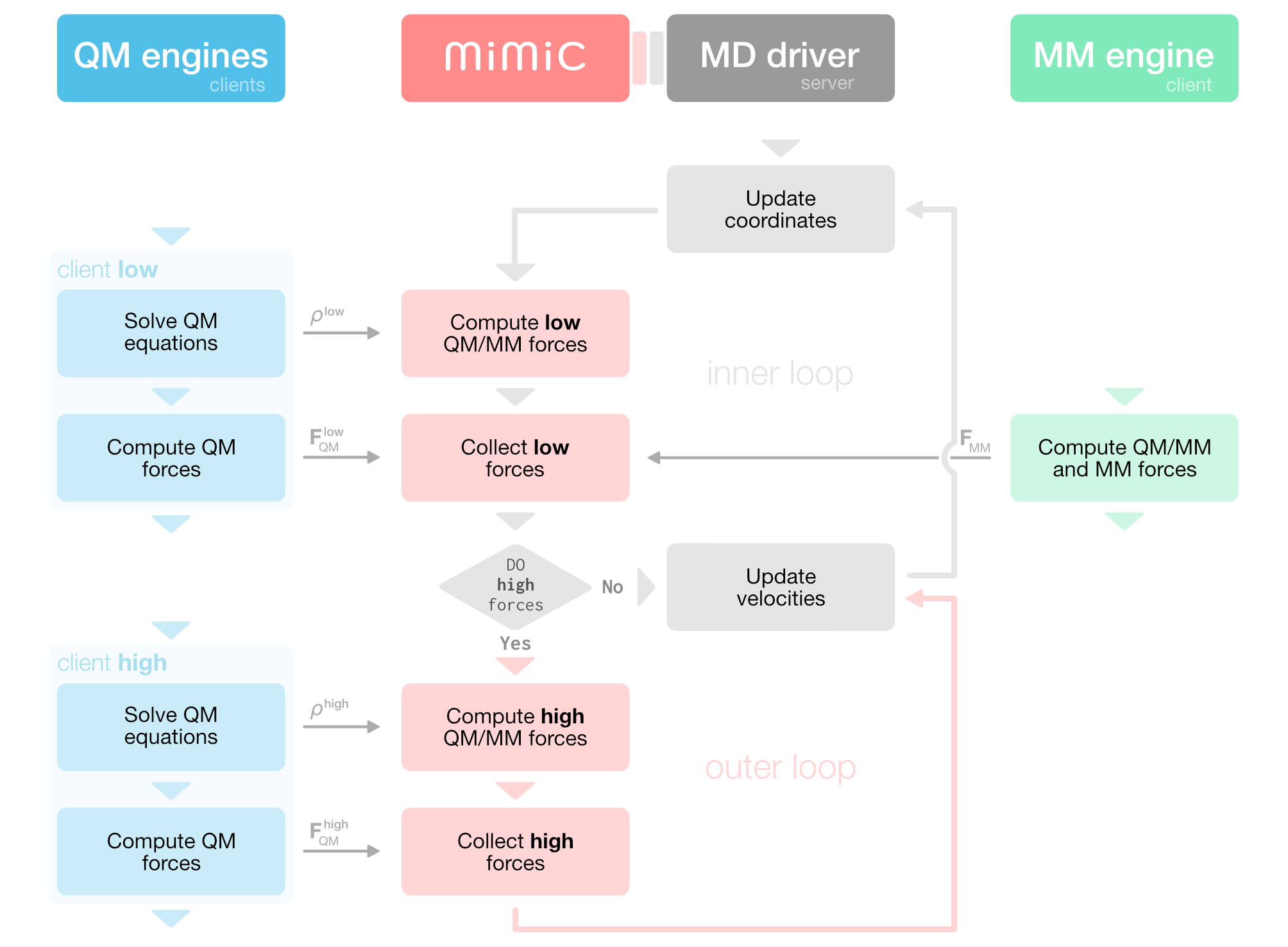}
\caption{Schematic workflow for MTS QM/MM MD using MiMiC.}
\label{fig:workflow_MTS}
\end{figure*}

MTS-generated trajectories are, by construction, at the high level of theory with the outer time step $\Delta t$, regardless of the low-level method used.
The choice of the latter only determines the magnitude of the correction and, hence, the oscillations of the slow force component over time.
These oscillations dictate the maximal ratio between the inner and outer steps that can be applied to generate stable dynamics.

\subsubsection*{Simulation Workflow}

Performing a full MTS MD step requires executing $n$ inner steps at a lower level of theory (as shown in Eq.~\ref{eq:MTS_propagator}) followed by a single high-level outer step.
Fig.~\ref{fig:workflow_MTS} depicts a schematic workflow of an MTS QM/MM MD algorithm using MiMiC.
At every inner step, MiMiC collects the low-level forces from the QM and MM clients and computes the low-level QM/MM force contributions (Eq.~\ref{eq:Flow_qmmm}), which together form the fast forces (Eq.~\ref{eq:F_fast}).
If the current step is not a multiple of $n$, velocities and coordinates are updated at the low level with the time step $\Delta t / n$; else, at every $n$-th step, the outer loop is also undertaken.
In the outer loop, MiMiC additionally collects the high-level QM forces, computes the high-level QM/MM force contributions (Eq.~\ref{eq:Fhigh_qmmm}), and then proceeds to calculate the slow forces (Eq.~\ref{eq:F_slow}).
These are used by the MD driver to update the velocities, thus completing a full MD cycle with the time step $\Delta t$ at the high level of theory. 
Note that with MiMiC, it is straightforward to use two different QM client programs for the low- and high-level forces.

\section{\label{sec:features}Features}
The first development version of the MiMiC framework was released in 2022 on the occasion of the CECAM Flagship School \emph{Multiscale Molecular Dynamics with MiMiC}\cite{mimic-school} held at the CECAM Headquarters at EPFL in Lausanne, Switzerland.
It implements the DFT-based electrostatic embedding QM/MM MD method illustrated in Sec.~\ref{sec:elec_emb}, using GROMACS~\cite{abraham2015,gromacs2015} as the MM client and CPMD~\cite{cpmd_free} as the QM client and MD driver.
Furthermore, it also supports MTS QM/MM MD (Sec.~\ref{sec:mts_md}) through CPMD, and, through the PLUMED library, QM/MM-based enhanced sampling simulations and free energy calculations\cite{Bonomi2009a, PLUMED2, Colon-Ramos2019}.

The \emph{CPMD} program\cite{cpmd_free}, distributed under the MIT license since 2023, presents a number of interesting features that made it our initial choice as a QM client and MD driver.
It is a highly parallelized PW/pseudopotential implementation of KS-DFT that provides access to a wide variety of exchange--correlation functionals\cite{Bircher2018, Bircher2018shedding, Villard2024}, together with a computationally efficient treatment of hybrid functionals\cite{Weber2014, Bircher2018exploring}, and second-order Møller--Plesset perturbation theory (MP2)\cite{Bircher2020}.
Moreover, it supports density functional perturbation theory \cite{putrino2000generalized}, linear response \cite{hutter2003excited} and real-time time-dependent DFT (TDDFT), and the restricted open-shell KS (ROKS) formalism\cite{frank1998molecular}.
As an MD driver, CPMD offers a broad range of features, as it implements both Born--Oppenheimer and Car--Parrinello MD, as well as nonadiabatic MD, using either the Ehrenfest scheme\cite{tavernelli2005molecular} or Tully's fewest-switches surface hopping\cite{tapavicza2007trajectory}.
The latter can also be performed in the presence of an explicit external field\cite{tavernelli2010mixed} with the possibility of performing local control simulations\cite{curchod2011local}.
In addition, nuclear quantum effects can be included within a path-integral formalism\cite{marx1996ab}.
CPMD also provides access to various accurate and efficient thermostats\cite{berendsen1984molecular, nose1984unified, nose1984molecular, hoover1985canonical, ceriotti2010colored} and an MTS acceleration algorithm via an extension of the microcanonical reversible reference system propagation algorithm (rRESPA) by \citeauthor{tuckerman1992reversible}\cite{tuckerman1992reversible}, which also allows performing NVT-MTS propagation\cite{martyna1996explicit, liberatore2018versatile}.

Our initial choice for an MM client was \emph{GROMACS}\cite{gromacs2015}, which is a free and open-source software suite for classical MD, distributed under LGPL version 2.1, whose developers aimed to provide the highest possible performance and efficiency on any hardware\cite{abraham2015}.
It implements a native heterogeneous parallelization setup using both CPUs and GPUs with the possibility to offload all force components to GPUs in single-precision calculations.
The latest versions\cite{pall2020heterogeneous} also introduced new direct GPU–GPU communication and extended GPU integration, enabling excellent performance across multiple GPUs and efficient multi-node parallelization.
Moreover, GROMACS supports a large variety of popular biomolecular force fields, including AMBER\cite{ponder2003force}, CHARMM\cite{brooks2009charmm}, GROMOS\cite{scott1999gromos}, and OPLS\cite{robertson2015improved}, and provides a wide array of well-documented pre- and post-processing tools.
New users can also benefit from an extensive range of tutorials, documentation, and an active user community.

\begin{figure}[b]
\centering
\includegraphics[width=0.48\textwidth]{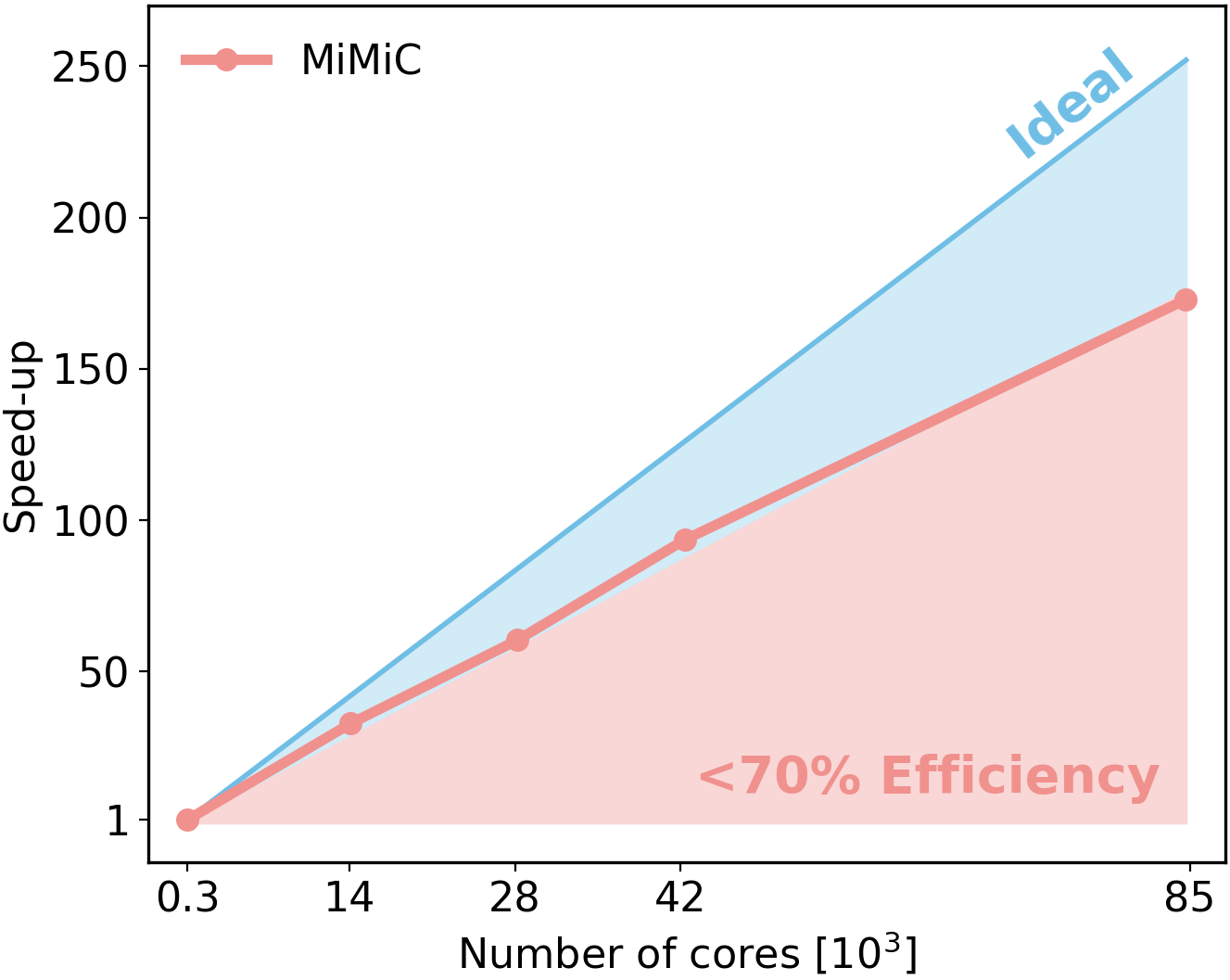}
\caption{Parallel scalability of a MiMiC-based QM/MM implementation using GROMACS as the MM client and CPMD as the QM client and MD driver.
The system comprised 130\:828 atoms, of which 142 were treated at the QM level using the hybrid B3LYP functional.
The speed-up is calculated based on the CPU time required for one MD step, normalized to a reference run using seven nodes.
Data from \citeauthor{drugdesign2023}\cite{drugdesign2023} who ran the simulations on the CPU partition of the JUWELS cluster at the Jülich Supercomputing Centre.}
\label{fig:idh1_scaling}
\end{figure}

Combining the benefits of these external programs, the initial MiMiC-based QM/MM implementation achieved high performance in MD simulations of large and complex biomolecular systems.
A recent MiMiC-based QM/MM MD study\cite{drugdesign2023} demonstrated strong scaling up to 84\:672 CPU cores of the JUWELS cluster~\cite{alvarez2021} at the Jülich Supercomputing Centre (corresponding to $\sim$77\% of the available nodes) while maintaining a parallel efficiency of about 70\% (see Fig.~\ref{fig:idh1_scaling}).
The study compared the attainable throughput for QM/MM MD of two systems containing different QM subsystem sizes.
It focused on the solvated isocitrate dehydrogenase-1 (IDH1) enzyme (130\:828 atoms), which contained 142 atoms in its QM subsystem, and the p38$\alpha$ mitogen-activated protein kinase (169\:550 atoms), which contained a smaller QM subsystem of 46 atoms. 
For both systems studied, two simulations were launched in which either the BLYP or the hybrid B3LYP exchange--correlation functional was used to treat the QM subsystem.
The QM/MM MD simulation of the system with a larger QM subsystem (IDH1) reached a throughput of 0.74 ps/day on 84\:672 CPU cores while employing the B3LYP hybrid functional.
In comparison, the simulation of the system with the smaller QM subsystem (p38$\alpha$) achieved a throughput of 4.8 ps/day with B3LYP and maintained a parallel efficiency above 70\% up to 12\:288 CPU cores.
Furthermore, using the BLYP functional, the large QM subsystem simulation realized a throughput of 5.4 ps/day using 5\:184 CPU cores, while the smaller QM subsystem saw a throughput of 21 ps/day using just 384 CPU cores.
These simulations demonstrate that the current MiMiC-based QM/MM implementation allows for subnanosecond-scale QM/MM MD thanks to the unhindered scaling capabilities of CPMD.
Given sufficient computational resources, converged free energy values of biological systems similar to those highlighted here could be attained.
For further examples of biomolecular system studies with MiMiC, including more information about IDH1, see Sec.~\ref{sec:applications}.

\section{\label{sec:upcoming_clients}Upcoming Features and Client Programs}
An important advantage of MiMiC is the possibility to implement new features directly within the framework itself, making them client-agnostic, and thus available to use together with every supported client program.
As a number of development endeavors are currently in progress, here we describe a few features that are expected to be available soon.

These efforts include the implementation of dynamically generated restrained electrostatic potential (D-RESP) charges\cite{laio2002d}, extending the original approach to also fit higher-order atom-centered multipoles.
Such an extension will also make it possible to implement different coupling schemes for electrostatic interactions, for example, by introducing a three-level approach\cite{laio2004variational}, which would significantly increase electrostatic embedding efficiency, particularly for elongated QM subsystems.
In addition, the implementation of D-RESP charges will also serve as an on-the-fly analysis tool for tracking ongoing changes in the electronic structure and will also enable charge-driven enhanced sampling\cite{sulpizi2005electron}.
Another important addition, building on D-RESP, will be the development of QM/MM force-matching schemes, such as those in Refs.~\citenum{Maurer2007} and \citenum{Doemer2014}, in which the total forces on the QM atoms, as well as the QM electrostatic potential and field, are extracted during a QM/MM MD simulation, for a set of sampled configurations.
These quantities are then used to parameterize standard or polarizable biophysical force fields for the QM subsystem.
Such QM-matched force fields can be employed to achieve more extensive sampling in a fully classical MD simulation of the whole system.

A different extension of capabilities in MiMiC will be based on enhancing the MTS workflow illustrated in Sec.~\ref{sec:mts_md}, which unlocks a plethora of possible combinations of different high- and low-level methods, including wave function, DFT, semiempirical, force field, and ML models\cite{mouvet2022recent}.
A future release of MiMiC will also include an adaptive ML-accelerated MTS QM/MM scheme in which the ML model is retrained on the fly whenever the system accesses new regions of configurational space (e.g., during a chemical reaction) where the original ML model has insufficient predictive power.
In addition, this adaptive scheme allows for a dynamic training set build-up during the run, thus circumventing the need for extensive and computationally demanding a priori generation of high-level reference data. 

Regarding the development of new interfaces to external client programs, we currently focus on achieving higher performance and higher accuracy for both QM and MM subsystems.
Thus, we strive to enable large-scale QM/MM MD simulations on current and future (pre-)exascale supercomputers by coupling to highly scalable GPU-enabled QM client programs, namely, CP2K~\cite{Kuhne2020}, DFT-FE~\cite{Motamarri2020, Das2022}, and Quantum ESPRESSO~\cite{Giannozzi2009, Carnimeo2023}, as well as making available more accurate wave-function methods, as implemented in CP2K and CFOUR~\cite{matthews2020coupled, cfourcode}.
As a further step towards higher accuracy, we are implementing a fully polarizable QM/MM method (see Sec.~\ref{sec:pol_emb}).
The pilot implementation will use CPMD as the MD driver and QM client and the high-performance Tinker-HP~\cite{thp1, thp2} package as the MM client.
However, the implementation in MiMiC will be general, and support for other clients, such as OpenMM~\cite{eastman2013openmm, eastman2017openmm, eastman2023openmm}, will follow soon after.

To achieve these goals, work on interfaces with several client programs is currently underway.
In the remainder of this section, we review their distinctive features, which make them ideal candidates for integration into the framework.
As new programs will be coupled to MiMiC, the pre-processing toolkit MiMiCPy will also be augmented accordingly.

\emph{CP2K}\cite{Kuhne2020} is a free and open-source software package, distributed under the GNU General Public License (GPL) version 2.0 or later, for simulations of solid-state and (bio)molecular systems, giving access to a broad range of electronic structure methods.
It focuses on ensuring excellent performance, thanks to algorithms developed for modern HPC systems, also including GPU support.
In particular, the electronic structure module Quickstep\cite{krack2004quickstep, vandevondele2005quickstep, Kuhne2020} provides an efficient infrastructure of integral routines and optimization algorithms for a wide spectrum of methods, including orbital-free DFT, KS-DFT, and MP2.
Quickstep is based on mixed Gaussian and PW (GPW) basis sets and their augmented extension (GAPW), where auxiliary PW basis sets are used within a GTO scheme.
Recently, CP2K's linear scaling DFT implementation was used for calculations on the SARS-CoV-2 spike proteins with almost 83 million atoms, for which it scaled up to 4\:400 NVIDIA A100 GPUs with a parallel efficiency of over 80\%\cite{schade2023breaking}.

\emph{DFT-FE}~\cite{Motamarri2020, Das2022} is a recently developed free and open-source code, distributed under LGPL version 2.1 or later, that implements real-space KS-DFT based on a spatially adaptive and systematically convergent higher-order spectral finite-element (FE) basis.
It accommodates both norm-conserving pseudopotentials and all-electron calculations, with ongoing implementation efforts to support ultrasoft projector-augmented-wave pseudopotentials.
DFT-FE enables fast and accurate large-scale DFT calculations on generic materials systems involving many tens of thousands of electrons on both many-core CPU and hybrid CPU-GPU architectures.
The good parallel scalability and efficient use of GPUs have enabled systematically convergent DFT calculations with low wall times, e.g., an ab initio MD step of systems containing 10\:000-20\:000 electrons can be completed in wall times of <1 minute~\cite{Das2022}.
In addition to the support of NVIDIA GPUs through CUDA, DFT-FE's GPU porting layer was recently extended to support AMD GPUs through HIP, with its exascale capability recently demonstrated through scaling up to 8\:000 GPU nodes (32\:000 AMD MI250X GPUs) of the Oak Ridge Leadership Computing Facility Frontier exascale supercomputer~\cite{das2023GordonBell}.

\emph{Quantum ESPRESSO}\cite{Giannozzi2020} is an integrated and free suite of programs, distributed under GPL version 2.0 or later, designed for first-principles electronic structure calculations and materials modeling at the nanoscale.
In recent years, it has been refactored and reorganized in multiple layers of code\cite{Carnimeo2023}, consisting of loosely coupled modules and libraries, which are tailored to be independently extensible and maintainable.
Recent versions consist of a core computational layer for DFT simulations based on PWs and pseudopotentials, containing a quantum-engine package to solve self-consistent KS equations for periodic systems (PWscf).
To achieve high performance on different architectures, Quantum ESPRESSO features a layered, fine-tuneable parallelization scheme based on MPI and OpenMP, as well as GPU support.
Porting to GPU has been achieved through CUDA Fortran and, to an increasing degree, directive-based programming models (OpenACC and OpenMP) for offloading to heterogeneous platforms, making it as architecture-agnostic as possible.
The optimal combination of all available parallelism levels currently allows the Quantum ESPRESSO suite to achieve a substantial increase in performance and efficiency\cite{Carnimeo2023}.

\emph{CFOUR}\cite{cfourcode, matthews2020coupled} is a program package, distributed freely for academic users, specialized in high-accuracy wave function-based methods, with applications in thermodynamic, spectroscopic, and kinetic phenomena of small to medium-sized molecules.
It implements Hartree--Fock (HF), multiconfigurational complete active space (CAS) SCF, and a multitude of single-reference post-HF methods.
These include Møller--Plesset perturbation theory up to fourth-order and coupled cluster (CC) approximations ranging from CC2 to CCSDTQ.
A particular strength of CFOUR is its ability to compute analytic energy derivatives for most of the implemented methods.
The program can also treat excited states via the equation-of-motion CC theory (EOM-CC) with analytic energy derivatives for some of the approximations.
An initial interface between CFOUR and MiMiC was presented recently\cite{cfour}, showcasing MP2, CCSD(T), and CASSCF within both ab initio QM and electrostatic embedding QM/MM MD.
This work also demonstrated the utility of the QM$^{\text{low}}$--QM$^{\text{high}}$ MTS implemented in CPMD by performing BLYP--CCSD(T) MTS MD simulations.
The next steps for this interface include further development of the hierarchical QM$^{\text{low}}$--QM$^{\text{high}}$ MTS schemes, also exploiting the ML acceleration capabilities.

\emph{Tinker-HP}\cite{thp1, thp2} is a powerful, massively MPI parallelized MD package, distributed freely for academic institutions, that supports simulations using standard fixed point-charge and advanced polarizable force fields.
One of its core strengths is the ability to run long timescale simulations of very large biomolecular systems using polarizable force fields.
This is facilitated by a parallel 3D spatial decomposition tailored for point-dipole polarizable models and efficient iterative and non-iterative polarization solvers.
Additionally, thanks to recent expansion to GPUs through OpenACC and CUDA, Tinker-HP performs efficiently on both CPU and GPU architectures\cite{thp2}.

\emph{OpenMM}\cite{eastman2013openmm, eastman2017openmm, eastman2023openmm} is a free and open source classical MD package, distributed under LGPL version 3.0 and MIT licenses, designed to be simple and easy to use, with a focus on extensibility.
It offers many common classical force fields, and users can easily add support for new types of force fields via plugins in a modular and extensible way.
Notable plugins include polarizable force fields (AMOEBA\cite{Ponder2010-nr} and Drude oscillator\cite{lamoureux2003modeling} models) and, more recently, ML potentials\cite{eastman2023openmm}.
Moreover, it also provides several custom force classes that allow users to specify arbitrary algebraic expressions for force components.
It performs efficiently on different hardware platforms, especially GPU-based architectures, for both single- and double-precision calculations.

\begin{figure}
\centering
\includegraphics[width=0.469\textwidth]{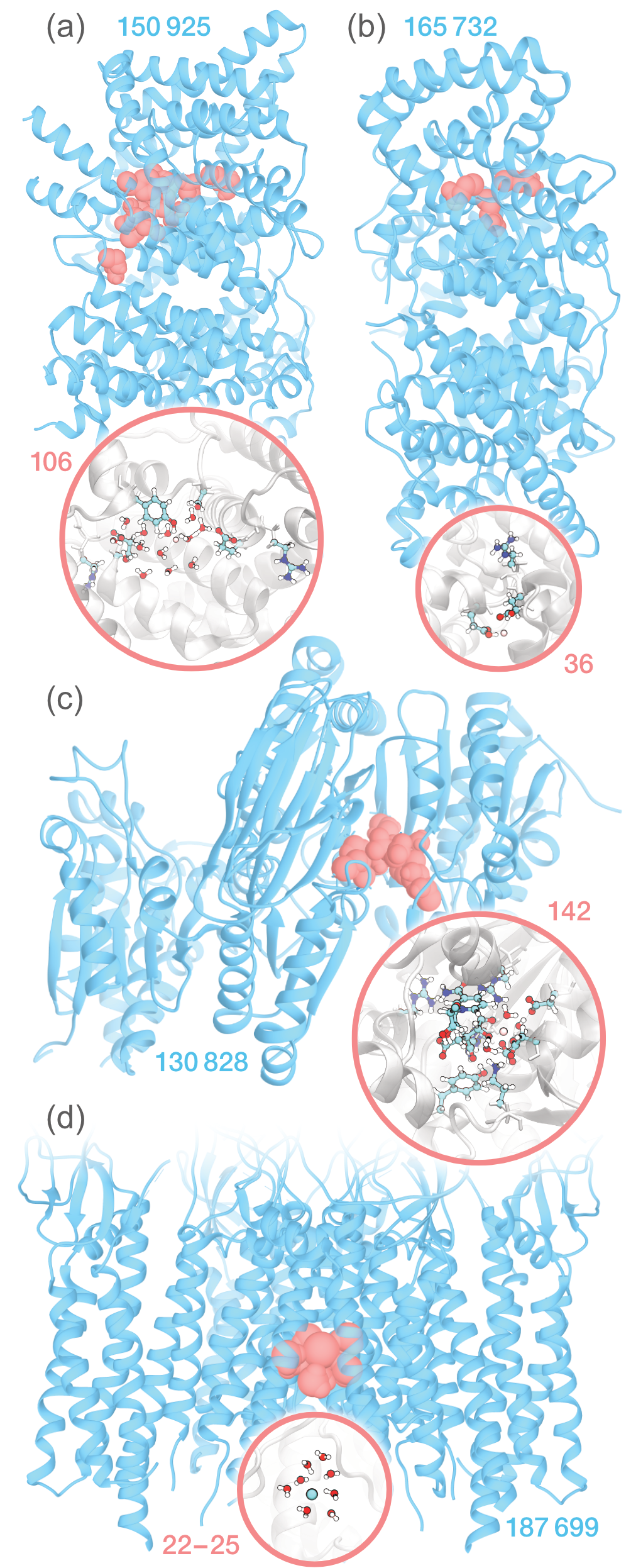}
\caption{Biologically relevant systems simulated using MiMiC: (a) the CLC-ec1 (Cl\textsuperscript{--}/H\textsuperscript{+}) antiporter protein \cite{mimic_jctc_hpc} (b) the CLC\textsuperscript{F} antiporter protein \cite{clc2021}; (c) the IDH1 enzyme \cite{drugdesign2023} and (d) the AMPA receptor\cite{glutamate2023}. QM subsystems are shown within the circular insets, with coral and blue numbers representing the number of QM and MM atoms, respectively.}
\label{fig:biosystems}
\end{figure}
\section{\label{sec:applications}Illustrative Applications}
Recent applications of MiMiC to study biological systems demonstrate its utility in multiscale modeling\cite{mimic_jctc_hpc}.
All results of the studies presented in this section were acquired using the development version of MiMiC that features an electrostatic embedding QM/MM implementation using CPMD and GROMACS.

One exemplary class of systems that has been extensively investigated through MiMiC are transporter proteins, which play a key role in cellular homeostasis. 
These systems are inherently multiscale in nature: transporter proteins are embedded in lipid bilayers, resulting in a large system, but they are often involved in the movement of ions, for which an accurate QM description is required. 
The first study focused on the CLC-type Cl\textsuperscript{$-$}/H\textsuperscript{+} transporter protein from \textit{Escherichia coli}, CLC-ec1, embedded in a solvated lipid bilayer\cite{clc2020} (Fig. \ref{fig:biosystems}a). 
The protein exchanges protons with chlorides and many other anions, but is inhibited by fluoride. \citeauthor{clc2020}\cite{clc2020} gained insight on the molecular basis of fluoride inhibition of the protein by MiMiC-based QM/MM MD and well-tempered metadynamics (via PLUMED\cite{PLUMED, PLUMED2}), at the BLYP and B3LYP levels of theory. 
Since the fluoride binding mechanism involves proton transfer phenomena, the use of a quantum mechanical approach is imperative\cite{clc2020}. 
The QM subsystem consisted of 106 atoms encompassing the glutamate gating region while the remainder of the protein, water, ions, and lipid bilayer (together 150\:925 atoms) were described by a fixed point-charge force field enabling electrostatic embedding. 
Subsequent MiMiC-based MD and metadynamics simulations considered a related protein, the CLC\textsuperscript{F} antiporter (solvated and embedded in a lipid bilayer for a total of 165\:732 atoms), though with a smaller QM subsystem of 36 atoms, using the same levels of theory to model the full exchange between anion and proton (Fig. \ref{fig:biosystems}b)\cite{clc2021, clc2024}. 
The simulations provided clues to the intriguing question of why a protein structurally similar to CLC-ec1, such as CLC\textsuperscript{F}, transports fluoride instead of being inhibited by it. 
More recently, MiMiC-based QM/MM MD simulations have been extended to DgoT, a bacterial homolog of human SLC17 organic anion transporters \cite{2023dgot,slc17_2024}.
Here, the simulations shed light on a key proton transfer process occurring in DgoT's transport cycle in which bond breaking and formation occur.

Ion channels represent another class of large membrane-protein systems. 
A recent study\cite{glutamate2023} employed MiMiC to investigate calcium conduction in the $\alpha$-amino-3-hydroxy-5-methyl-4-isoxazolepropionic acid receptors (AMPARs), neurotransmitter-activated cation channels ubiquitously expressed in vertebrate brains (Fig. \ref{fig:biosystems}d). 
In particular, a Ca\textsuperscript{2+} ion and its first hydration shell were treated at the BLYP or B3LYP levels, resulting in a QM subsystem of between 22 and 25 atoms depending on the simulation, while all other water molecules, as well as protein and lipids, were described classically for a total of 187\:699 atoms. 
In this study, thanks to the QM/MM partitioning, the large electric field of the protein was modeled, as well as the structure and dynamics of calcium-bound water molecules.
Moreover, the study provided data to assess the accuracy of a newly developed force field for Ca\textsuperscript{2+}, an ion that is notoriously difficult to describe at the force field level.

MiMiC-based QM/MM simulations used in drug design studies have pointed to the limitations of classical force field descriptions in the prediction of important pharmacological quantities, such as the ligand unbinding rate constant, $k_\mathrm{off}$, and its inverse, the residence time.
Accurate estimation of $k_\mathrm{off}$ requires a precise calculation of the free energy transition state and not just of binding free energy and affinity\cite{ligand_binding}. 
A study in 2020\cite{koff2020} investigated the residence time of iperoxo targeting the human muscarinic acetylcholine receptor 2 (M\textsubscript{2}), yielding insight into potential therapeutics aimed at these receptors, which regulate several key biological processes.
This study combined force field metadynamics with MiMiC-based QM/MM simulations, in which the QM subsystem consisted of the ligand, key amino acid side chains, and nearby water molecules described at the B3LYP level.

A more recent study\cite{drugdesign2023} further highlights the potentiality of MiMiC for drug design.
It focused on the IDH1 enzyme (Fig. \ref{fig:biosystems}c), which is a crucial actor within the Krebs cycle of cellular respiration as it serves as a catalyst in the conversion of isocitrate to $\alpha$-ketoglutarate \cite{idh1_ref}. 
The IDH1 active site includes a Mg\textsuperscript{2+} ion, the NADH+ cofactor, the ligand, and important amino acid residues that all together make up the Michaelis complex. 
To investigate the enzymatic reaction, which invariably involves bond breaking and formation processes, quantum mechanics is required. 
The QM subsystem encompasses the active site, comprising 142 atoms, and the MM subsystem includes the remainder of the enzyme, water, and ions (all represented through a fixed point-charge force field) for a total of 130\:828 atoms.
The conclusions drawn from these simulations turned out to be consistent with previous studies on the same system.

\section{\label{sec:outlook}Outlook}

We presently devote the majority of our efforts towards multiscale simulations capable of exploiting the vast computational resources available on current and upcoming exascale supercomputers\cite{Remmel_CEN, Slava_WIREs_2021, Martin_2022, Beck_2024}.
The MiMiC framework is prepared for the future, not only in terms of scalability and extensibility, but also in its capability to run client programs independently and concurrently on separate resources, making it ideal for modular, heterogeneous HPC architectures\cite{carpenter_2022_6090425, suarez_2022_6508394, carrasco2024multiscale}.
Specifically, each program can use the type and amount of resources that are optimal for its performance, and the different programs can easily run on different partitions or modules of a supercomputer.
Hybrid CPU/GPU technology is already prevalent on most state-of-the-art supercomputers, while modular architectures have been adopted by multiple HPC centers.
A prominent example is the first European exascale supercomputer JUPITER that will feature multiple modules, including a general-purpose CPU partition and a high-performance GPU-accelerated partition\cite{jupiter_technical}.
Moreover, it is expected that future technologies, such as quantum or neuromorphic computing, will eventually be integrated into traditional HPC architectures.
As these novel computing technologies become available, MiMiC is virtually ready to incorporate them into its simulation workflows.

In terms of new features, we aim to extend the range of multiscale techniques to address scientific challenges beyond the scope of current QM/MM methods.
One of our development directions is to introduce additional versatility in the QM subsystem treatment by enabling larger, more accurate, and yet efficient simulations through fragment-based methods and multilayered QM/QM/MM models.
To model larger systems over longer time scales, we will also consider CG and continuum models (CM), enabling, for example, the implementation of MM/CG, QM/MM/CG, and QM/MM/CM schemes.
Another promising way to achieve more efficient calculations is the introduction of ML-based force fields\cite{Unke2021, Kocer2022}, which will give access to, e.g., ML/MM simulations.
Along the same lines is the development of ML-based schemes where a subsystem is described by an ML model that is retrained on the fly based on, e.g., QM calculations.
Among the other desired extensions is the implementation of schemes that allow subsystems to adjust to changes over the course of a simulation, such as adaptive QM/MM\cite{Mato2021}.
Finally, we aim to extend MiMiC beyond the scope of biomolecular systems by implementing the necessary features for systems within the material science domain.
Here, multiscale simulations pose different challenges, but they also open up numerous opportunities to provide a fundamental understanding of various processes for which current models would not be applicable.

We are confident that our efforts to push the performance of multiscale simulations through advanced parallel algorithms and methods, in conjunction with the advances in computational chemistry software interfaced with MiMiC, will eventually open up new avenues of exploration to address exceptionally challenging scientific problems.

\begin{acknowledgments}
The authors are grateful to Viacheslav Bolnykh for his essential foundational contributions during the initial phase of the project.
The authors also thank Till Kirsch, Michele Cascella, and Jürgen Gauss for the CFOUR interface, and Maria Gabriella Chiariello, Florian Schackert, and Wenping Lyu for their contributions to the studies showcased in the Illustrative Applications Section.
BR, DM, EI, and PC thank the Helmholtz European Partnering program (\emph{Innovative high-performance computing approaches for molecular neuromedicine}) for funding.
PC thanks the Deutsche Forschungsgemeinschaft (DFG) RU2518 DynIon and the computing time granted by the JARA Vergabegremium on the JARA partition of the supercomputer JURECA at Forschungszentrum Jülich.
UR acknowledges funding from the Swiss National Science Foundation via the NCCR MUST and individual grants Nos 200020-185092 and 200020-219440, as well as computing time from the Swiss National Computing Centre CSCS.
JMHO gratefully acknowledges the financial support from VILLUM FONDEN (Grant No.\ VIL29478).
\end{acknowledgments}

\section*{Author Declarations}
\subsection{Conflict of Interest}
The authors have no conflicts to disclose.

\section*{Data Availability}
Data sharing is not applicable to this article as no new data were created or analyzed in this study.


\bibliography{manuscript}

\begin{thebibliography}{132}%
\makeatletter
\providecommand \@ifxundefined [1]{%
 \@ifx{#1\undefined}
}%
\providecommand \@ifnum [1]{%
 \ifnum #1\expandafter \@firstoftwo
 \else \expandafter \@secondoftwo
 \fi
}%
\providecommand \@ifx [1]{%
 \ifx #1\expandafter \@firstoftwo
 \else \expandafter \@secondoftwo
 \fi
}%
\providecommand \natexlab [1]{#1}%
\providecommand \enquote  [1]{``#1''}%
\providecommand \bibnamefont  [1]{#1}%
\providecommand \bibfnamefont [1]{#1}%
\providecommand \citenamefont [1]{#1}%
\providecommand \href@noop [0]{\@secondoftwo}%
\providecommand \href [0]{\begingroup \@sanitize@url \@href}%
\providecommand \@href[1]{\@@startlink{#1}\@@href}%
\providecommand \@@href[1]{\endgroup#1\@@endlink}%
\providecommand \@sanitize@url [0]{\catcode `\\12\catcode `\$12\catcode `\&12\catcode `\#12\catcode `\^12\catcode `\_12\catcode `\%12\relax}%
\providecommand \@@startlink[1]{}%
\providecommand \@@endlink[0]{}%
\providecommand \url  [0]{\begingroup\@sanitize@url \@url }%
\providecommand \@url [1]{\endgroup\@href {#1}{\urlprefix }}%
\providecommand \urlprefix  [0]{URL }%
\providecommand \Eprint [0]{\href }%
\providecommand \doibase [0]{http://dx.doi.org/}%
\providecommand \selectlanguage [0]{\@gobble}%
\providecommand \bibinfo  [0]{\@secondoftwo}%
\providecommand \bibfield  [0]{\@secondoftwo}%
\providecommand \translation [1]{[#1]}%
\providecommand \BibitemOpen [0]{}%
\providecommand \bibitemStop [0]{}%
\providecommand \bibitemNoStop [0]{.\EOS\space}%
\providecommand \EOS [0]{\spacefactor3000\relax}%
\providecommand \BibitemShut  [1]{\csname bibitem#1\endcsname}%
\let\auto@bib@innerbib\@empty
\bibitem [{\citenamefont {Shinoda}, \citenamefont {DeVane},\ and\ \citenamefont {Klein}(2012)}]{Klein_2012}%
  \BibitemOpen
  \bibfield  {author} {\bibinfo {author} {\bibfnamefont {W.}~\bibnamefont {Shinoda}}, \bibinfo {author} {\bibfnamefont {R.}~\bibnamefont {DeVane}}, \ and\ \bibinfo {author} {\bibfnamefont {M.~L.}\ \bibnamefont {Klein}},\ }\bibfield  {title} {\enquote {\bibinfo {title} {{Computer simulation studies of self-assembling macromolecules}},}\ }\href {\doibase 10.1016/j.sbi.2012.01.011} {\bibfield  {journal} {\bibinfo  {journal} {Curr. Opin. Struct. Biol.}\ }\textbf {\bibinfo {volume} {22}},\ \bibinfo {pages} {175--186} (\bibinfo {year} {2012})}\BibitemShut {NoStop}%
\bibitem [{\citenamefont {Kjølbye}\ \emph {et~al.}(2022)\citenamefont {Kjølbye}, \citenamefont {Pereira}, \citenamefont {Bartocci}, \citenamefont {Pannuzzo}, \citenamefont {Albani}, \citenamefont {Marchetto}, \citenamefont {Jiménez-García}, \citenamefont {Martin}, \citenamefont {Rossetti}, \citenamefont {Cecchini}, \citenamefont {Wu}, \citenamefont {Monticelli},\ and\ \citenamefont {Souza}}]{Pereira_2022}%
  \BibitemOpen
  \bibfield  {author} {\bibinfo {author} {\bibfnamefont {L.~R.}\ \bibnamefont {Kjølbye}}, \bibinfo {author} {\bibfnamefont {G.~P.}\ \bibnamefont {Pereira}}, \bibinfo {author} {\bibfnamefont {A.}~\bibnamefont {Bartocci}}, \bibinfo {author} {\bibfnamefont {M.}~\bibnamefont {Pannuzzo}}, \bibinfo {author} {\bibfnamefont {S.}~\bibnamefont {Albani}}, \bibinfo {author} {\bibfnamefont {A.}~\bibnamefont {Marchetto}}, \bibinfo {author} {\bibfnamefont {B.}~\bibnamefont {Jiménez-García}}, \bibinfo {author} {\bibfnamefont {J.}~\bibnamefont {Martin}}, \bibinfo {author} {\bibfnamefont {G.}~\bibnamefont {Rossetti}}, \bibinfo {author} {\bibfnamefont {M.}~\bibnamefont {Cecchini}}, \bibinfo {author} {\bibfnamefont {S.}~\bibnamefont {Wu}}, \bibinfo {author} {\bibfnamefont {L.}~\bibnamefont {Monticelli}}, \ and\ \bibinfo {author} {\bibfnamefont {P.~C.~T.}\ \bibnamefont {Souza}},\ }\bibfield  {title} {\enquote {\bibinfo {title} {{Towards design of drugs and delivery systems with the Martini coarse-grained model}},}\ }\href
  {\doibase 10.1017/qrd.2022.16} {\bibfield  {journal} {\bibinfo  {journal} {QRB Discovery}\ }\textbf {\bibinfo {volume} {3}},\ \bibinfo {pages} {e19} (\bibinfo {year} {2022})}\BibitemShut {NoStop}%
\bibitem [{\citenamefont {Jin}\ \emph {et~al.}(2022)\citenamefont {Jin}, \citenamefont {Pak}, \citenamefont {Durumeric}, \citenamefont {Loose},\ and\ \citenamefont {Voth}}]{Voth_2022}%
  \BibitemOpen
  \bibfield  {author} {\bibinfo {author} {\bibfnamefont {J.}~\bibnamefont {Jin}}, \bibinfo {author} {\bibfnamefont {A.~J.}\ \bibnamefont {Pak}}, \bibinfo {author} {\bibfnamefont {A.~E.~P.}\ \bibnamefont {Durumeric}}, \bibinfo {author} {\bibfnamefont {T.~D.}\ \bibnamefont {Loose}}, \ and\ \bibinfo {author} {\bibfnamefont {G.~A.}\ \bibnamefont {Voth}},\ }\bibfield  {title} {\enquote {\bibinfo {title} {{Bottom-up Coarse-Graining: Principles and Perspectives}},}\ }\href {\doibase 10.1021/acs.jctc.2c00643} {\bibfield  {journal} {\bibinfo  {journal} {J. Chem. Theory Comput.}\ }\textbf {\bibinfo {volume} {18}},\ \bibinfo {pages} {5759--5791} (\bibinfo {year} {2022})}\BibitemShut {NoStop}%
\bibitem [{\citenamefont {Borges-Araújo}\ \emph {et~al.}(2023)\citenamefont {Borges-Araújo}, \citenamefont {Patmanidis}, \citenamefont {Singh}, \citenamefont {Santos}, \citenamefont {Sieradzan}, \citenamefont {Vanni}, \citenamefont {Czaplewski}, \citenamefont {Pantano}, \citenamefont {Shinoda}, \citenamefont {Monticelli}, \citenamefont {Liwo}, \citenamefont {Marrink},\ and\ \citenamefont {Souza}}]{Souza_2023}%
  \BibitemOpen
  \bibfield  {author} {\bibinfo {author} {\bibfnamefont {L.}~\bibnamefont {Borges-Araújo}}, \bibinfo {author} {\bibfnamefont {I.}~\bibnamefont {Patmanidis}}, \bibinfo {author} {\bibfnamefont {A.~P.}\ \bibnamefont {Singh}}, \bibinfo {author} {\bibfnamefont {L.~H.~S.}\ \bibnamefont {Santos}}, \bibinfo {author} {\bibfnamefont {A.~K.}\ \bibnamefont {Sieradzan}}, \bibinfo {author} {\bibfnamefont {S.}~\bibnamefont {Vanni}}, \bibinfo {author} {\bibfnamefont {C.}~\bibnamefont {Czaplewski}}, \bibinfo {author} {\bibfnamefont {S.}~\bibnamefont {Pantano}}, \bibinfo {author} {\bibfnamefont {W.}~\bibnamefont {Shinoda}}, \bibinfo {author} {\bibfnamefont {L.}~\bibnamefont {Monticelli}}, \bibinfo {author} {\bibfnamefont {A.}~\bibnamefont {Liwo}}, \bibinfo {author} {\bibfnamefont {S.~J.}\ \bibnamefont {Marrink}}, \ and\ \bibinfo {author} {\bibfnamefont {P.~C.~T.}\ \bibnamefont {Souza}},\ }\bibfield  {title} {\enquote {\bibinfo {title} {{Pragmatic Coarse-Graining of Proteins: Models and Applications}},}\ }\href {\doibase
  10.1021/acs.jctc.3c00733} {\bibfield  {journal} {\bibinfo  {journal} {J. Chem. Theory Comput.}\ }\textbf {\bibinfo {volume} {19}},\ \bibinfo {pages} {7112--7135} (\bibinfo {year} {2023})}\BibitemShut {NoStop}%
\bibitem [{\citenamefont {Warshel}\ and\ \citenamefont {Levitt}(1976)}]{warshel1976theoretical}%
  \BibitemOpen
  \bibfield  {author} {\bibinfo {author} {\bibfnamefont {A.}~\bibnamefont {Warshel}}\ and\ \bibinfo {author} {\bibfnamefont {M.}~\bibnamefont {Levitt}},\ }\bibfield  {title} {\enquote {\bibinfo {title} {{Theoretical studies of enzymic reactions: Dielectric, electrostatic and steric stabilization of the carbonium ion in the reaction of lysozyme}},}\ }\href {\doibase 10.1016/0022-2836(76)90311-9} {\bibfield  {journal} {\bibinfo  {journal} {J. Mol. Biol.}\ }\textbf {\bibinfo {volume} {103}},\ \bibinfo {pages} {227--249} (\bibinfo {year} {1976})}\BibitemShut {NoStop}%
\bibitem [{\citenamefont {Singh}\ and\ \citenamefont {Kollman}(1986)}]{Singh_QMMM}%
  \BibitemOpen
  \bibfield  {author} {\bibinfo {author} {\bibfnamefont {U.~C.}\ \bibnamefont {Singh}}\ and\ \bibinfo {author} {\bibfnamefont {P.~A.}\ \bibnamefont {Kollman}},\ }\bibfield  {title} {\enquote {\bibinfo {title} {{A combined ab initio quantum mechanical and molecular mechanical method for carrying out simulations on complex molecular systems: Applications to the CH$_3$Cl + Cl$^{-}$ exchange reaction and gas phase protonation of polyethers}},}\ }\href {\doibase 10.1002/jcc.540070604} {\bibfield  {journal} {\bibinfo  {journal} {J. Comput. Chem.}\ }\textbf {\bibinfo {volume} {7}},\ \bibinfo {pages} {718--730} (\bibinfo {year} {1986})}\BibitemShut {NoStop}%
\bibitem [{\citenamefont {Field}, \citenamefont {Bash},\ and\ \citenamefont {Karplus}(1990)}]{Karplus_QMMM}%
  \BibitemOpen
  \bibfield  {author} {\bibinfo {author} {\bibfnamefont {M.~J.}\ \bibnamefont {Field}}, \bibinfo {author} {\bibfnamefont {P.~A.}\ \bibnamefont {Bash}}, \ and\ \bibinfo {author} {\bibfnamefont {M.}~\bibnamefont {Karplus}},\ }\bibfield  {title} {\enquote {\bibinfo {title} {{A combined quantum mechanical and molecular mechanical potential for molecular dynamics simulations}},}\ }\href {\doibase 10.1002/jcc.540110605} {\bibfield  {journal} {\bibinfo  {journal} {J. Comput. Chem.}\ }\textbf {\bibinfo {volume} {11}},\ \bibinfo {pages} {700--733} (\bibinfo {year} {1990})}\BibitemShut {NoStop}%
\bibitem [{\citenamefont {Senn}\ and\ \citenamefont {Thiel}(2009)}]{Senn_QMMM}%
  \BibitemOpen
  \bibfield  {author} {\bibinfo {author} {\bibfnamefont {H.~M.}\ \bibnamefont {Senn}}\ and\ \bibinfo {author} {\bibfnamefont {W.}~\bibnamefont {Thiel}},\ }\bibfield  {title} {\enquote {\bibinfo {title} {{QM/MM Methods for Biomolecular Systems}},}\ }\href {\doibase 10.1002/anie.200802019} {\bibfield  {journal} {\bibinfo  {journal} {Angew. Chem. Int. Ed.}\ }\textbf {\bibinfo {volume} {48}},\ \bibinfo {pages} {1198--1229} (\bibinfo {year} {2009})}\BibitemShut {NoStop}%
\bibitem [{\citenamefont {Brunk}\ and\ \citenamefont {Rothlisberger}(2015)}]{Rothlisberger_review}%
  \BibitemOpen
  \bibfield  {author} {\bibinfo {author} {\bibfnamefont {E.}~\bibnamefont {Brunk}}\ and\ \bibinfo {author} {\bibfnamefont {U.}~\bibnamefont {Rothlisberger}},\ }\bibfield  {title} {\enquote {\bibinfo {title} {{Mixed Quantum Mechanical/Molecular Mechanical Molecular Dynamics Simulations of Biological Systems in Ground and Electronically Excited States}},}\ }\href {\doibase 10.1021/cr500628b} {\bibfield  {journal} {\bibinfo  {journal} {Chem. Rev.}\ }\textbf {\bibinfo {volume} {115}},\ \bibinfo {pages} {6217--6263} (\bibinfo {year} {2015})}\BibitemShut {NoStop}%
\bibitem [{\citenamefont {Morzan}\ \emph {et~al.}(2018)\citenamefont {Morzan}, \citenamefont {Alonso~de Armiño}, \citenamefont {Foglia}, \citenamefont {Ramírez}, \citenamefont {González~Lebrero}, \citenamefont {Scherlis},\ and\ \citenamefont {Estrin}}]{Estrin_review}%
  \BibitemOpen
  \bibfield  {author} {\bibinfo {author} {\bibfnamefont {U.~N.}\ \bibnamefont {Morzan}}, \bibinfo {author} {\bibfnamefont {D.~J.}\ \bibnamefont {Alonso~de Armiño}}, \bibinfo {author} {\bibfnamefont {N.~O.}\ \bibnamefont {Foglia}}, \bibinfo {author} {\bibfnamefont {F.}~\bibnamefont {Ramírez}}, \bibinfo {author} {\bibfnamefont {M.~C.}\ \bibnamefont {González~Lebrero}}, \bibinfo {author} {\bibfnamefont {D.~A.}\ \bibnamefont {Scherlis}}, \ and\ \bibinfo {author} {\bibfnamefont {D.~A.}\ \bibnamefont {Estrin}},\ }\bibfield  {title} {\enquote {\bibinfo {title} {{Spectroscopy in Complex Environments from QM--MM Simulations}},}\ }\href {\doibase 10.1021/acs.chemrev.8b00026} {\bibfield  {journal} {\bibinfo  {journal} {Chem. Rev.}\ }\textbf {\bibinfo {volume} {118}},\ \bibinfo {pages} {4071--4113} (\bibinfo {year} {2018})}\BibitemShut {NoStop}%
\bibitem [{\citenamefont {Lipparini}\ and\ \citenamefont {Mennucci}(2021)}]{Mennucci_review}%
  \BibitemOpen
  \bibfield  {author} {\bibinfo {author} {\bibfnamefont {F.}~\bibnamefont {Lipparini}}\ and\ \bibinfo {author} {\bibfnamefont {B.}~\bibnamefont {Mennucci}},\ }\bibfield  {title} {\enquote {\bibinfo {title} {{Hybrid QM/classical models: Methodological advances and new applications}},}\ }\href {\doibase 10.1063/5.0064075} {\bibfield  {journal} {\bibinfo  {journal} {Chem. Phys. Rev.}\ }\textbf {\bibinfo {volume} {2}} (\bibinfo {year} {2021}),\ 10.1063/5.0064075}\BibitemShut {NoStop}%
\bibitem [{\citenamefont {Götz}, \citenamefont {Clark},\ and\ \citenamefont {Walker}(2014)}]{AMBER_QM_MM_MD_interface}%
  \BibitemOpen
  \bibfield  {author} {\bibinfo {author} {\bibfnamefont {A.~W.}\ \bibnamefont {Götz}}, \bibinfo {author} {\bibfnamefont {M.~A.}\ \bibnamefont {Clark}}, \ and\ \bibinfo {author} {\bibfnamefont {R.~C.}\ \bibnamefont {Walker}},\ }\bibfield  {title} {\enquote {\bibinfo {title} {{An extensible interface for QM/MM molecular dynamics simulations with AMBER}},}\ }\href {\doibase 10.1002/jcc.23444} {\bibfield  {journal} {\bibinfo  {journal} {J. Comput. Chem.}\ }\textbf {\bibinfo {volume} {35}},\ \bibinfo {pages} {95--108} (\bibinfo {year} {2014})}\BibitemShut {NoStop}%
\bibitem [{\citenamefont {Melo}\ \emph {et~al.}(2018)\citenamefont {Melo}, \citenamefont {Bernardi}, \citenamefont {Rudack}, \citenamefont {Scheurer}, \citenamefont {Riplinger}, \citenamefont {Phillips}, \citenamefont {Maia}, \citenamefont {Rocha}, \citenamefont {Ribeiro}, \citenamefont {Stone}, \citenamefont {Neese}, \citenamefont {Schulten},\ and\ \citenamefont {Luthey-Schulten}}]{NAMD_goes_quantum}%
  \BibitemOpen
  \bibfield  {author} {\bibinfo {author} {\bibfnamefont {M.~C.~R.}\ \bibnamefont {Melo}}, \bibinfo {author} {\bibfnamefont {R.~C.}\ \bibnamefont {Bernardi}}, \bibinfo {author} {\bibfnamefont {T.}~\bibnamefont {Rudack}}, \bibinfo {author} {\bibfnamefont {M.}~\bibnamefont {Scheurer}}, \bibinfo {author} {\bibfnamefont {C.}~\bibnamefont {Riplinger}}, \bibinfo {author} {\bibfnamefont {J.~C.}\ \bibnamefont {Phillips}}, \bibinfo {author} {\bibfnamefont {J.~D.~C.}\ \bibnamefont {Maia}}, \bibinfo {author} {\bibfnamefont {G.~B.}\ \bibnamefont {Rocha}}, \bibinfo {author} {\bibfnamefont {J.~V.}\ \bibnamefont {Ribeiro}}, \bibinfo {author} {\bibfnamefont {J.~E.}\ \bibnamefont {Stone}}, \bibinfo {author} {\bibfnamefont {F.}~\bibnamefont {Neese}}, \bibinfo {author} {\bibfnamefont {K.}~\bibnamefont {Schulten}}, \ and\ \bibinfo {author} {\bibfnamefont {Z.}~\bibnamefont {Luthey-Schulten}},\ }\bibfield  {title} {\enquote {\bibinfo {title} {{NAMD goes quantum: an integrative suite for hybrid simulations}},}\ }\href {\doibase
  10.1038/nmeth.4638} {\bibfield  {journal} {\bibinfo  {journal} {Nat. Methods}\ }\textbf {\bibinfo {volume} {15}},\ \bibinfo {pages} {351--354} (\bibinfo {year} {2018})}\BibitemShut {NoStop}%
\bibitem [{\citenamefont {Cruzeiro}\ \emph {et~al.}(2023)\citenamefont {Cruzeiro}, \citenamefont {Wang}, \citenamefont {Pieri}, \citenamefont {Hohenstein},\ and\ \citenamefont {Martínez}}]{TCPB_2023}%
  \BibitemOpen
  \bibfield  {author} {\bibinfo {author} {\bibfnamefont {V.~W.~D.}\ \bibnamefont {Cruzeiro}}, \bibinfo {author} {\bibfnamefont {Y.}~\bibnamefont {Wang}}, \bibinfo {author} {\bibfnamefont {E.}~\bibnamefont {Pieri}}, \bibinfo {author} {\bibfnamefont {E.~G.}\ \bibnamefont {Hohenstein}}, \ and\ \bibinfo {author} {\bibfnamefont {T.~J.}\ \bibnamefont {Martínez}},\ }\bibfield  {title} {\enquote {\bibinfo {title} {{TeraChem protocol buffers (TCPB): Accelerating QM and QM/MM simulations with a client--server model}},}\ }\href {\doibase 10.1063/5.0130886} {\bibfield  {journal} {\bibinfo  {journal} {J. Chem. Phys.}\ }\textbf {\bibinfo {volume} {158}},\ \bibinfo {pages} {044801} (\bibinfo {year} {2023})}\BibitemShut {NoStop}%
\bibitem [{\citenamefont {Woodcock}\ \emph {et~al.}(2011)\citenamefont {Woodcock}, \citenamefont {Miller}, \citenamefont {Hodoscek}, \citenamefont {Okur}, \citenamefont {Larkin}, \citenamefont {Ponder},\ and\ \citenamefont {Brooks}}]{MSCALE}%
  \BibitemOpen
  \bibfield  {author} {\bibinfo {author} {\bibfnamefont {H.~L.}\ \bibnamefont {Woodcock}}, \bibinfo {author} {\bibfnamefont {B.~T.}\ \bibnamefont {Miller}}, \bibinfo {author} {\bibfnamefont {M.}~\bibnamefont {Hodoscek}}, \bibinfo {author} {\bibfnamefont {A.}~\bibnamefont {Okur}}, \bibinfo {author} {\bibfnamefont {J.~D.}\ \bibnamefont {Larkin}}, \bibinfo {author} {\bibfnamefont {J.~W.}\ \bibnamefont {Ponder}}, \ and\ \bibinfo {author} {\bibfnamefont {B.~R.}\ \bibnamefont {Brooks}},\ }\bibfield  {title} {\enquote {\bibinfo {title} {{MSCALE: A General Utility for Multiscale Modeling}},}\ }\href {\doibase 10.1021/ct100738h} {\bibfield  {journal} {\bibinfo  {journal} {J. Chem. Theory Comput.}\ }\textbf {\bibinfo {volume} {7}},\ \bibinfo {pages} {1208--1219} (\bibinfo {year} {2011})}\BibitemShut {NoStop}%
\bibitem [{\citenamefont {Ma}\ \emph {et~al.}(2015)\citenamefont {Ma}, \citenamefont {Martin-Samos}, \citenamefont {Fabris}, \citenamefont {Laio},\ and\ \citenamefont {Piccinin}}]{QMMMW}%
  \BibitemOpen
  \bibfield  {author} {\bibinfo {author} {\bibfnamefont {C.}~\bibnamefont {Ma}}, \bibinfo {author} {\bibfnamefont {L.}~\bibnamefont {Martin-Samos}}, \bibinfo {author} {\bibfnamefont {S.}~\bibnamefont {Fabris}}, \bibinfo {author} {\bibfnamefont {A.}~\bibnamefont {Laio}}, \ and\ \bibinfo {author} {\bibfnamefont {S.}~\bibnamefont {Piccinin}},\ }\bibfield  {title} {\enquote {\bibinfo {title} {{QMMMW: A wrapper for QM/MM simulations with Quantum ESPRESSO and LAMMPS}},}\ }\href {\doibase 10.1016/j.cpc.2015.04.024} {\bibfield  {journal} {\bibinfo  {journal} {Comput. Phys. Commun.}\ }\textbf {\bibinfo {volume} {195}},\ \bibinfo {pages} {191--198} (\bibinfo {year} {2015})}\BibitemShut {NoStop}%
\bibitem [{\citenamefont {Torras}, \citenamefont {Deumens},\ and\ \citenamefont {Trickey}(2006)}]{PUPIL_soft_integration}%
  \BibitemOpen
  \bibfield  {author} {\bibinfo {author} {\bibfnamefont {J.}~\bibnamefont {Torras}}, \bibinfo {author} {\bibfnamefont {E.}~\bibnamefont {Deumens}}, \ and\ \bibinfo {author} {\bibfnamefont {S.~B.}\ \bibnamefont {Trickey}},\ }\bibfield  {title} {\enquote {\bibinfo {title} {{Software Integration in Multi-scale Simulations: the PUPIL System}},}\ }\href {\doibase 10.1007/s10820-006-9011-3} {\bibfield  {journal} {\bibinfo  {journal} {J. Comput.-Aided Mater. Des.}\ }\textbf {\bibinfo {volume} {13}},\ \bibinfo {pages} {201--212} (\bibinfo {year} {2006})}\BibitemShut {NoStop}%
\bibitem [{\citenamefont {Torras}\ \emph {et~al.}(2015)\citenamefont {Torras}, \citenamefont {Roberts}, \citenamefont {Seabra},\ and\ \citenamefont {Trickey}}]{PUPIL_soft_integration2}%
  \BibitemOpen
  \bibfield  {author} {\bibinfo {author} {\bibfnamefont {J.}~\bibnamefont {Torras}}, \bibinfo {author} {\bibfnamefont {B.~P.}\ \bibnamefont {Roberts}}, \bibinfo {author} {\bibfnamefont {G.~M.}\ \bibnamefont {Seabra}}, \ and\ \bibinfo {author} {\bibfnamefont {S.~B.}\ \bibnamefont {Trickey}},\ }\enquote {\bibinfo {title} {{PUPIL: A Software Integration System for Multi-Scale QM/MM-MD Simulations and Its Application to Biomolecular Systems}},}\ in\ \href {\doibase 10.1016/bs.apcsb.2015.06.002} {\emph {\bibinfo {booktitle} {Combined Quantum Mechanical and Molecular Mechanical Modelling of Biomolecular Interactions}}},\ \bibinfo {series} {Adv. Protein Chem. Struct. Biol.}, Vol.\ \bibinfo {volume} {100},\ \bibinfo {editor} {edited by\ \bibinfo {editor} {\bibfnamefont {T.}~\bibnamefont {Karabencheva-Christova}}}\ (\bibinfo  {publisher} {Elsevier},\ \bibinfo {year} {2015})\ pp.\ \bibinfo {pages} {1--31}\BibitemShut {NoStop}%
\bibitem [{\citenamefont {Řezáč}(2016)}]{Cuby}%
  \BibitemOpen
  \bibfield  {author} {\bibinfo {author} {\bibfnamefont {J.}~\bibnamefont {Řezáč}},\ }\bibfield  {title} {\enquote {\bibinfo {title} {{Cuby: An integrative framework for computational chemistry}},}\ }\href {\doibase 10.1002/jcc.24312} {\bibfield  {journal} {\bibinfo  {journal} {J. Comput. Chem.}\ }\textbf {\bibinfo {volume} {37}},\ \bibinfo {pages} {1230--1237} (\bibinfo {year} {2016})}\BibitemShut {NoStop}%
\bibitem [{\citenamefont {Larsen}\ \emph {et~al.}(2017)\citenamefont {Larsen}, \citenamefont {Mortensen}, \citenamefont {Blomqvist}, \citenamefont {Castelli}, \citenamefont {Christensen}, \citenamefont {Dułak}, \citenamefont {Friis}, \citenamefont {Groves}, \citenamefont {Hammer}, \citenamefont {Hargus}, \citenamefont {Hermes}, \citenamefont {Jennings}, \citenamefont {Jensen}, \citenamefont {Kermode}, \citenamefont {Kitchin}, \citenamefont {Kolsbjerg}, \citenamefont {Kubal}, \citenamefont {Kaasbjerg}, \citenamefont {Lysgaard}, \citenamefont {Maronsson}, \citenamefont {Maxson}, \citenamefont {Olsen}, \citenamefont {Pastewka}, \citenamefont {Peterson}, \citenamefont {Rostgaard}, \citenamefont {Schiøtz}, \citenamefont {Schütt}, \citenamefont {Strange}, \citenamefont {Thygesen}, \citenamefont {Vegge}, \citenamefont {Vilhelmsen}, \citenamefont {Walter}, \citenamefont {Zeng},\ and\ \citenamefont {Jacobsen}}]{ASE}%
  \BibitemOpen
  \bibfield  {author} {\bibinfo {author} {\bibfnamefont {A.~H.}\ \bibnamefont {Larsen}}, \bibinfo {author} {\bibfnamefont {J.~J.}\ \bibnamefont {Mortensen}}, \bibinfo {author} {\bibfnamefont {J.}~\bibnamefont {Blomqvist}}, \bibinfo {author} {\bibfnamefont {I.~E.}\ \bibnamefont {Castelli}}, \bibinfo {author} {\bibfnamefont {R.}~\bibnamefont {Christensen}}, \bibinfo {author} {\bibfnamefont {M.}~\bibnamefont {Dułak}}, \bibinfo {author} {\bibfnamefont {J.}~\bibnamefont {Friis}}, \bibinfo {author} {\bibfnamefont {M.~N.}\ \bibnamefont {Groves}}, \bibinfo {author} {\bibfnamefont {B.}~\bibnamefont {Hammer}}, \bibinfo {author} {\bibfnamefont {C.}~\bibnamefont {Hargus}}, \bibinfo {author} {\bibfnamefont {E.~D.}\ \bibnamefont {Hermes}}, \bibinfo {author} {\bibfnamefont {P.~C.}\ \bibnamefont {Jennings}}, \bibinfo {author} {\bibfnamefont {P.~B.}\ \bibnamefont {Jensen}}, \bibinfo {author} {\bibfnamefont {J.}~\bibnamefont {Kermode}}, \bibinfo {author} {\bibfnamefont {J.~R.}\ \bibnamefont {Kitchin}}, \bibinfo {author}
  {\bibfnamefont {E.~L.}\ \bibnamefont {Kolsbjerg}}, \bibinfo {author} {\bibfnamefont {J.}~\bibnamefont {Kubal}}, \bibinfo {author} {\bibfnamefont {K.}~\bibnamefont {Kaasbjerg}}, \bibinfo {author} {\bibfnamefont {S.}~\bibnamefont {Lysgaard}}, \bibinfo {author} {\bibfnamefont {J.~B.}\ \bibnamefont {Maronsson}}, \bibinfo {author} {\bibfnamefont {T.}~\bibnamefont {Maxson}}, \bibinfo {author} {\bibfnamefont {T.}~\bibnamefont {Olsen}}, \bibinfo {author} {\bibfnamefont {L.}~\bibnamefont {Pastewka}}, \bibinfo {author} {\bibfnamefont {A.}~\bibnamefont {Peterson}}, \bibinfo {author} {\bibfnamefont {C.}~\bibnamefont {Rostgaard}}, \bibinfo {author} {\bibfnamefont {J.}~\bibnamefont {Schiøtz}}, \bibinfo {author} {\bibfnamefont {O.}~\bibnamefont {Schütt}}, \bibinfo {author} {\bibfnamefont {M.}~\bibnamefont {Strange}}, \bibinfo {author} {\bibfnamefont {K.~S.}\ \bibnamefont {Thygesen}}, \bibinfo {author} {\bibfnamefont {T.}~\bibnamefont {Vegge}}, \bibinfo {author} {\bibfnamefont {L.}~\bibnamefont {Vilhelmsen}}, \bibinfo
  {author} {\bibfnamefont {M.}~\bibnamefont {Walter}}, \bibinfo {author} {\bibfnamefont {Z.}~\bibnamefont {Zeng}}, \ and\ \bibinfo {author} {\bibfnamefont {K.~W.}\ \bibnamefont {Jacobsen}},\ }\bibfield  {title} {\enquote {\bibinfo {title} {{The atomic simulation environment---a Python library for working with atoms}},}\ }\href {\doibase 10.1088/1361-648X/aa680e} {\bibfield  {journal} {\bibinfo  {journal} {J. Phys.: Condens. Matter.}\ }\textbf {\bibinfo {volume} {29}},\ \bibinfo {pages} {273002} (\bibinfo {year} {2017})}\BibitemShut {NoStop}%
\bibitem [{\citenamefont {Altoè}\ \emph {et~al.}(2007)\citenamefont {Altoè}, \citenamefont {Stenta}, \citenamefont {Bottoni},\ and\ \citenamefont {Garavelli}}]{Alto2007}%
  \BibitemOpen
  \bibfield  {author} {\bibinfo {author} {\bibfnamefont {P.}~\bibnamefont {Altoè}}, \bibinfo {author} {\bibfnamefont {M.}~\bibnamefont {Stenta}}, \bibinfo {author} {\bibfnamefont {A.}~\bibnamefont {Bottoni}}, \ and\ \bibinfo {author} {\bibfnamefont {M.}~\bibnamefont {Garavelli}},\ }\bibfield  {title} {\enquote {\bibinfo {title} {{A tunable QM/MM approach to chemical reactivity, structure and physico-chemical properties prediction}},}\ }\href {\doibase 10.1007/s00214-007-0275-9} {\bibfield  {journal} {\bibinfo  {journal} {Theor. Chem. Acc.}\ }\textbf {\bibinfo {volume} {118}},\ \bibinfo {pages} {219--240} (\bibinfo {year} {2007})}\BibitemShut {NoStop}%
\bibitem [{\citenamefont {Weingart}\ \emph {et~al.}(2018)\citenamefont {Weingart}, \citenamefont {Nenov}, \citenamefont {Altoè}, \citenamefont {Rivalta}, \citenamefont {Segarra-Martí}, \citenamefont {Dokukina},\ and\ \citenamefont {Garavelli}}]{Weingart2018}%
  \BibitemOpen
  \bibfield  {author} {\bibinfo {author} {\bibfnamefont {O.}~\bibnamefont {Weingart}}, \bibinfo {author} {\bibfnamefont {A.}~\bibnamefont {Nenov}}, \bibinfo {author} {\bibfnamefont {P.}~\bibnamefont {Altoè}}, \bibinfo {author} {\bibfnamefont {I.}~\bibnamefont {Rivalta}}, \bibinfo {author} {\bibfnamefont {J.}~\bibnamefont {Segarra-Martí}}, \bibinfo {author} {\bibfnamefont {I.}~\bibnamefont {Dokukina}}, \ and\ \bibinfo {author} {\bibfnamefont {M.}~\bibnamefont {Garavelli}},\ }\bibfield  {title} {\enquote {\bibinfo {title} {{COBRAMM 2.0 --- A software interface for tailoring molecular electronic structure calculations and running nanoscale (QM/MM) simulations}},}\ }\href {\doibase 10.1007/s00894-018-3769-6} {\bibfield  {journal} {\bibinfo  {journal} {J. Mol. Model.}\ }\textbf {\bibinfo {volume} {24}} (\bibinfo {year} {2018}),\ 10.1007/s00894-018-3769-6}\BibitemShut {NoStop}%
\bibitem [{\citenamefont {Kratz}\ \emph {et~al.}(2016)\citenamefont {Kratz}, \citenamefont {Walker}, \citenamefont {Lagardère}, \citenamefont {Lipparini}, \citenamefont {Piquemal},\ and\ \citenamefont {Andrés~Cisneros}}]{LICHEM}%
  \BibitemOpen
  \bibfield  {author} {\bibinfo {author} {\bibfnamefont {E.~G.}\ \bibnamefont {Kratz}}, \bibinfo {author} {\bibfnamefont {A.~R.}\ \bibnamefont {Walker}}, \bibinfo {author} {\bibfnamefont {L.}~\bibnamefont {Lagardère}}, \bibinfo {author} {\bibfnamefont {F.}~\bibnamefont {Lipparini}}, \bibinfo {author} {\bibfnamefont {J.-P.}\ \bibnamefont {Piquemal}}, \ and\ \bibinfo {author} {\bibfnamefont {G.}~\bibnamefont {Andrés~Cisneros}},\ }\bibfield  {title} {\enquote {\bibinfo {title} {{LICHEM: A QM/MM program for simulations with multipolar and polarizable force fields}},}\ }\href {\doibase 10.1002/jcc.24295} {\bibfield  {journal} {\bibinfo  {journal} {J. Comput. Chem.}\ }\textbf {\bibinfo {volume} {37}},\ \bibinfo {pages} {1019--1029} (\bibinfo {year} {2016})}\BibitemShut {NoStop}%
\bibitem [{\citenamefont {G\"{o}kcan}, \citenamefont {Vázquez-Montelongo},\ and\ \citenamefont {Cisneros}(2019)}]{LICHEM2}%
  \BibitemOpen
  \bibfield  {author} {\bibinfo {author} {\bibfnamefont {H.}~\bibnamefont {G\"{o}kcan}}, \bibinfo {author} {\bibfnamefont {E.~A.}\ \bibnamefont {Vázquez-Montelongo}}, \ and\ \bibinfo {author} {\bibfnamefont {G.~A.}\ \bibnamefont {Cisneros}},\ }\bibfield  {title} {\enquote {\bibinfo {title} {{LICHEM 1.1: Recent Improvements and New Capabilities}},}\ }\href {\doibase 10.1021/acs.jctc.9b00028} {\bibfield  {journal} {\bibinfo  {journal} {J. Chem. Theory Comput.}\ }\textbf {\bibinfo {volume} {15}},\ \bibinfo {pages} {3056--3065} (\bibinfo {year} {2019})}\BibitemShut {NoStop}%
\bibitem [{\citenamefont {Zhang}\ \emph {et~al.}(2019)\citenamefont {Zhang}, \citenamefont {Altarawy}, \citenamefont {Barnes}, \citenamefont {Turney},\ and\ \citenamefont {Schaefer}}]{Janus}%
  \BibitemOpen
  \bibfield  {author} {\bibinfo {author} {\bibfnamefont {B.}~\bibnamefont {Zhang}}, \bibinfo {author} {\bibfnamefont {D.}~\bibnamefont {Altarawy}}, \bibinfo {author} {\bibfnamefont {T.}~\bibnamefont {Barnes}}, \bibinfo {author} {\bibfnamefont {J.~M.}\ \bibnamefont {Turney}}, \ and\ \bibinfo {author} {\bibfnamefont {H.~F.~I.}\ \bibnamefont {Schaefer}},\ }\bibfield  {title} {\enquote {\bibinfo {title} {{Janus: An Extensible Open-Source Software Package for Adaptive QM/MM Methods}},}\ }\href {\doibase 10.1021/acs.jctc.9b00182} {\bibfield  {journal} {\bibinfo  {journal} {J. Chem. Theory Comput.}\ }\textbf {\bibinfo {volume} {15}},\ \bibinfo {pages} {4362--4373} (\bibinfo {year} {2019})}\BibitemShut {NoStop}%
\bibitem [{\citenamefont {Barnes}\ \emph {et~al.}(2021)\citenamefont {Barnes}, \citenamefont {Marin-Rimoldi}, \citenamefont {Ellis},\ and\ \citenamefont {Crawford}}]{Molssi_driver}%
  \BibitemOpen
  \bibfield  {author} {\bibinfo {author} {\bibfnamefont {T.~A.}\ \bibnamefont {Barnes}}, \bibinfo {author} {\bibfnamefont {E.}~\bibnamefont {Marin-Rimoldi}}, \bibinfo {author} {\bibfnamefont {S.}~\bibnamefont {Ellis}}, \ and\ \bibinfo {author} {\bibfnamefont {T.~D.}\ \bibnamefont {Crawford}},\ }\bibfield  {title} {\enquote {\bibinfo {title} {{The MolSSI Driver Interface Project: A framework for standardized, on-the-fly interoperability between computational molecular sciences codes}},}\ }\href {\doibase 10.1016/j.cpc.2020.107688} {\bibfield  {journal} {\bibinfo  {journal} {Comput. Phys. Commun.}\ }\textbf {\bibinfo {volume} {261}},\ \bibinfo {pages} {107688} (\bibinfo {year} {2021})}\BibitemShut {NoStop}%
\bibitem [{\citenamefont {Martí}(2021)}]{QM3}%
  \BibitemOpen
  \bibfield  {author} {\bibinfo {author} {\bibfnamefont {S.}~\bibnamefont {Martí}},\ }\bibfield  {title} {\enquote {\bibinfo {title} {{QMCube (QM\textsuperscript{3}): An all-purpose suite for multiscale QM/MM calculations}},}\ }\href {\doibase 10.1002/jcc.26465} {\bibfield  {journal} {\bibinfo  {journal} {J. Comput. Chem.}\ }\textbf {\bibinfo {volume} {42}},\ \bibinfo {pages} {447--457} (\bibinfo {year} {2021})}\BibitemShut {NoStop}%
\bibitem [{\citenamefont {Metz}\ \emph {et~al.}(2014)\citenamefont {Metz}, \citenamefont {Kästner}, \citenamefont {Sokol}, \citenamefont {Keal},\ and\ \citenamefont {Sherwood}}]{ChemShell_QM_MM}%
  \BibitemOpen
  \bibfield  {author} {\bibinfo {author} {\bibfnamefont {S.}~\bibnamefont {Metz}}, \bibinfo {author} {\bibfnamefont {J.}~\bibnamefont {Kästner}}, \bibinfo {author} {\bibfnamefont {A.~A.}\ \bibnamefont {Sokol}}, \bibinfo {author} {\bibfnamefont {T.~W.}\ \bibnamefont {Keal}}, \ and\ \bibinfo {author} {\bibfnamefont {P.}~\bibnamefont {Sherwood}},\ }\bibfield  {title} {\enquote {\bibinfo {title} {{ChemShell---a modular software package for QM/MM simulations}},}\ }\href {\doibase 10.1002/wcms.1163} {\bibfield  {journal} {\bibinfo  {journal} {Wiley Interdiscip. Rev. Comput. Mol. Sci.}\ }\textbf {\bibinfo {volume} {4}},\ \bibinfo {pages} {101--110} (\bibinfo {year} {2014})}\BibitemShut {NoStop}%
\bibitem [{\citenamefont {Lu}\ \emph {et~al.}(2019)\citenamefont {Lu}, \citenamefont {Farrow}, \citenamefont {Fayon}, \citenamefont {Logsdail}, \citenamefont {Sokol}, \citenamefont {Catlow}, \citenamefont {Sherwood},\ and\ \citenamefont {Keal}}]{ChemShell_QM_MM_redevelopment}%
  \BibitemOpen
  \bibfield  {author} {\bibinfo {author} {\bibfnamefont {Y.}~\bibnamefont {Lu}}, \bibinfo {author} {\bibfnamefont {M.~R.}\ \bibnamefont {Farrow}}, \bibinfo {author} {\bibfnamefont {P.}~\bibnamefont {Fayon}}, \bibinfo {author} {\bibfnamefont {A.~J.}\ \bibnamefont {Logsdail}}, \bibinfo {author} {\bibfnamefont {A.~A.}\ \bibnamefont {Sokol}}, \bibinfo {author} {\bibfnamefont {C.~R.~A.}\ \bibnamefont {Catlow}}, \bibinfo {author} {\bibfnamefont {P.}~\bibnamefont {Sherwood}}, \ and\ \bibinfo {author} {\bibfnamefont {T.~W.}\ \bibnamefont {Keal}},\ }\bibfield  {title} {\enquote {\bibinfo {title} {{Open-Source, Python-Based Redevelopment of the ChemShell Multiscale QM/MM Environment}},}\ }\href {\doibase 10.1021/acs.jctc.8b01036} {\bibfield  {journal} {\bibinfo  {journal} {J. Chem. Theory Comput.}\ }\textbf {\bibinfo {volume} {15}},\ \bibinfo {pages} {1317--1328} (\bibinfo {year} {2019})},\ \bibinfo {note} {pMID: 30511845}\BibitemShut {NoStop}%
\bibitem [{\citenamefont {Lu}\ \emph {et~al.}(2023)\citenamefont {Lu}, \citenamefont {Sen}, \citenamefont {Yong}, \citenamefont {Gunn}, \citenamefont {Purton}, \citenamefont {Guan}, \citenamefont {Desmoutier}, \citenamefont {Abdul~Nasir}, \citenamefont {Zhang}, \citenamefont {Zhu}, \citenamefont {Hou}, \citenamefont {Jackson-Masters}, \citenamefont {Watts}, \citenamefont {Hanson}, \citenamefont {Thomas}, \citenamefont {Jayawardena}, \citenamefont {Logsdail}, \citenamefont {Woodley}, \citenamefont {Senn}, \citenamefont {Sherwood}, \citenamefont {Catlow}, \citenamefont {Sokol},\ and\ \citenamefont {Keal}}]{Lu2023}%
  \BibitemOpen
  \bibfield  {author} {\bibinfo {author} {\bibfnamefont {Y.}~\bibnamefont {Lu}}, \bibinfo {author} {\bibfnamefont {K.}~\bibnamefont {Sen}}, \bibinfo {author} {\bibfnamefont {C.}~\bibnamefont {Yong}}, \bibinfo {author} {\bibfnamefont {D.~S.~D.}\ \bibnamefont {Gunn}}, \bibinfo {author} {\bibfnamefont {J.~A.}\ \bibnamefont {Purton}}, \bibinfo {author} {\bibfnamefont {J.}~\bibnamefont {Guan}}, \bibinfo {author} {\bibfnamefont {A.}~\bibnamefont {Desmoutier}}, \bibinfo {author} {\bibfnamefont {J.}~\bibnamefont {Abdul~Nasir}}, \bibinfo {author} {\bibfnamefont {X.}~\bibnamefont {Zhang}}, \bibinfo {author} {\bibfnamefont {L.}~\bibnamefont {Zhu}}, \bibinfo {author} {\bibfnamefont {Q.}~\bibnamefont {Hou}}, \bibinfo {author} {\bibfnamefont {J.}~\bibnamefont {Jackson-Masters}}, \bibinfo {author} {\bibfnamefont {S.}~\bibnamefont {Watts}}, \bibinfo {author} {\bibfnamefont {R.}~\bibnamefont {Hanson}}, \bibinfo {author} {\bibfnamefont {H.~N.}\ \bibnamefont {Thomas}}, \bibinfo {author} {\bibfnamefont {O.}~\bibnamefont
  {Jayawardena}}, \bibinfo {author} {\bibfnamefont {A.~J.}\ \bibnamefont {Logsdail}}, \bibinfo {author} {\bibfnamefont {S.~M.}\ \bibnamefont {Woodley}}, \bibinfo {author} {\bibfnamefont {H.~M.}\ \bibnamefont {Senn}}, \bibinfo {author} {\bibfnamefont {P.}~\bibnamefont {Sherwood}}, \bibinfo {author} {\bibfnamefont {C.~R.~A.}\ \bibnamefont {Catlow}}, \bibinfo {author} {\bibfnamefont {A.~A.}\ \bibnamefont {Sokol}}, \ and\ \bibinfo {author} {\bibfnamefont {T.~W.}\ \bibnamefont {Keal}},\ }\bibfield  {title} {\enquote {\bibinfo {title} {{Multiscale QM/MM modelling of catalytic systems with ChemShell}},}\ }\href {\doibase 10.1039/d3cp00648d} {\bibfield  {journal} {\bibinfo  {journal} {Phys. Chem. Chem. Phys.}\ }\textbf {\bibinfo {volume} {25}},\ \bibinfo {pages} {21816--21835} (\bibinfo {year} {2023})}\BibitemShut {NoStop}%
\bibitem [{\citenamefont {Lin}\ \emph {et~al.}(2024)\citenamefont {Lin}, \citenamefont {Zhang}, \citenamefont {Pezeshki}, \citenamefont {Duster}, \citenamefont {Wang}, \citenamefont {Wu}, \citenamefont {Zheng}, \citenamefont {Gagliardi},\ and\ \citenamefont {Truhlar}}]{QMMM2023}%
  \BibitemOpen
  \bibfield  {author} {\bibinfo {author} {\bibfnamefont {H.}~\bibnamefont {Lin}}, \bibinfo {author} {\bibfnamefont {Y.}~\bibnamefont {Zhang}}, \bibinfo {author} {\bibfnamefont {S.}~\bibnamefont {Pezeshki}}, \bibinfo {author} {\bibfnamefont {A.~W.}\ \bibnamefont {Duster}}, \bibinfo {author} {\bibfnamefont {B.}~\bibnamefont {Wang}}, \bibinfo {author} {\bibfnamefont {X.-P.}\ \bibnamefont {Wu}}, \bibinfo {author} {\bibfnamefont {S.-W.}\ \bibnamefont {Zheng}}, \bibinfo {author} {\bibfnamefont {L.}~\bibnamefont {Gagliardi}}, \ and\ \bibinfo {author} {\bibfnamefont {D.~G.}\ \bibnamefont {Truhlar}},\ }\bibfield  {title} {\enquote {\bibinfo {title} {{QMMM 2023: A program for combined quantum mechanical and molecular mechanical modeling and simulations}},}\ }\href {\doibase 10.1016/j.cpc.2023.108987} {\bibfield  {journal} {\bibinfo  {journal} {Comput. Phys. Commun.}\ }\textbf {\bibinfo {volume} {295}},\ \bibinfo {pages} {108987} (\bibinfo {year} {2024})}\BibitemShut {NoStop}%
\bibitem [{\citenamefont {Isborn}\ \emph {et~al.}(2012)\citenamefont {Isborn}, \citenamefont {Götz}, \citenamefont {Clark}, \citenamefont {Walker},\ and\ \citenamefont {Martínez}}]{Isborn_jctc}%
  \BibitemOpen
  \bibfield  {author} {\bibinfo {author} {\bibfnamefont {C.~M.}\ \bibnamefont {Isborn}}, \bibinfo {author} {\bibfnamefont {A.~W.}\ \bibnamefont {Götz}}, \bibinfo {author} {\bibfnamefont {M.~A.}\ \bibnamefont {Clark}}, \bibinfo {author} {\bibfnamefont {R.~C.}\ \bibnamefont {Walker}}, \ and\ \bibinfo {author} {\bibfnamefont {T.~J.}\ \bibnamefont {Martínez}},\ }\bibfield  {title} {\enquote {\bibinfo {title} {{Electronic Absorption Spectra from MM and ab Initio QM/MM Molecular Dynamics: Environmental Effects on the Absorption Spectrum of Photoactive Yellow Protein}},}\ }\href {\doibase 10.1021/ct3006826} {\bibfield  {journal} {\bibinfo  {journal} {J. Chem. Theory Comput.}\ }\textbf {\bibinfo {volume} {8}},\ \bibinfo {pages} {5092--5106} (\bibinfo {year} {2012})}\BibitemShut {NoStop}%
\bibitem [{\citenamefont {Cruzeiro}\ \emph {et~al.}(2021)\citenamefont {Cruzeiro}, \citenamefont {Manathunga}, \citenamefont {Merz},\ and\ \citenamefont {Götz}}]{QUICK_AMBER}%
  \BibitemOpen
  \bibfield  {author} {\bibinfo {author} {\bibfnamefont {V.~W.~D.}\ \bibnamefont {Cruzeiro}}, \bibinfo {author} {\bibfnamefont {M.}~\bibnamefont {Manathunga}}, \bibinfo {author} {\bibfnamefont {K.~M.~J.}\ \bibnamefont {Merz}}, \ and\ \bibinfo {author} {\bibfnamefont {A.~W.}\ \bibnamefont {Götz}},\ }\bibfield  {title} {\enquote {\bibinfo {title} {{Open-Source Multi-GPU-Accelerated QM/MM Simulations with AMBER and QUICK}},}\ }\href {\doibase 10.1021/acs.jcim.1c00169} {\bibfield  {journal} {\bibinfo  {journal} {J. Chem. Inf. Model.}\ }\textbf {\bibinfo {volume} {61}},\ \bibinfo {pages} {2109--2115} (\bibinfo {year} {2021})}\BibitemShut {NoStop}%
\bibitem [{\citenamefont {Olsen}\ \emph {et~al.}(2019)\citenamefont {Olsen}, \citenamefont {Bolnykh}, \citenamefont {Meloni}, \citenamefont {Ippoliti}, \citenamefont {Bircher}, \citenamefont {Carloni},\ and\ \citenamefont {Rothlisberger}}]{olsen2019mimic}%
  \BibitemOpen
  \bibfield  {author} {\bibinfo {author} {\bibfnamefont {J.~M.~H.}\ \bibnamefont {Olsen}}, \bibinfo {author} {\bibfnamefont {V.}~\bibnamefont {Bolnykh}}, \bibinfo {author} {\bibfnamefont {S.}~\bibnamefont {Meloni}}, \bibinfo {author} {\bibfnamefont {E.}~\bibnamefont {Ippoliti}}, \bibinfo {author} {\bibfnamefont {M.~P.}\ \bibnamefont {Bircher}}, \bibinfo {author} {\bibfnamefont {P.}~\bibnamefont {Carloni}}, \ and\ \bibinfo {author} {\bibfnamefont {U.}~\bibnamefont {Rothlisberger}},\ }\bibfield  {title} {\enquote {\bibinfo {title} {{MiMiC: A Novel Framework for Multiscale Modeling in Computational Chemistry}},}\ }\href {\doibase 10.1021/acs.jctc.9b00093} {\bibfield  {journal} {\bibinfo  {journal} {J. Chem. Theory Comput.}\ }\textbf {\bibinfo {volume} {15}},\ \bibinfo {pages} {3810--3823} (\bibinfo {year} {2019})}\BibitemShut {NoStop}%
\bibitem [{\citenamefont {Bolnykh}\ \emph {et~al.}(2019)\citenamefont {Bolnykh}, \citenamefont {Olsen}, \citenamefont {Meloni}, \citenamefont {Bircher}, \citenamefont {Ippoliti}, \citenamefont {Carloni},\ and\ \citenamefont {Rothlisberger}}]{mimic_jctc_hpc}%
  \BibitemOpen
  \bibfield  {author} {\bibinfo {author} {\bibfnamefont {V.}~\bibnamefont {Bolnykh}}, \bibinfo {author} {\bibfnamefont {J.~M.~H.}\ \bibnamefont {Olsen}}, \bibinfo {author} {\bibfnamefont {S.}~\bibnamefont {Meloni}}, \bibinfo {author} {\bibfnamefont {M.~P.}\ \bibnamefont {Bircher}}, \bibinfo {author} {\bibfnamefont {E.}~\bibnamefont {Ippoliti}}, \bibinfo {author} {\bibfnamefont {P.}~\bibnamefont {Carloni}}, \ and\ \bibinfo {author} {\bibfnamefont {U.}~\bibnamefont {Rothlisberger}},\ }\bibfield  {title} {\enquote {\bibinfo {title} {{Extreme Scalability of DFT-Based QM/MM MD Simulations Using MiMiC}},}\ }\href {\doibase 10.1021/acs.jctc.9b00424} {\bibfield  {journal} {\bibinfo  {journal} {J. Chem. Theory Comput.}\ }\textbf {\bibinfo {volume} {15}},\ \bibinfo {pages} {5601--5613} (\bibinfo {year} {2019})}\BibitemShut {NoStop}%
\bibitem [{\citenamefont {Raghavan}\ \emph {et~al.}(2023{\natexlab{a}})\citenamefont {Raghavan}, \citenamefont {Schackert}, \citenamefont {Levy}, \citenamefont {Johnson}, \citenamefont {Ippoliti}, \citenamefont {Mandelli}, \citenamefont {Olsen}, \citenamefont {Rothlisberger},\ and\ \citenamefont {Carloni}}]{raghavan2023mimicpy}%
  \BibitemOpen
  \bibfield  {author} {\bibinfo {author} {\bibfnamefont {B.}~\bibnamefont {Raghavan}}, \bibinfo {author} {\bibfnamefont {F.~K.}\ \bibnamefont {Schackert}}, \bibinfo {author} {\bibfnamefont {A.}~\bibnamefont {Levy}}, \bibinfo {author} {\bibfnamefont {S.~K.}\ \bibnamefont {Johnson}}, \bibinfo {author} {\bibfnamefont {E.}~\bibnamefont {Ippoliti}}, \bibinfo {author} {\bibfnamefont {D.}~\bibnamefont {Mandelli}}, \bibinfo {author} {\bibfnamefont {J.~M.~H.}\ \bibnamefont {Olsen}}, \bibinfo {author} {\bibfnamefont {U.}~\bibnamefont {Rothlisberger}}, \ and\ \bibinfo {author} {\bibfnamefont {P.}~\bibnamefont {Carloni}},\ }\bibfield  {title} {\enquote {\bibinfo {title} {{MiMiCPy: An Efficient Toolkit for MiMiC-Based QM/MM Simulations}},}\ }\href {\doibase 10.1021/acs.jcim.2c01620} {\bibfield  {journal} {\bibinfo  {journal} {J. Chem. Inf. Model.}\ }\textbf {\bibinfo {volume} {63}},\ \bibinfo {pages} {1406--1412} (\bibinfo {year} {2023}{\natexlab{a}})}\BibitemShut {NoStop}%
\bibitem [{\citenamefont {Humphrey}, \citenamefont {Dalke},\ and\ \citenamefont {Schulten}(1996)}]{humphrey1996vmd}%
  \BibitemOpen
  \bibfield  {author} {\bibinfo {author} {\bibfnamefont {W.}~\bibnamefont {Humphrey}}, \bibinfo {author} {\bibfnamefont {A.}~\bibnamefont {Dalke}}, \ and\ \bibinfo {author} {\bibfnamefont {K.}~\bibnamefont {Schulten}},\ }\bibfield  {title} {\enquote {\bibinfo {title} {{VMD: Visual molecular dynamics}},}\ }\href {\doibase 10.1016/0263-7855(96)00018-5} {\bibfield  {journal} {\bibinfo  {journal} {J. Mol. Graphics}\ }\textbf {\bibinfo {volume} {14}},\ \bibinfo {pages} {33--38} (\bibinfo {year} {1996})}\BibitemShut {NoStop}%
\bibitem [{\citenamefont {{Schr\"odinger, LLC}}(2015)}]{schrodinger2015pymol}%
  \BibitemOpen
  \bibfield  {author} {\bibinfo {author} {\bibnamefont {{Schr\"odinger, LLC}}},\ }\href@noop {} {\enquote {\bibinfo {title} {{The {PyMOL} Molecular Graphics System, Version~1.8}},}\ } (\bibinfo {year} {2015})\BibitemShut {NoStop}%
\bibitem [{mim(2024)}]{mimic-projects}%
  \BibitemOpen
  \href@noop {} {\enquote {\bibinfo {title} {{MiMiC Project on GitLab}},}\ }\bibinfo {howpublished} {\url{https://gitlab.com/mimic-project}} (\bibinfo {year} {2024}),\ \bibinfo {note} {date accessed: 2024-03-27}\BibitemShut {NoStop}%
\bibitem [{\citenamefont {Olsen}\ \emph {et~al.}(2022)\citenamefont {Olsen}, \citenamefont {Bolnykh}, \citenamefont {Meloni}, \citenamefont {Ippoliti}, \citenamefont {Carloni},\ and\ \citenamefont {Röthlisberger}}]{olsen_2022_7304688}%
  \BibitemOpen
  \bibfield  {author} {\bibinfo {author} {\bibfnamefont {J.~M.~H.}\ \bibnamefont {Olsen}}, \bibinfo {author} {\bibfnamefont {V.}~\bibnamefont {Bolnykh}}, \bibinfo {author} {\bibfnamefont {S.}~\bibnamefont {Meloni}}, \bibinfo {author} {\bibfnamefont {E.}~\bibnamefont {Ippoliti}}, \bibinfo {author} {\bibfnamefont {P.}~\bibnamefont {Carloni}}, \ and\ \bibinfo {author} {\bibfnamefont {U.}~\bibnamefont {Röthlisberger}},\ }\bibfield  {title} {\enquote {\bibinfo {title} {{MiMiC: A Framework for Multiscale Modeling in Computational Chemistry}},}\ }\href {\doibase 10.5281/zenodo.5024022} {\bibfield  {journal} {\bibinfo  {journal} {{Zenodo}}\ } (\bibinfo {year} {2022}),\ 10.5281/zenodo.5024022}\BibitemShut {NoStop}%
\bibitem [{\citenamefont {Bolnykh}\ \emph {et~al.}(2023)\citenamefont {Bolnykh}, \citenamefont {Olsen}, \citenamefont {Meloni}, \citenamefont {Ippoliti}, \citenamefont {Malapally}, \citenamefont {Röthlisberger},\ and\ \citenamefont {Carloni}}]{bolnykh_2023_7497400}%
  \BibitemOpen
  \bibfield  {author} {\bibinfo {author} {\bibfnamefont {V.}~\bibnamefont {Bolnykh}}, \bibinfo {author} {\bibfnamefont {J.~M.~H.}\ \bibnamefont {Olsen}}, \bibinfo {author} {\bibfnamefont {S.}~\bibnamefont {Meloni}}, \bibinfo {author} {\bibfnamefont {E.}~\bibnamefont {Ippoliti}}, \bibinfo {author} {\bibfnamefont {N.}~\bibnamefont {Malapally}}, \bibinfo {author} {\bibfnamefont {U.}~\bibnamefont {Röthlisberger}}, \ and\ \bibinfo {author} {\bibfnamefont {P.}~\bibnamefont {Carloni}},\ }\bibfield  {title} {\enquote {\bibinfo {title} {{MiMiC Communication Library}},}\ }\href {\doibase 10.5281/zenodo.5035084} {\bibfield  {journal} {\bibinfo  {journal} {{Zenodo}}\ } (\bibinfo {year} {2023}),\ 10.5281/zenodo.5035084}\BibitemShut {NoStop}%
\bibitem [{\citenamefont {Preston-Werner}(2013)}]{semver}%
  \BibitemOpen
  \bibfield  {author} {\bibinfo {author} {\bibfnamefont {T.}~\bibnamefont {Preston-Werner}},\ }\href@noop {} {\enquote {\bibinfo {title} {{Semantic Versioning 2.0.0}},}\ }\bibinfo {howpublished} {\url{https://semver.org/}} (\bibinfo {year} {2013})\BibitemShut {NoStop}%
\bibitem [{\citenamefont {Laio}, \citenamefont {VandeVondele},\ and\ \citenamefont {Rothlisberger}(2002{\natexlab{a}})}]{laio2002hamiltonian}%
  \BibitemOpen
  \bibfield  {author} {\bibinfo {author} {\bibfnamefont {A.}~\bibnamefont {Laio}}, \bibinfo {author} {\bibfnamefont {J.}~\bibnamefont {VandeVondele}}, \ and\ \bibinfo {author} {\bibfnamefont {U.}~\bibnamefont {Rothlisberger}},\ }\bibfield  {title} {\enquote {\bibinfo {title} {{A Hamiltonian electrostatic coupling scheme for hybrid Car--Parrinello molecular dynamics simulations}},}\ }\href {\doibase 10.1063/1.1462041} {\bibfield  {journal} {\bibinfo  {journal} {J. Chem. Phys.}\ }\textbf {\bibinfo {volume} {116}},\ \bibinfo {pages} {6941--6947} (\bibinfo {year} {2002}{\natexlab{a}})}\BibitemShut {NoStop}%
\bibitem [{\citenamefont {Raymond}(1991)}]{Raymond1991}%
  \BibitemOpen
  \bibfield  {author} {\bibinfo {author} {\bibfnamefont {X.}~\bibnamefont {Raymond}},\ }\href@noop {} {\emph {\bibinfo {title} {Elementary introduction to the theory of pseudodifferential operators}}},\ Studies in Advanced Mathematics\ (\bibinfo  {publisher} {CRC Press},\ \bibinfo {address} {Boca Raton, FL},\ \bibinfo {year} {1991})\BibitemShut {NoStop}%
\bibitem [{\citenamefont {Bondanza}\ \emph {et~al.}(2020)\citenamefont {Bondanza}, \citenamefont {Nottoli}, \citenamefont {Cupellini}, \citenamefont {Lipparini},\ and\ \citenamefont {Mennucci}}]{Bondanza2020-lp}%
  \BibitemOpen
  \bibfield  {author} {\bibinfo {author} {\bibfnamefont {M.}~\bibnamefont {Bondanza}}, \bibinfo {author} {\bibfnamefont {M.}~\bibnamefont {Nottoli}}, \bibinfo {author} {\bibfnamefont {L.}~\bibnamefont {Cupellini}}, \bibinfo {author} {\bibfnamefont {F.}~\bibnamefont {Lipparini}}, \ and\ \bibinfo {author} {\bibfnamefont {B.}~\bibnamefont {Mennucci}},\ }\bibfield  {title} {\enquote {\bibinfo {title} {{Polarizable embedding QM/MM: The future gold standard for complex (bio)systems?}}}\ }\href {\doibase 10.1039/D0CP02119A} {\bibfield  {journal} {\bibinfo  {journal} {Phys. Chem. Chem. Phys.}\ }\textbf {\bibinfo {volume} {22}},\ \bibinfo {pages} {14433--14448} (\bibinfo {year} {2020})}\BibitemShut {NoStop}%
\bibitem [{\citenamefont {Ponder}\ \emph {et~al.}(2010)\citenamefont {Ponder}, \citenamefont {Wu}, \citenamefont {Ren}, \citenamefont {Pande}, \citenamefont {Chodera}, \citenamefont {Schnieders}, \citenamefont {Haque}, \citenamefont {Mobley}, \citenamefont {Lambrecht}, \citenamefont {Distasio}, \citenamefont {Head-Gordon}, \citenamefont {Clark}, \citenamefont {Johnson},\ and\ \citenamefont {Head-Gordon}}]{Ponder2010-nr}%
  \BibitemOpen
  \bibfield  {author} {\bibinfo {author} {\bibfnamefont {J.~W.}\ \bibnamefont {Ponder}}, \bibinfo {author} {\bibfnamefont {C.}~\bibnamefont {Wu}}, \bibinfo {author} {\bibfnamefont {P.}~\bibnamefont {Ren}}, \bibinfo {author} {\bibfnamefont {V.~S.}\ \bibnamefont {Pande}}, \bibinfo {author} {\bibfnamefont {J.~D.}\ \bibnamefont {Chodera}}, \bibinfo {author} {\bibfnamefont {M.~J.}\ \bibnamefont {Schnieders}}, \bibinfo {author} {\bibfnamefont {I.}~\bibnamefont {Haque}}, \bibinfo {author} {\bibfnamefont {D.~L.}\ \bibnamefont {Mobley}}, \bibinfo {author} {\bibfnamefont {D.~S.}\ \bibnamefont {Lambrecht}}, \bibinfo {author} {\bibfnamefont {R.~A.}\ \bibnamefont {Distasio}}, \bibinfo {author} {\bibfnamefont {M.}~\bibnamefont {Head-Gordon}}, \bibinfo {author} {\bibfnamefont {G.~N.~I.}\ \bibnamefont {Clark}}, \bibinfo {author} {\bibfnamefont {M.~E.}\ \bibnamefont {Johnson}}, \ and\ \bibinfo {author} {\bibfnamefont {T.}~\bibnamefont {Head-Gordon}},\ }\bibfield  {title} {\enquote {\bibinfo {title} {{Current Status of the
  AMOEBA Polarizable Force Field}},}\ }\href {\doibase 10.1021/jp910674d} {\bibfield  {journal} {\bibinfo  {journal} {J. Phys. Chem. B}\ }\textbf {\bibinfo {volume} {114}},\ \bibinfo {pages} {2549--2564} (\bibinfo {year} {2010})}\BibitemShut {NoStop}%
\bibitem [{\citenamefont {Loco}\ \emph {et~al.}(2017)\citenamefont {Loco}, \citenamefont {Lagardère}, \citenamefont {Caprasecca}, \citenamefont {Lipparini}, \citenamefont {Mennucci},\ and\ \citenamefont {Piquemal}}]{locohybridqmmmmd}%
  \BibitemOpen
  \bibfield  {author} {\bibinfo {author} {\bibfnamefont {D.}~\bibnamefont {Loco}}, \bibinfo {author} {\bibfnamefont {L.}~\bibnamefont {Lagardère}}, \bibinfo {author} {\bibfnamefont {S.}~\bibnamefont {Caprasecca}}, \bibinfo {author} {\bibfnamefont {F.}~\bibnamefont {Lipparini}}, \bibinfo {author} {\bibfnamefont {B.}~\bibnamefont {Mennucci}}, \ and\ \bibinfo {author} {\bibfnamefont {J.-P.}\ \bibnamefont {Piquemal}},\ }\bibfield  {title} {\enquote {\bibinfo {title} {{Hybrid QM/MM Molecular Dynamics with AMOEBA Polarizable Embedding}},}\ }\href {\doibase 10.1021/acs.jctc.7b00572} {\bibfield  {journal} {\bibinfo  {journal} {J. Chem. Theory Comput.}\ }\textbf {\bibinfo {volume} {13}},\ \bibinfo {pages} {4025--4033} (\bibinfo {year} {2017})}\BibitemShut {NoStop}%
\bibitem [{\citenamefont {Nottoli}\ and\ \citenamefont {Lipparini}(2020)}]{Nottoli2020-ty}%
  \BibitemOpen
  \bibfield  {author} {\bibinfo {author} {\bibfnamefont {M.}~\bibnamefont {Nottoli}}\ and\ \bibinfo {author} {\bibfnamefont {F.}~\bibnamefont {Lipparini}},\ }\bibfield  {title} {\enquote {\bibinfo {title} {{General formulation of polarizable embedding models and of their coupling}},}\ }\href {\doibase 10.1063/5.0035165} {\bibfield  {journal} {\bibinfo  {journal} {J. Chem. Phys.}\ }\textbf {\bibinfo {volume} {153}},\ \bibinfo {pages} {224108} (\bibinfo {year} {2020})}\BibitemShut {NoStop}%
\bibitem [{\citenamefont {Thole}(1981)}]{Thole1981-po}%
  \BibitemOpen
  \bibfield  {author} {\bibinfo {author} {\bibfnamefont {B.~T.}\ \bibnamefont {Thole}},\ }\bibfield  {title} {\enquote {\bibinfo {title} {{Molecular polarizabilities calculated with a modified dipole interaction}},}\ }\href {\doibase 10.1016/0301-0104(81)85176-2} {\bibfield  {journal} {\bibinfo  {journal} {Chem. Phys.}\ }\textbf {\bibinfo {volume} {59}},\ \bibinfo {pages} {341--350} (\bibinfo {year} {1981})}\BibitemShut {NoStop}%
\bibitem [{\citenamefont {Tuckerman}, \citenamefont {Berne},\ and\ \citenamefont {Martyna}(1992)}]{tuckerman1992reversible}%
  \BibitemOpen
  \bibfield  {author} {\bibinfo {author} {\bibfnamefont {M.}~\bibnamefont {Tuckerman}}, \bibinfo {author} {\bibfnamefont {B.~J.}\ \bibnamefont {Berne}}, \ and\ \bibinfo {author} {\bibfnamefont {G.~J.}\ \bibnamefont {Martyna}},\ }\bibfield  {title} {\enquote {\bibinfo {title} {{Reversible multiple time scale molecular dynamics}},}\ }\href {\doibase 10.1063/1.463137} {\bibfield  {journal} {\bibinfo  {journal} {J. Chem. Phys.}\ }\textbf {\bibinfo {volume} {97}},\ \bibinfo {pages} {1990--2001} (\bibinfo {year} {1992})}\BibitemShut {NoStop}%
\bibitem [{\citenamefont {Steele}(2013)}]{steele2013communication}%
  \BibitemOpen
  \bibfield  {author} {\bibinfo {author} {\bibfnamefont {R.~P.}\ \bibnamefont {Steele}},\ }\bibfield  {title} {\enquote {\bibinfo {title} {{Multiple-timestep ab initio molecular dynamics with electron correlation}},}\ }\href@noop {} {\bibfield  {journal} {\bibinfo  {journal} {J. Chem. Phys.}\ }\textbf {\bibinfo {volume} {139}},\ \bibinfo {pages} {011102} (\bibinfo {year} {2013})}\BibitemShut {NoStop}%
\bibitem [{\citenamefont {Liberatore}, \citenamefont {Meli},\ and\ \citenamefont {Rothlisberger}(2018)}]{liberatore2018versatile}%
  \BibitemOpen
  \bibfield  {author} {\bibinfo {author} {\bibfnamefont {E.}~\bibnamefont {Liberatore}}, \bibinfo {author} {\bibfnamefont {R.}~\bibnamefont {Meli}}, \ and\ \bibinfo {author} {\bibfnamefont {U.}~\bibnamefont {Rothlisberger}},\ }\bibfield  {title} {\enquote {\bibinfo {title} {{A Versatile Multiple Time Step Scheme for Efficient ab Initio Molecular Dynamics Simulations}},}\ }\href {\doibase 10.1021/acs.jctc.7b01189} {\bibfield  {journal} {\bibinfo  {journal} {J. Chem. Theory Comput.}\ }\textbf {\bibinfo {volume} {14}},\ \bibinfo {pages} {2834--2842} (\bibinfo {year} {2018})}\BibitemShut {NoStop}%
\bibitem [{\citenamefont {Mouvet}\ \emph {et~al.}(2022)\citenamefont {Mouvet}, \citenamefont {Villard}, \citenamefont {Bolnykh},\ and\ \citenamefont {Rothlisberger}}]{mouvet2022recent}%
  \BibitemOpen
  \bibfield  {author} {\bibinfo {author} {\bibfnamefont {F.}~\bibnamefont {Mouvet}}, \bibinfo {author} {\bibfnamefont {J.}~\bibnamefont {Villard}}, \bibinfo {author} {\bibfnamefont {V.}~\bibnamefont {Bolnykh}}, \ and\ \bibinfo {author} {\bibfnamefont {U.}~\bibnamefont {Rothlisberger}},\ }\bibfield  {title} {\enquote {\bibinfo {title} {{Recent Advances in First-Principles Based Molecular Dynamics}},}\ }\href {\doibase 10.1021/acs.accounts.1c00503} {\bibfield  {journal} {\bibinfo  {journal} {Acc. Chem. Res.}\ }\textbf {\bibinfo {volume} {55}},\ \bibinfo {pages} {221--230} (\bibinfo {year} {2022})}\BibitemShut {NoStop}%
\bibitem [{mim(2022)}]{mimic-school}%
  \BibitemOpen
  \href@noop {} {\enquote {\bibinfo {title} {{CECAM Flagship School: Multiscale Molecular Dynamics with MiMiC}},}\ }\bibinfo {howpublished} {\url{https://www.cecam.org/workshop-details/1119}} (\bibinfo {year} {2022}),\ \bibinfo {note} {date accessed: 2024-03-27}\BibitemShut {NoStop}%
\bibitem [{\citenamefont {Abraham}\ \emph {et~al.}(2015)\citenamefont {Abraham}, \citenamefont {Murtola}, \citenamefont {Schulz}, \citenamefont {Páll}, \citenamefont {Smith}, \citenamefont {Hess},\ and\ \citenamefont {Lindahl}}]{abraham2015}%
  \BibitemOpen
  \bibfield  {author} {\bibinfo {author} {\bibfnamefont {M.~J.}\ \bibnamefont {Abraham}}, \bibinfo {author} {\bibfnamefont {T.}~\bibnamefont {Murtola}}, \bibinfo {author} {\bibfnamefont {R.}~\bibnamefont {Schulz}}, \bibinfo {author} {\bibfnamefont {S.}~\bibnamefont {Páll}}, \bibinfo {author} {\bibfnamefont {J.~C.}\ \bibnamefont {Smith}}, \bibinfo {author} {\bibfnamefont {B.}~\bibnamefont {Hess}}, \ and\ \bibinfo {author} {\bibfnamefont {E.}~\bibnamefont {Lindahl}},\ }\bibfield  {title} {\enquote {\bibinfo {title} {{GROMACS: High performance molecular simulations through multi-level parallelism from laptops to supercomputers}},}\ }\href {\doibase 10.1016/j.softx.2015.06.001} {\bibfield  {journal} {\bibinfo  {journal} {SoftwareX}\ }\textbf {\bibinfo {volume} {1-2}},\ \bibinfo {pages} {19--25} (\bibinfo {year} {2015})}\BibitemShut {NoStop}%
\bibitem [{\citenamefont {Páll}\ \emph {et~al.}(2015)\citenamefont {Páll}, \citenamefont {Abraham}, \citenamefont {Kutzner}, \citenamefont {Hess},\ and\ \citenamefont {Lindahl}}]{gromacs2015}%
  \BibitemOpen
  \bibfield  {author} {\bibinfo {author} {\bibfnamefont {S.}~\bibnamefont {Páll}}, \bibinfo {author} {\bibfnamefont {M.~J.}\ \bibnamefont {Abraham}}, \bibinfo {author} {\bibfnamefont {C.}~\bibnamefont {Kutzner}}, \bibinfo {author} {\bibfnamefont {B.}~\bibnamefont {Hess}}, \ and\ \bibinfo {author} {\bibfnamefont {E.}~\bibnamefont {Lindahl}},\ }\bibfield  {title} {\enquote {\bibinfo {title} {{Tackling Exascale Software Challenges in Molecular Dynamics Simulations with GROMACS}},}\ }in\ \href {\doibase 10.1007/978-3-319-15976-8_1} {\emph {\bibinfo {booktitle} {Solving Software Challenges for Exascale}}}\ (\bibinfo  {publisher} {Springer International Publishing},\ \bibinfo {year} {2015})\ pp.\ \bibinfo {pages} {3--27}\BibitemShut {NoStop}%
\bibitem [{cpm(2023)}]{cpmd_free}%
  \BibitemOpen
  \href@noop {} {\enquote {\bibinfo {title} {{CPMD, copyright 1990-2023 by IBM Corp. and copyright 1994-2001 by Max Planck Institute, Stuttgart.}}}\ }\bibinfo {howpublished} {\url{http://www.cpmd.org/}} (\bibinfo {year} {2023}),\ \bibinfo {note} {date accessed: 2024-03-27}\BibitemShut {NoStop}%
\bibitem [{\citenamefont {Bonomi}\ \emph {et~al.}(2009)\citenamefont {Bonomi}, \citenamefont {Branduardi}, \citenamefont {Bussi}, \citenamefont {Camilloni}, \citenamefont {Provasi}, \citenamefont {Raiteri}, \citenamefont {Donadio}, \citenamefont {Marinelli}, \citenamefont {Pietrucci}, \citenamefont {Broglia},\ and\ \citenamefont {Parrinello}}]{Bonomi2009a}%
  \BibitemOpen
  \bibfield  {author} {\bibinfo {author} {\bibfnamefont {M.}~\bibnamefont {Bonomi}}, \bibinfo {author} {\bibfnamefont {D.}~\bibnamefont {Branduardi}}, \bibinfo {author} {\bibfnamefont {G.}~\bibnamefont {Bussi}}, \bibinfo {author} {\bibfnamefont {C.}~\bibnamefont {Camilloni}}, \bibinfo {author} {\bibfnamefont {D.}~\bibnamefont {Provasi}}, \bibinfo {author} {\bibfnamefont {P.}~\bibnamefont {Raiteri}}, \bibinfo {author} {\bibfnamefont {D.}~\bibnamefont {Donadio}}, \bibinfo {author} {\bibfnamefont {F.}~\bibnamefont {Marinelli}}, \bibinfo {author} {\bibfnamefont {F.}~\bibnamefont {Pietrucci}}, \bibinfo {author} {\bibfnamefont {R.~A.}\ \bibnamefont {Broglia}}, \ and\ \bibinfo {author} {\bibfnamefont {M.}~\bibnamefont {Parrinello}},\ }\bibfield  {title} {\enquote {\bibinfo {title} {{PLUMED: A portable plugin for free-energy calculations with molecular dynamics}},}\ }\href {\doibase 10.1016/j.cpc.2009.05.011} {\bibfield  {journal} {\bibinfo  {journal} {Comput. Phys. Commun.}\ }\textbf {\bibinfo {volume} {180}},\
  \bibinfo {pages} {1961--1972} (\bibinfo {year} {2009})}\BibitemShut {NoStop}%
\bibitem [{\citenamefont {Tribello}\ \emph {et~al.}(2014)\citenamefont {Tribello}, \citenamefont {Bonomi}, \citenamefont {Branduardi}, \citenamefont {Camilloni},\ and\ \citenamefont {Bussi}}]{PLUMED2}%
  \BibitemOpen
  \bibfield  {author} {\bibinfo {author} {\bibfnamefont {G.~A.}\ \bibnamefont {Tribello}}, \bibinfo {author} {\bibfnamefont {M.}~\bibnamefont {Bonomi}}, \bibinfo {author} {\bibfnamefont {D.}~\bibnamefont {Branduardi}}, \bibinfo {author} {\bibfnamefont {C.}~\bibnamefont {Camilloni}}, \ and\ \bibinfo {author} {\bibfnamefont {G.}~\bibnamefont {Bussi}},\ }\bibfield  {title} {\enquote {\bibinfo {title} {{PLUMED 2: New feathers for an old bird}},}\ }\href {\doibase 10.1016/j.cpc.2013.09.018} {\bibfield  {journal} {\bibinfo  {journal} {Comput. Phys. Commun.}\ }\textbf {\bibinfo {volume} {185}},\ \bibinfo {pages} {604--613} (\bibinfo {year} {2014})}\BibitemShut {NoStop}%
\bibitem [{\citenamefont {Col{\'{o}}n-Ramos}\ \emph {et~al.}(2019)\citenamefont {Col{\'{o}}n-Ramos}, \citenamefont {{La Riviere}}, \citenamefont {Shroff},\ and\ \citenamefont {Oldenbourg}}]{Colon-Ramos2019}%
  \BibitemOpen
  \bibfield  {author} {\bibinfo {author} {\bibfnamefont {D.~A.}\ \bibnamefont {Col{\'{o}}n-Ramos}}, \bibinfo {author} {\bibfnamefont {P.}~\bibnamefont {{La Riviere}}}, \bibinfo {author} {\bibfnamefont {H.}~\bibnamefont {Shroff}}, \ and\ \bibinfo {author} {\bibfnamefont {R.}~\bibnamefont {Oldenbourg}},\ }\bibfield  {title} {\enquote {\bibinfo {title} {{Promoting transparency and reproducibility in enhanced molecular simulations}},}\ }\href {\doibase 10.1038/s41592-019-0506-8} {\bibfield  {journal} {\bibinfo  {journal} {Nat. Methods}\ }\textbf {\bibinfo {volume} {16}},\ \bibinfo {pages} {670--673} (\bibinfo {year} {2019})}\BibitemShut {NoStop}%
\bibitem [{\citenamefont {Bircher}\ and\ \citenamefont {Rothlisberger}(2018{\natexlab{a}})}]{Bircher2018}%
  \BibitemOpen
  \bibfield  {author} {\bibinfo {author} {\bibfnamefont {M.~P.}\ \bibnamefont {Bircher}}\ and\ \bibinfo {author} {\bibfnamefont {U.}~\bibnamefont {Rothlisberger}},\ }\bibfield  {title} {\enquote {\bibinfo {title} {{Plane-Wave Implementation and Performance of à-la-Carte Coulomb-Attenuated Exchange-Correlation Functionals for Predicting Optical Excitation Energies in Some Notorious Cases}},}\ }\href {\doibase 10.1021/acs.jctc.8b00069} {\bibfield  {journal} {\bibinfo  {journal} {J. Chem. Theory Comput.}\ }\textbf {\bibinfo {volume} {14}},\ \bibinfo {pages} {3184--3195} (\bibinfo {year} {2018}{\natexlab{a}})}\BibitemShut {NoStop}%
\bibitem [{\citenamefont {Bircher}, \citenamefont {López-Tarifa},\ and\ \citenamefont {Rothlisberger}(2018)}]{Bircher2018shedding}%
  \BibitemOpen
  \bibfield  {author} {\bibinfo {author} {\bibfnamefont {M.~P.}\ \bibnamefont {Bircher}}, \bibinfo {author} {\bibfnamefont {P.}~\bibnamefont {López-Tarifa}}, \ and\ \bibinfo {author} {\bibfnamefont {U.}~\bibnamefont {Rothlisberger}},\ }\bibfield  {title} {\enquote {\bibinfo {title} {{Shedding Light on the Basis Set Dependence of the Minnesota Functionals: Differences Between Plane Waves, Slater Functions, and Gaussians}},}\ }\href {\doibase 10.1021/acs.jctc.8b00897} {\bibfield  {journal} {\bibinfo  {journal} {J. Chem. Theory Comput.}\ }\textbf {\bibinfo {volume} {15}},\ \bibinfo {pages} {557--571} (\bibinfo {year} {2018})}\BibitemShut {NoStop}%
\bibitem [{\citenamefont {Villard}, \citenamefont {Bircher},\ and\ \citenamefont {Rothlisberger}(2024)}]{Villard2024}%
  \BibitemOpen
  \bibfield  {author} {\bibinfo {author} {\bibfnamefont {J.}~\bibnamefont {Villard}}, \bibinfo {author} {\bibfnamefont {M.~P.}\ \bibnamefont {Bircher}}, \ and\ \bibinfo {author} {\bibfnamefont {U.}~\bibnamefont {Rothlisberger}},\ }\bibfield  {title} {\enquote {\bibinfo {title} {{Structure and dynamics of liquid water from ab initio simulations: adding Minnesota density functionals to Jacob’s ladder}},}\ }\href {\doibase 10.1039/d3sc05828j} {\bibfield  {journal} {\bibinfo  {journal} {Chem. Sci.}\ } (\bibinfo {year} {2024}),\ 10.1039/d3sc05828j}\BibitemShut {NoStop}%
\bibitem [{\citenamefont {Weber}\ \emph {et~al.}(2014)\citenamefont {Weber}, \citenamefont {Bekas}, \citenamefont {Laino}, \citenamefont {Curioni}, \citenamefont {Bertsch},\ and\ \citenamefont {Futral}}]{Weber2014}%
  \BibitemOpen
  \bibfield  {author} {\bibinfo {author} {\bibfnamefont {V.}~\bibnamefont {Weber}}, \bibinfo {author} {\bibfnamefont {C.}~\bibnamefont {Bekas}}, \bibinfo {author} {\bibfnamefont {T.}~\bibnamefont {Laino}}, \bibinfo {author} {\bibfnamefont {A.}~\bibnamefont {Curioni}}, \bibinfo {author} {\bibfnamefont {A.}~\bibnamefont {Bertsch}}, \ and\ \bibinfo {author} {\bibfnamefont {S.}~\bibnamefont {Futral}},\ }\bibfield  {title} {\enquote {\bibinfo {title} {{Shedding Light on Lithium/Air Batteries Using Millions of Threads on the BG/Q Supercomputer}},}\ }in\ \href {\doibase 10.1109/ipdps.2014.81} {\emph {\bibinfo {booktitle} {28th International Parallel and Distributed Processing Symposium}}}\ (\bibinfo  {publisher} {IEEE},\ \bibinfo {year} {2014})\BibitemShut {NoStop}%
\bibitem [{\citenamefont {Bircher}\ and\ \citenamefont {Rothlisberger}(2018{\natexlab{b}})}]{Bircher2018exploring}%
  \BibitemOpen
  \bibfield  {author} {\bibinfo {author} {\bibfnamefont {M.~P.}\ \bibnamefont {Bircher}}\ and\ \bibinfo {author} {\bibfnamefont {U.}~\bibnamefont {Rothlisberger}},\ }\bibfield  {title} {\enquote {\bibinfo {title} {{Exploiting Coordinate Scaling Relations To Accelerate Exact Exchange Calculations}},}\ }\href {\doibase 10.1021/acs.jpclett.8b01620} {\bibfield  {journal} {\bibinfo  {journal} {J. Phys. Chem. Lett.}\ }\textbf {\bibinfo {volume} {9}},\ \bibinfo {pages} {3886--3890} (\bibinfo {year} {2018}{\natexlab{b}})}\BibitemShut {NoStop}%
\bibitem [{\citenamefont {Bircher}, \citenamefont {Villard},\ and\ \citenamefont {Rothlisberger}(2020)}]{Bircher2020}%
  \BibitemOpen
  \bibfield  {author} {\bibinfo {author} {\bibfnamefont {M.~P.}\ \bibnamefont {Bircher}}, \bibinfo {author} {\bibfnamefont {J.}~\bibnamefont {Villard}}, \ and\ \bibinfo {author} {\bibfnamefont {U.}~\bibnamefont {Rothlisberger}},\ }\bibfield  {title} {\enquote {\bibinfo {title} {{Efficient Treatment of Correlation Energies at the Basis-Set Limit by Monte Carlo Summation of Continuum States}},}\ }\href {\doibase 10.1021/acs.jctc.0c00724} {\bibfield  {journal} {\bibinfo  {journal} {J. Chem. Theory Comput.}\ }\textbf {\bibinfo {volume} {16}},\ \bibinfo {pages} {6550--6559} (\bibinfo {year} {2020})}\BibitemShut {NoStop}%
\bibitem [{\citenamefont {Putrino}, \citenamefont {Sebastiani},\ and\ \citenamefont {Parrinello}(2000)}]{putrino2000generalized}%
  \BibitemOpen
  \bibfield  {author} {\bibinfo {author} {\bibfnamefont {A.}~\bibnamefont {Putrino}}, \bibinfo {author} {\bibfnamefont {D.}~\bibnamefont {Sebastiani}}, \ and\ \bibinfo {author} {\bibfnamefont {M.}~\bibnamefont {Parrinello}},\ }\bibfield  {title} {\enquote {\bibinfo {title} {{Generalized variational density functional perturbation theory}},}\ }\href {\doibase 10.1063/1.1312830} {\bibfield  {journal} {\bibinfo  {journal} {J. Chem. Phys.}\ }\textbf {\bibinfo {volume} {113}},\ \bibinfo {pages} {7102--7109} (\bibinfo {year} {2000})}\BibitemShut {NoStop}%
\bibitem [{\citenamefont {Hutter}(2003)}]{hutter2003excited}%
  \BibitemOpen
  \bibfield  {author} {\bibinfo {author} {\bibfnamefont {J.}~\bibnamefont {Hutter}},\ }\bibfield  {title} {\enquote {\bibinfo {title} {{Excited state nuclear forces from the Tamm--Dancoff approximation to time-dependent density functional theory within the plane wave basis set framework}},}\ }\href {\doibase 10.1063/1.1540109} {\bibfield  {journal} {\bibinfo  {journal} {J. Chem. Phys.}\ }\textbf {\bibinfo {volume} {118}},\ \bibinfo {pages} {3928--3934} (\bibinfo {year} {2003})}\BibitemShut {NoStop}%
\bibitem [{\citenamefont {Frank}\ \emph {et~al.}(1998)\citenamefont {Frank}, \citenamefont {Hutter}, \citenamefont {Marx},\ and\ \citenamefont {Parrinello}}]{frank1998molecular}%
  \BibitemOpen
  \bibfield  {author} {\bibinfo {author} {\bibfnamefont {I.}~\bibnamefont {Frank}}, \bibinfo {author} {\bibfnamefont {J.}~\bibnamefont {Hutter}}, \bibinfo {author} {\bibfnamefont {D.}~\bibnamefont {Marx}}, \ and\ \bibinfo {author} {\bibfnamefont {M.}~\bibnamefont {Parrinello}},\ }\bibfield  {title} {\enquote {\bibinfo {title} {{Molecular dynamics in low-spin excited states}},}\ }\href {\doibase 10.1063/1.475804} {\bibfield  {journal} {\bibinfo  {journal} {J. Chem. Phys.}\ }\textbf {\bibinfo {volume} {108}},\ \bibinfo {pages} {4060--4069} (\bibinfo {year} {1998})}\BibitemShut {NoStop}%
\bibitem [{\citenamefont {Tavernelli}, \citenamefont {R{\"o}hrig},\ and\ \citenamefont {Rothlisberger}(2005)}]{tavernelli2005molecular}%
  \BibitemOpen
  \bibfield  {author} {\bibinfo {author} {\bibfnamefont {I.}~\bibnamefont {Tavernelli}}, \bibinfo {author} {\bibfnamefont {U.~F.}\ \bibnamefont {R{\"o}hrig}}, \ and\ \bibinfo {author} {\bibfnamefont {U.}~\bibnamefont {Rothlisberger}},\ }\bibfield  {title} {\enquote {\bibinfo {title} {{Molecular dynamics in electronically excited states using time-dependent density functional theory}},}\ }\href {\doibase 10.1080/00268970512331339378} {\bibfield  {journal} {\bibinfo  {journal} {Mol. Phys.}\ }\textbf {\bibinfo {volume} {103}},\ \bibinfo {pages} {963--981} (\bibinfo {year} {2005})}\BibitemShut {NoStop}%
\bibitem [{\citenamefont {Tapavicza}, \citenamefont {Tavernelli},\ and\ \citenamefont {Rothlisberger}(2007)}]{tapavicza2007trajectory}%
  \BibitemOpen
  \bibfield  {author} {\bibinfo {author} {\bibfnamefont {E.}~\bibnamefont {Tapavicza}}, \bibinfo {author} {\bibfnamefont {I.}~\bibnamefont {Tavernelli}}, \ and\ \bibinfo {author} {\bibfnamefont {U.}~\bibnamefont {Rothlisberger}},\ }\bibfield  {title} {\enquote {\bibinfo {title} {{Trajectory surface hopping within linear response time-dependent density-functional theory}},}\ }\href {\doibase 10.1103/PhysRevLett.98.023001} {\bibfield  {journal} {\bibinfo  {journal} {Phys. Rev. Lett.}\ }\textbf {\bibinfo {volume} {98}},\ \bibinfo {pages} {023001} (\bibinfo {year} {2007})}\BibitemShut {NoStop}%
\bibitem [{\citenamefont {Tavernelli}, \citenamefont {Curchod},\ and\ \citenamefont {Rothlisberger}(2010)}]{tavernelli2010mixed}%
  \BibitemOpen
  \bibfield  {author} {\bibinfo {author} {\bibfnamefont {I.}~\bibnamefont {Tavernelli}}, \bibinfo {author} {\bibfnamefont {B.~F.}\ \bibnamefont {Curchod}}, \ and\ \bibinfo {author} {\bibfnamefont {U.}~\bibnamefont {Rothlisberger}},\ }\bibfield  {title} {\enquote {\bibinfo {title} {{Mixed quantum-classical dynamics with time-dependent external fields: A time-dependent density-functional-theory approach}},}\ }\href {\doibase 10.1103/PhysRevA.81.052508} {\bibfield  {journal} {\bibinfo  {journal} {Phys. Rev. A}\ }\textbf {\bibinfo {volume} {81}},\ \bibinfo {pages} {052508} (\bibinfo {year} {2010})}\BibitemShut {NoStop}%
\bibitem [{\citenamefont {Curchod}\ \emph {et~al.}(2011)\citenamefont {Curchod}, \citenamefont {Penfold}, \citenamefont {Rothlisberger},\ and\ \citenamefont {Tavernelli}}]{curchod2011local}%
  \BibitemOpen
  \bibfield  {author} {\bibinfo {author} {\bibfnamefont {B.~F.}\ \bibnamefont {Curchod}}, \bibinfo {author} {\bibfnamefont {T.~J.}\ \bibnamefont {Penfold}}, \bibinfo {author} {\bibfnamefont {U.}~\bibnamefont {Rothlisberger}}, \ and\ \bibinfo {author} {\bibfnamefont {I.}~\bibnamefont {Tavernelli}},\ }\bibfield  {title} {\enquote {\bibinfo {title} {{Local control theory in trajectory-based nonadiabatic dynamics}},}\ }\href {\doibase 10.1103/PhysRevA.84.042507} {\bibfield  {journal} {\bibinfo  {journal} {Phys. Rev. A}\ }\textbf {\bibinfo {volume} {84}},\ \bibinfo {pages} {042507} (\bibinfo {year} {2011})}\BibitemShut {NoStop}%
\bibitem [{\citenamefont {Marx}\ and\ \citenamefont {Parrinello}(1996)}]{marx1996ab}%
  \BibitemOpen
  \bibfield  {author} {\bibinfo {author} {\bibfnamefont {D.}~\bibnamefont {Marx}}\ and\ \bibinfo {author} {\bibfnamefont {M.}~\bibnamefont {Parrinello}},\ }\bibfield  {title} {\enquote {\bibinfo {title} {{Ab initio path integral molecular dynamics: Basic ideas}},}\ }\href {\doibase 10.1063/1.471221} {\bibfield  {journal} {\bibinfo  {journal} {J. Chem. Phys.}\ }\textbf {\bibinfo {volume} {104}},\ \bibinfo {pages} {4077--4082} (\bibinfo {year} {1996})}\BibitemShut {NoStop}%
\bibitem [{\citenamefont {Berendsen}\ \emph {et~al.}(1984)\citenamefont {Berendsen}, \citenamefont {Postma}, \citenamefont {Van~Gunsteren}, \citenamefont {DiNola},\ and\ \citenamefont {Haak}}]{berendsen1984molecular}%
  \BibitemOpen
  \bibfield  {author} {\bibinfo {author} {\bibfnamefont {H.~J.}\ \bibnamefont {Berendsen}}, \bibinfo {author} {\bibfnamefont {J.~v.}\ \bibnamefont {Postma}}, \bibinfo {author} {\bibfnamefont {W.~F.}\ \bibnamefont {Van~Gunsteren}}, \bibinfo {author} {\bibfnamefont {A.}~\bibnamefont {DiNola}}, \ and\ \bibinfo {author} {\bibfnamefont {J.~R.}\ \bibnamefont {Haak}},\ }\bibfield  {title} {\enquote {\bibinfo {title} {{Molecular dynamics with coupling to an external bath}},}\ }\href {\doibase 10.1063/1.448118} {\bibfield  {journal} {\bibinfo  {journal} {J. Chem. Phys.}\ }\textbf {\bibinfo {volume} {81}},\ \bibinfo {pages} {3684--3690} (\bibinfo {year} {1984})}\BibitemShut {NoStop}%
\bibitem [{\citenamefont {Nos{\'e}}(1984{\natexlab{a}})}]{nose1984unified}%
  \BibitemOpen
  \bibfield  {author} {\bibinfo {author} {\bibfnamefont {S.}~\bibnamefont {Nos{\'e}}},\ }\bibfield  {title} {\enquote {\bibinfo {title} {{A unified formulation of the constant temperature molecular dynamics methods}},}\ }\href {\doibase 10.1063/1.447334} {\bibfield  {journal} {\bibinfo  {journal} {J. Chem. Phys.}\ }\textbf {\bibinfo {volume} {81}},\ \bibinfo {pages} {511--519} (\bibinfo {year} {1984}{\natexlab{a}})}\BibitemShut {NoStop}%
\bibitem [{\citenamefont {Nos{\'e}}(1984{\natexlab{b}})}]{nose1984molecular}%
  \BibitemOpen
  \bibfield  {author} {\bibinfo {author} {\bibfnamefont {S.}~\bibnamefont {Nos{\'e}}},\ }\bibfield  {title} {\enquote {\bibinfo {title} {{A molecular dynamics method for simulations in the canonical ensemble}},}\ }\href {\doibase 10.1080/00268978400101201} {\bibfield  {journal} {\bibinfo  {journal} {Mol. Phys.}\ }\textbf {\bibinfo {volume} {52}},\ \bibinfo {pages} {255--268} (\bibinfo {year} {1984}{\natexlab{b}})}\BibitemShut {NoStop}%
\bibitem [{\citenamefont {Hoover}(1985)}]{hoover1985canonical}%
  \BibitemOpen
  \bibfield  {author} {\bibinfo {author} {\bibfnamefont {W.~G.}\ \bibnamefont {Hoover}},\ }\bibfield  {title} {\enquote {\bibinfo {title} {{Canonical dynamics: Equilibrium phase-space distributions}},}\ }\href {\doibase 10.1103/PhysRevA.31.1695} {\bibfield  {journal} {\bibinfo  {journal} {Phys. Rev. A}\ }\textbf {\bibinfo {volume} {31}},\ \bibinfo {pages} {1695} (\bibinfo {year} {1985})}\BibitemShut {NoStop}%
\bibitem [{\citenamefont {Ceriotti}, \citenamefont {Bussi},\ and\ \citenamefont {Parrinello}(2010)}]{ceriotti2010colored}%
  \BibitemOpen
  \bibfield  {author} {\bibinfo {author} {\bibfnamefont {M.}~\bibnamefont {Ceriotti}}, \bibinfo {author} {\bibfnamefont {G.}~\bibnamefont {Bussi}}, \ and\ \bibinfo {author} {\bibfnamefont {M.}~\bibnamefont {Parrinello}},\ }\bibfield  {title} {\enquote {\bibinfo {title} {Colored-noise thermostats {\`a} la carte},}\ }\href {\doibase 10.1021/ct900563s} {\bibfield  {journal} {\bibinfo  {journal} {J. Chem. Theory Comput.}\ }\textbf {\bibinfo {volume} {6}},\ \bibinfo {pages} {1170--1180} (\bibinfo {year} {2010})}\BibitemShut {NoStop}%
\bibitem [{\citenamefont {Martyna}\ \emph {et~al.}(1996)\citenamefont {Martyna}, \citenamefont {Tuckerman}, \citenamefont {Tobias},\ and\ \citenamefont {Klein}}]{martyna1996explicit}%
  \BibitemOpen
  \bibfield  {author} {\bibinfo {author} {\bibfnamefont {G.~J.}\ \bibnamefont {Martyna}}, \bibinfo {author} {\bibfnamefont {M.~E.}\ \bibnamefont {Tuckerman}}, \bibinfo {author} {\bibfnamefont {D.~J.}\ \bibnamefont {Tobias}}, \ and\ \bibinfo {author} {\bibfnamefont {M.~L.}\ \bibnamefont {Klein}},\ }\bibfield  {title} {\enquote {\bibinfo {title} {{Explicit reversible integrators for extended systems dynamics}},}\ }\href {\doibase 10.1080/00268979600100761} {\bibfield  {journal} {\bibinfo  {journal} {Mol. Phys.}\ }\textbf {\bibinfo {volume} {87}},\ \bibinfo {pages} {1117--1157} (\bibinfo {year} {1996})}\BibitemShut {NoStop}%
\bibitem [{\citenamefont {P{\'a}ll}\ \emph {et~al.}(2020)\citenamefont {P{\'a}ll}, \citenamefont {Zhmurov}, \citenamefont {Bauer}, \citenamefont {Abraham}, \citenamefont {Lundborg}, \citenamefont {Gray}, \citenamefont {Hess},\ and\ \citenamefont {Lindahl}}]{pall2020heterogeneous}%
  \BibitemOpen
  \bibfield  {author} {\bibinfo {author} {\bibfnamefont {S.}~\bibnamefont {P{\'a}ll}}, \bibinfo {author} {\bibfnamefont {A.}~\bibnamefont {Zhmurov}}, \bibinfo {author} {\bibfnamefont {P.}~\bibnamefont {Bauer}}, \bibinfo {author} {\bibfnamefont {M.}~\bibnamefont {Abraham}}, \bibinfo {author} {\bibfnamefont {M.}~\bibnamefont {Lundborg}}, \bibinfo {author} {\bibfnamefont {A.}~\bibnamefont {Gray}}, \bibinfo {author} {\bibfnamefont {B.}~\bibnamefont {Hess}}, \ and\ \bibinfo {author} {\bibfnamefont {E.}~\bibnamefont {Lindahl}},\ }\bibfield  {title} {\enquote {\bibinfo {title} {{Heterogeneous parallelization and acceleration of molecular dynamics simulations in GROMACS}},}\ }\href {\doibase 10.1063/5.0018516} {\bibfield  {journal} {\bibinfo  {journal} {J. Chem. Phys.}\ }\textbf {\bibinfo {volume} {153}},\ \bibinfo {pages} {134110} (\bibinfo {year} {2020})}\BibitemShut {NoStop}%
\bibitem [{\citenamefont {Ponder}\ and\ \citenamefont {Case}(2003)}]{ponder2003force}%
  \BibitemOpen
  \bibfield  {author} {\bibinfo {author} {\bibfnamefont {J.~W.}\ \bibnamefont {Ponder}}\ and\ \bibinfo {author} {\bibfnamefont {D.~A.}\ \bibnamefont {Case}},\ }\bibfield  {title} {\enquote {\bibinfo {title} {{Force fields for protein simulations}},}\ }\href {\doibase 10.1016/S0065-3233(03)66002-X} {\bibfield  {journal} {\bibinfo  {journal} {Adv. Protein Chem.}\ }\textbf {\bibinfo {volume} {66}},\ \bibinfo {pages} {27--85} (\bibinfo {year} {2003})}\BibitemShut {NoStop}%
\bibitem [{\citenamefont {Brooks}\ \emph {et~al.}(2009)\citenamefont {Brooks}, \citenamefont {Brooks}, \citenamefont {Mackerell}, \citenamefont {Nilsson}, \citenamefont {Petrella}, \citenamefont {Roux}, \citenamefont {Won}, \citenamefont {Archontis}, \citenamefont {Bartels}, \citenamefont {Boresch}, \citenamefont {Caflisch}, \citenamefont {Caves}, \citenamefont {Cui}, \citenamefont {Dinner}, \citenamefont {Feig}, \citenamefont {Fischer}, \citenamefont {Gao}, \citenamefont {Hodoscek}, \citenamefont {Im}, \citenamefont {Kuczera}, \citenamefont {Lazaridis}, \citenamefont {Ma}, \citenamefont {Ovchinnikov}, \citenamefont {Paci}, \citenamefont {Pastor}, \citenamefont {Post}, \citenamefont {Pu}, \citenamefont {Schaefer}, \citenamefont {Tidor}, \citenamefont {Venable}, \citenamefont {Woodcock}, \citenamefont {Wu}, \citenamefont {Yang}, \citenamefont {York},\ and\ \citenamefont {Karplus}}]{brooks2009charmm}%
  \BibitemOpen
  \bibfield  {author} {\bibinfo {author} {\bibfnamefont {B.~R.}\ \bibnamefont {Brooks}}, \bibinfo {author} {\bibfnamefont {C.~L.}\ \bibnamefont {Brooks}}, \bibinfo {author} {\bibfnamefont {A.~D.}\ \bibnamefont {Mackerell}}, \bibinfo {author} {\bibfnamefont {L.}~\bibnamefont {Nilsson}}, \bibinfo {author} {\bibfnamefont {R.~J.}\ \bibnamefont {Petrella}}, \bibinfo {author} {\bibfnamefont {B.}~\bibnamefont {Roux}}, \bibinfo {author} {\bibfnamefont {Y.}~\bibnamefont {Won}}, \bibinfo {author} {\bibfnamefont {G.}~\bibnamefont {Archontis}}, \bibinfo {author} {\bibfnamefont {C.}~\bibnamefont {Bartels}}, \bibinfo {author} {\bibfnamefont {S.}~\bibnamefont {Boresch}}, \bibinfo {author} {\bibfnamefont {A.}~\bibnamefont {Caflisch}}, \bibinfo {author} {\bibfnamefont {L.}~\bibnamefont {Caves}}, \bibinfo {author} {\bibfnamefont {Q.}~\bibnamefont {Cui}}, \bibinfo {author} {\bibfnamefont {A.~R.}\ \bibnamefont {Dinner}}, \bibinfo {author} {\bibfnamefont {M.}~\bibnamefont {Feig}}, \bibinfo {author} {\bibfnamefont
  {S.}~\bibnamefont {Fischer}}, \bibinfo {author} {\bibfnamefont {J.}~\bibnamefont {Gao}}, \bibinfo {author} {\bibfnamefont {M.}~\bibnamefont {Hodoscek}}, \bibinfo {author} {\bibfnamefont {W.}~\bibnamefont {Im}}, \bibinfo {author} {\bibfnamefont {K.}~\bibnamefont {Kuczera}}, \bibinfo {author} {\bibfnamefont {T.}~\bibnamefont {Lazaridis}}, \bibinfo {author} {\bibfnamefont {J.}~\bibnamefont {Ma}}, \bibinfo {author} {\bibfnamefont {V.}~\bibnamefont {Ovchinnikov}}, \bibinfo {author} {\bibfnamefont {E.}~\bibnamefont {Paci}}, \bibinfo {author} {\bibfnamefont {R.~W.}\ \bibnamefont {Pastor}}, \bibinfo {author} {\bibfnamefont {C.~B.}\ \bibnamefont {Post}}, \bibinfo {author} {\bibfnamefont {J.~Z.}\ \bibnamefont {Pu}}, \bibinfo {author} {\bibfnamefont {M.}~\bibnamefont {Schaefer}}, \bibinfo {author} {\bibfnamefont {B.}~\bibnamefont {Tidor}}, \bibinfo {author} {\bibfnamefont {R.~M.}\ \bibnamefont {Venable}}, \bibinfo {author} {\bibfnamefont {H.~L.}\ \bibnamefont {Woodcock}}, \bibinfo {author} {\bibfnamefont
  {X.}~\bibnamefont {Wu}}, \bibinfo {author} {\bibfnamefont {W.}~\bibnamefont {Yang}}, \bibinfo {author} {\bibfnamefont {D.~M.}\ \bibnamefont {York}}, \ and\ \bibinfo {author} {\bibfnamefont {M.}~\bibnamefont {Karplus}},\ }\bibfield  {title} {\enquote {\bibinfo {title} {{CHARMM: The biomolecular simulation program}},}\ }\href {\doibase 10.1002/jcc.21287} {\bibfield  {journal} {\bibinfo  {journal} {J. Comput. Chem.}\ }\textbf {\bibinfo {volume} {30}},\ \bibinfo {pages} {1545–1614} (\bibinfo {year} {2009})}\BibitemShut {NoStop}%
\bibitem [{\citenamefont {Scott}\ \emph {et~al.}(1999)\citenamefont {Scott}, \citenamefont {H{\"u}nenberger}, \citenamefont {Tironi}, \citenamefont {Mark}, \citenamefont {Billeter}, \citenamefont {Fennen}, \citenamefont {Torda}, \citenamefont {Huber}, \citenamefont {Kr{\"u}ger},\ and\ \citenamefont {Van~Gunsteren}}]{scott1999gromos}%
  \BibitemOpen
  \bibfield  {author} {\bibinfo {author} {\bibfnamefont {W.~R.}\ \bibnamefont {Scott}}, \bibinfo {author} {\bibfnamefont {P.~H.}\ \bibnamefont {H{\"u}nenberger}}, \bibinfo {author} {\bibfnamefont {I.~G.}\ \bibnamefont {Tironi}}, \bibinfo {author} {\bibfnamefont {A.~E.}\ \bibnamefont {Mark}}, \bibinfo {author} {\bibfnamefont {S.~R.}\ \bibnamefont {Billeter}}, \bibinfo {author} {\bibfnamefont {J.}~\bibnamefont {Fennen}}, \bibinfo {author} {\bibfnamefont {A.~E.}\ \bibnamefont {Torda}}, \bibinfo {author} {\bibfnamefont {T.}~\bibnamefont {Huber}}, \bibinfo {author} {\bibfnamefont {P.}~\bibnamefont {Kr{\"u}ger}}, \ and\ \bibinfo {author} {\bibfnamefont {W.~F.}\ \bibnamefont {Van~Gunsteren}},\ }\bibfield  {title} {\enquote {\bibinfo {title} {{The GROMOS Biomolecular Simulation Program Package}},}\ }\href {\doibase 10.1021/jp984217f} {\bibfield  {journal} {\bibinfo  {journal} {J. Phys. Chem. A}\ }\textbf {\bibinfo {volume} {103}},\ \bibinfo {pages} {3596--3607} (\bibinfo {year} {1999})}\BibitemShut {NoStop}%
\bibitem [{\citenamefont {Robertson}, \citenamefont {Tirado-Rives},\ and\ \citenamefont {Jorgensen}(2015)}]{robertson2015improved}%
  \BibitemOpen
  \bibfield  {author} {\bibinfo {author} {\bibfnamefont {M.~J.}\ \bibnamefont {Robertson}}, \bibinfo {author} {\bibfnamefont {J.}~\bibnamefont {Tirado-Rives}}, \ and\ \bibinfo {author} {\bibfnamefont {W.~L.}\ \bibnamefont {Jorgensen}},\ }\bibfield  {title} {\enquote {\bibinfo {title} {{Improved Peptide and Protein Torsional Energetics with the OPLS-AA Force Field}},}\ }\href {\doibase 10.1021/acs.jctc.5b00356} {\bibfield  {journal} {\bibinfo  {journal} {J. Chem. Theory Comput.}\ }\textbf {\bibinfo {volume} {11}},\ \bibinfo {pages} {3499--3509} (\bibinfo {year} {2015})}\BibitemShut {NoStop}%
\bibitem [{\citenamefont {Raghavan}\ \emph {et~al.}(2023{\natexlab{b}})\citenamefont {Raghavan}, \citenamefont {Paulikat}, \citenamefont {Ahmad}, \citenamefont {Callea}, \citenamefont {Rizzi}, \citenamefont {Ippoliti}, \citenamefont {Mandelli}, \citenamefont {Bonati}, \citenamefont {De~Vivo},\ and\ \citenamefont {Carloni}}]{drugdesign2023}%
  \BibitemOpen
  \bibfield  {author} {\bibinfo {author} {\bibfnamefont {B.}~\bibnamefont {Raghavan}}, \bibinfo {author} {\bibfnamefont {M.}~\bibnamefont {Paulikat}}, \bibinfo {author} {\bibfnamefont {K.}~\bibnamefont {Ahmad}}, \bibinfo {author} {\bibfnamefont {L.}~\bibnamefont {Callea}}, \bibinfo {author} {\bibfnamefont {A.}~\bibnamefont {Rizzi}}, \bibinfo {author} {\bibfnamefont {E.}~\bibnamefont {Ippoliti}}, \bibinfo {author} {\bibfnamefont {D.}~\bibnamefont {Mandelli}}, \bibinfo {author} {\bibfnamefont {L.}~\bibnamefont {Bonati}}, \bibinfo {author} {\bibfnamefont {M.}~\bibnamefont {De~Vivo}}, \ and\ \bibinfo {author} {\bibfnamefont {P.}~\bibnamefont {Carloni}},\ }\bibfield  {title} {\enquote {\bibinfo {title} {{Drug Design in the Exascale Era: A Perspective from Massively Parallel QM/MM Simulations}},}\ }\href {\doibase 10.1021/acs.jcim.3c00557} {\bibfield  {journal} {\bibinfo  {journal} {J. Chem. Inf. Model.}\ }\textbf {\bibinfo {volume} {63}},\ \bibinfo {pages} {3647--3658} (\bibinfo {year}
  {2023}{\natexlab{b}})}\BibitemShut {NoStop}%
\bibitem [{\citenamefont {Alvarez}(2021)}]{alvarez2021}%
  \BibitemOpen
  \bibfield  {author} {\bibinfo {author} {\bibfnamefont {D.}~\bibnamefont {Alvarez}},\ }\bibfield  {title} {\enquote {\bibinfo {title} {{JUWELS Cluster and Booster: Exascale Pathfinder with Modular Supercomputing Architecture at Juelich Supercomputing Centre}},}\ }\href {\doibase 10.17815/jlsrf-7-183} {\bibfield  {journal} {\bibinfo  {journal} {JLSRF}\ }\textbf {\bibinfo {volume} {7}},\ \bibinfo {pages} {A183} (\bibinfo {year} {2021})}\BibitemShut {NoStop}%
\bibitem [{\citenamefont {Laio}, \citenamefont {VandeVondele},\ and\ \citenamefont {Rothlisberger}(2002{\natexlab{b}})}]{laio2002d}%
  \BibitemOpen
  \bibfield  {author} {\bibinfo {author} {\bibfnamefont {A.}~\bibnamefont {Laio}}, \bibinfo {author} {\bibfnamefont {J.}~\bibnamefont {VandeVondele}}, \ and\ \bibinfo {author} {\bibfnamefont {U.}~\bibnamefont {Rothlisberger}},\ }\bibfield  {title} {\enquote {\bibinfo {title} {{D-RESP: Dynamically Generated Electrostatic Potential Derived Charges from Quantum Mechanics/Molecular Mechanics Simulations}},}\ }\href {\doibase 10.1021/jp0143138} {\bibfield  {journal} {\bibinfo  {journal} {J. Phys. Chem. B}\ }\textbf {\bibinfo {volume} {106}},\ \bibinfo {pages} {7300--7307} (\bibinfo {year} {2002}{\natexlab{b}})}\BibitemShut {NoStop}%
\bibitem [{\citenamefont {Laio}\ \emph {et~al.}(2004)\citenamefont {Laio}, \citenamefont {Gervasio}, \citenamefont {VandeVondele}, \citenamefont {Sulpizi},\ and\ \citenamefont {Rothlisberger}}]{laio2004variational}%
  \BibitemOpen
  \bibfield  {author} {\bibinfo {author} {\bibfnamefont {A.}~\bibnamefont {Laio}}, \bibinfo {author} {\bibfnamefont {F.~L.}\ \bibnamefont {Gervasio}}, \bibinfo {author} {\bibfnamefont {J.}~\bibnamefont {VandeVondele}}, \bibinfo {author} {\bibfnamefont {M.}~\bibnamefont {Sulpizi}}, \ and\ \bibinfo {author} {\bibfnamefont {U.}~\bibnamefont {Rothlisberger}},\ }\bibfield  {title} {\enquote {\bibinfo {title} {{A Variational Definition of Electrostatic Potential Derived Charges}},}\ }\href {\doibase 10.1021/jp0496405} {\bibfield  {journal} {\bibinfo  {journal} {J. Phys. Chem. B}\ }\textbf {\bibinfo {volume} {108}},\ \bibinfo {pages} {7963--7968} (\bibinfo {year} {2004})}\BibitemShut {NoStop}%
\bibitem [{\citenamefont {Sulpizi}, \citenamefont {Rothlisberger},\ and\ \citenamefont {Laio}(2005)}]{sulpizi2005electron}%
  \BibitemOpen
  \bibfield  {author} {\bibinfo {author} {\bibfnamefont {M.}~\bibnamefont {Sulpizi}}, \bibinfo {author} {\bibfnamefont {U.}~\bibnamefont {Rothlisberger}}, \ and\ \bibinfo {author} {\bibfnamefont {A.}~\bibnamefont {Laio}},\ }\bibfield  {title} {\enquote {\bibinfo {title} {{Electron transfer induced dissociation of chloro-cyano-benzene radical anion: Driving chemical reactions via charge restraints}},}\ }\href {\doibase 10.1142/S0219633605001957} {\bibfield  {journal} {\bibinfo  {journal} {J. Theor. Comput. Chem.}\ }\textbf {\bibinfo {volume} {4}},\ \bibinfo {pages} {985--999} (\bibinfo {year} {2005})}\BibitemShut {NoStop}%
\bibitem [{\citenamefont {Maurer}\ \emph {et~al.}(2007)\citenamefont {Maurer}, \citenamefont {Laio}, \citenamefont {Hugosson}, \citenamefont {Colombo},\ and\ \citenamefont {Rothlisberger}}]{Maurer2007}%
  \BibitemOpen
  \bibfield  {author} {\bibinfo {author} {\bibfnamefont {P.}~\bibnamefont {Maurer}}, \bibinfo {author} {\bibfnamefont {A.}~\bibnamefont {Laio}}, \bibinfo {author} {\bibfnamefont {H.~W.}\ \bibnamefont {Hugosson}}, \bibinfo {author} {\bibfnamefont {M.~C.}\ \bibnamefont {Colombo}}, \ and\ \bibinfo {author} {\bibfnamefont {U.}~\bibnamefont {Rothlisberger}},\ }\bibfield  {title} {\enquote {\bibinfo {title} {{Automated Parametrization of Biomolecular Force Fields from Quantum Mechanics/Molecular Mechanics (QM/MM) Simulations through Force Matching}},}\ }\href {\doibase 10.1021/ct600284f} {\bibfield  {journal} {\bibinfo  {journal} {J. Chem. Theory Comput.}\ }\textbf {\bibinfo {volume} {3}},\ \bibinfo {pages} {628--639} (\bibinfo {year} {2007})}\BibitemShut {NoStop}%
\bibitem [{\citenamefont {Doemer}\ \emph {et~al.}(2014)\citenamefont {Doemer}, \citenamefont {Maurer}, \citenamefont {Campomanes}, \citenamefont {Tavernelli},\ and\ \citenamefont {Rothlisberger}}]{Doemer2014}%
  \BibitemOpen
  \bibfield  {author} {\bibinfo {author} {\bibfnamefont {M.}~\bibnamefont {Doemer}}, \bibinfo {author} {\bibfnamefont {P.}~\bibnamefont {Maurer}}, \bibinfo {author} {\bibfnamefont {P.}~\bibnamefont {Campomanes}}, \bibinfo {author} {\bibfnamefont {I.}~\bibnamefont {Tavernelli}}, \ and\ \bibinfo {author} {\bibfnamefont {U.}~\bibnamefont {Rothlisberger}},\ }\bibfield  {title} {\enquote {\bibinfo {title} {{Generalized QM/MM Force Matching Approach Applied to the 11-cis Protonated Schiff Base Chromophore of Rhodopsin}},}\ }\href {\doibase 10.1021/ct400697n} {\bibfield  {journal} {\bibinfo  {journal} {J. Chem. Theory Comput.}\ }\textbf {\bibinfo {volume} {10}},\ \bibinfo {pages} {412--422} (\bibinfo {year} {2014})}\BibitemShut {NoStop}%
\bibitem [{\citenamefont {K\"{u}hne}\ \emph {et~al.}(2020)\citenamefont {K\"{u}hne}, \citenamefont {Iannuzzi}, \citenamefont {Del~Ben}, \citenamefont {Rybkin}, \citenamefont {Seewald}, \citenamefont {Stein}, \citenamefont {Laino}, \citenamefont {Khaliullin}, \citenamefont {Sch\"{u}tt}, \citenamefont {Schiffmann}, \citenamefont {Golze}, \citenamefont {Wilhelm}, \citenamefont {Chulkov}, \citenamefont {Bani-Hashemian}, \citenamefont {Weber}, \citenamefont {Borštnik}, \citenamefont {Taillefumier}, \citenamefont {Jakobovits}, \citenamefont {Lazzaro}, \citenamefont {Pabst}, \citenamefont {M\"{u}ller}, \citenamefont {Schade}, \citenamefont {Guidon}, \citenamefont {Andermatt}, \citenamefont {Holmberg}, \citenamefont {Schenter}, \citenamefont {Hehn}, \citenamefont {Bussy}, \citenamefont {Belleflamme}, \citenamefont {Tabacchi}, \citenamefont {Gl\"{o}ß}, \citenamefont {Lass}, \citenamefont {Bethune}, \citenamefont {Mundy}, \citenamefont {Plessl}, \citenamefont {Watkins}, \citenamefont {VandeVondele}, \citenamefont
  {Krack},\ and\ \citenamefont {Hutter}}]{Kuhne2020}%
  \BibitemOpen
  \bibfield  {author} {\bibinfo {author} {\bibfnamefont {T.~D.}\ \bibnamefont {K\"{u}hne}}, \bibinfo {author} {\bibfnamefont {M.}~\bibnamefont {Iannuzzi}}, \bibinfo {author} {\bibfnamefont {M.}~\bibnamefont {Del~Ben}}, \bibinfo {author} {\bibfnamefont {V.~V.}\ \bibnamefont {Rybkin}}, \bibinfo {author} {\bibfnamefont {P.}~\bibnamefont {Seewald}}, \bibinfo {author} {\bibfnamefont {F.}~\bibnamefont {Stein}}, \bibinfo {author} {\bibfnamefont {T.}~\bibnamefont {Laino}}, \bibinfo {author} {\bibfnamefont {R.~Z.}\ \bibnamefont {Khaliullin}}, \bibinfo {author} {\bibfnamefont {O.}~\bibnamefont {Sch\"{u}tt}}, \bibinfo {author} {\bibfnamefont {F.}~\bibnamefont {Schiffmann}}, \bibinfo {author} {\bibfnamefont {D.}~\bibnamefont {Golze}}, \bibinfo {author} {\bibfnamefont {J.}~\bibnamefont {Wilhelm}}, \bibinfo {author} {\bibfnamefont {S.}~\bibnamefont {Chulkov}}, \bibinfo {author} {\bibfnamefont {M.~H.}\ \bibnamefont {Bani-Hashemian}}, \bibinfo {author} {\bibfnamefont {V.}~\bibnamefont {Weber}}, \bibinfo {author}
  {\bibfnamefont {U.}~\bibnamefont {Borštnik}}, \bibinfo {author} {\bibfnamefont {M.}~\bibnamefont {Taillefumier}}, \bibinfo {author} {\bibfnamefont {A.~S.}\ \bibnamefont {Jakobovits}}, \bibinfo {author} {\bibfnamefont {A.}~\bibnamefont {Lazzaro}}, \bibinfo {author} {\bibfnamefont {H.}~\bibnamefont {Pabst}}, \bibinfo {author} {\bibfnamefont {T.}~\bibnamefont {M\"{u}ller}}, \bibinfo {author} {\bibfnamefont {R.}~\bibnamefont {Schade}}, \bibinfo {author} {\bibfnamefont {M.}~\bibnamefont {Guidon}}, \bibinfo {author} {\bibfnamefont {S.}~\bibnamefont {Andermatt}}, \bibinfo {author} {\bibfnamefont {N.}~\bibnamefont {Holmberg}}, \bibinfo {author} {\bibfnamefont {G.~K.}\ \bibnamefont {Schenter}}, \bibinfo {author} {\bibfnamefont {A.}~\bibnamefont {Hehn}}, \bibinfo {author} {\bibfnamefont {A.}~\bibnamefont {Bussy}}, \bibinfo {author} {\bibfnamefont {F.}~\bibnamefont {Belleflamme}}, \bibinfo {author} {\bibfnamefont {G.}~\bibnamefont {Tabacchi}}, \bibinfo {author} {\bibfnamefont {A.}~\bibnamefont {Gl\"{o}ß}}, \bibinfo
  {author} {\bibfnamefont {M.}~\bibnamefont {Lass}}, \bibinfo {author} {\bibfnamefont {I.}~\bibnamefont {Bethune}}, \bibinfo {author} {\bibfnamefont {C.~J.}\ \bibnamefont {Mundy}}, \bibinfo {author} {\bibfnamefont {C.}~\bibnamefont {Plessl}}, \bibinfo {author} {\bibfnamefont {M.}~\bibnamefont {Watkins}}, \bibinfo {author} {\bibfnamefont {J.}~\bibnamefont {VandeVondele}}, \bibinfo {author} {\bibfnamefont {M.}~\bibnamefont {Krack}}, \ and\ \bibinfo {author} {\bibfnamefont {J.}~\bibnamefont {Hutter}},\ }\bibfield  {title} {\enquote {\bibinfo {title} {{CP2K: An electronic structure and molecular dynamics software package - Quickstep: Efficient and accurate electronic structure calculations}},}\ }\href {\doibase 10.1063/5.0007045} {\bibfield  {journal} {\bibinfo  {journal} {J. Chem. Phys.}\ }\textbf {\bibinfo {volume} {152}},\ \bibinfo {pages} {194103} (\bibinfo {year} {2020})}\BibitemShut {NoStop}%
\bibitem [{\citenamefont {Motamarri}\ \emph {et~al.}(2020)\citenamefont {Motamarri}, \citenamefont {Das}, \citenamefont {Rudraraju}, \citenamefont {Ghosh}, \citenamefont {Davydov},\ and\ \citenamefont {Gavini}}]{Motamarri2020}%
  \BibitemOpen
  \bibfield  {author} {\bibinfo {author} {\bibfnamefont {P.}~\bibnamefont {Motamarri}}, \bibinfo {author} {\bibfnamefont {S.}~\bibnamefont {Das}}, \bibinfo {author} {\bibfnamefont {S.}~\bibnamefont {Rudraraju}}, \bibinfo {author} {\bibfnamefont {K.}~\bibnamefont {Ghosh}}, \bibinfo {author} {\bibfnamefont {D.}~\bibnamefont {Davydov}}, \ and\ \bibinfo {author} {\bibfnamefont {V.}~\bibnamefont {Gavini}},\ }\bibfield  {title} {\enquote {\bibinfo {title} {{DFT-FE – A massively parallel adaptive finite-element code for large-scale density functional theory calculations}},}\ }\href {\doibase 10.1016/j.cpc.2019.07.016} {\bibfield  {journal} {\bibinfo  {journal} {Comput. Phys. Commun.}\ }\textbf {\bibinfo {volume} {246}},\ \bibinfo {pages} {106853} (\bibinfo {year} {2020})}\BibitemShut {NoStop}%
\bibitem [{\citenamefont {Das}\ \emph {et~al.}(2022)\citenamefont {Das}, \citenamefont {Motamarri}, \citenamefont {Subramanian}, \citenamefont {Rogers},\ and\ \citenamefont {Gavini}}]{Das2022}%
  \BibitemOpen
  \bibfield  {author} {\bibinfo {author} {\bibfnamefont {S.}~\bibnamefont {Das}}, \bibinfo {author} {\bibfnamefont {P.}~\bibnamefont {Motamarri}}, \bibinfo {author} {\bibfnamefont {V.}~\bibnamefont {Subramanian}}, \bibinfo {author} {\bibfnamefont {D.~M.}\ \bibnamefont {Rogers}}, \ and\ \bibinfo {author} {\bibfnamefont {V.}~\bibnamefont {Gavini}},\ }\bibfield  {title} {\enquote {\bibinfo {title} {{DFT-FE 1.0: A massively parallel hybrid CPU-GPU density functional theory code using finite-element discretization}},}\ }\href {\doibase 10.1016/j.cpc.2022.108473} {\bibfield  {journal} {\bibinfo  {journal} {Comput. Phys. Commun.}\ }\textbf {\bibinfo {volume} {280}},\ \bibinfo {pages} {108473} (\bibinfo {year} {2022})}\BibitemShut {NoStop}%
\bibitem [{\citenamefont {Giannozzi}\ \emph {et~al.}(2009)\citenamefont {Giannozzi}, \citenamefont {Baroni}, \citenamefont {Bonini}, \citenamefont {Calandra}, \citenamefont {Car}, \citenamefont {Cavazzoni}, \citenamefont {Ceresoli}, \citenamefont {Chiarotti}, \citenamefont {Cococcioni}, \citenamefont {Dabo}, \citenamefont {Dal~Corso}, \citenamefont {de~Gironcoli}, \citenamefont {Fabris}, \citenamefont {Fratesi}, \citenamefont {Gebauer}, \citenamefont {Gerstmann}, \citenamefont {Gougoussis}, \citenamefont {Kokalj}, \citenamefont {Lazzeri}, \citenamefont {Martin-Samos}, \citenamefont {Marzari}, \citenamefont {Mauri}, \citenamefont {Mazzarello}, \citenamefont {Paolini}, \citenamefont {Pasquarello}, \citenamefont {Paulatto}, \citenamefont {Sbraccia}, \citenamefont {Scandolo}, \citenamefont {Sclauzero}, \citenamefont {Seitsonen}, \citenamefont {Smogunov}, \citenamefont {Umari},\ and\ \citenamefont {Wentzcovitch}}]{Giannozzi2009}%
  \BibitemOpen
  \bibfield  {author} {\bibinfo {author} {\bibfnamefont {P.}~\bibnamefont {Giannozzi}}, \bibinfo {author} {\bibfnamefont {S.}~\bibnamefont {Baroni}}, \bibinfo {author} {\bibfnamefont {N.}~\bibnamefont {Bonini}}, \bibinfo {author} {\bibfnamefont {M.}~\bibnamefont {Calandra}}, \bibinfo {author} {\bibfnamefont {R.}~\bibnamefont {Car}}, \bibinfo {author} {\bibfnamefont {C.}~\bibnamefont {Cavazzoni}}, \bibinfo {author} {\bibfnamefont {D.}~\bibnamefont {Ceresoli}}, \bibinfo {author} {\bibfnamefont {G.~L.}\ \bibnamefont {Chiarotti}}, \bibinfo {author} {\bibfnamefont {M.}~\bibnamefont {Cococcioni}}, \bibinfo {author} {\bibfnamefont {I.}~\bibnamefont {Dabo}}, \bibinfo {author} {\bibfnamefont {A.}~\bibnamefont {Dal~Corso}}, \bibinfo {author} {\bibfnamefont {S.}~\bibnamefont {de~Gironcoli}}, \bibinfo {author} {\bibfnamefont {S.}~\bibnamefont {Fabris}}, \bibinfo {author} {\bibfnamefont {G.}~\bibnamefont {Fratesi}}, \bibinfo {author} {\bibfnamefont {R.}~\bibnamefont {Gebauer}}, \bibinfo {author} {\bibfnamefont
  {U.}~\bibnamefont {Gerstmann}}, \bibinfo {author} {\bibfnamefont {C.}~\bibnamefont {Gougoussis}}, \bibinfo {author} {\bibfnamefont {A.}~\bibnamefont {Kokalj}}, \bibinfo {author} {\bibfnamefont {M.}~\bibnamefont {Lazzeri}}, \bibinfo {author} {\bibfnamefont {L.}~\bibnamefont {Martin-Samos}}, \bibinfo {author} {\bibfnamefont {N.}~\bibnamefont {Marzari}}, \bibinfo {author} {\bibfnamefont {F.}~\bibnamefont {Mauri}}, \bibinfo {author} {\bibfnamefont {R.}~\bibnamefont {Mazzarello}}, \bibinfo {author} {\bibfnamefont {S.}~\bibnamefont {Paolini}}, \bibinfo {author} {\bibfnamefont {A.}~\bibnamefont {Pasquarello}}, \bibinfo {author} {\bibfnamefont {L.}~\bibnamefont {Paulatto}}, \bibinfo {author} {\bibfnamefont {C.}~\bibnamefont {Sbraccia}}, \bibinfo {author} {\bibfnamefont {S.}~\bibnamefont {Scandolo}}, \bibinfo {author} {\bibfnamefont {G.}~\bibnamefont {Sclauzero}}, \bibinfo {author} {\bibfnamefont {A.~P.}\ \bibnamefont {Seitsonen}}, \bibinfo {author} {\bibfnamefont {A.}~\bibnamefont {Smogunov}}, \bibinfo {author}
  {\bibfnamefont {P.}~\bibnamefont {Umari}}, \ and\ \bibinfo {author} {\bibfnamefont {R.~M.}\ \bibnamefont {Wentzcovitch}},\ }\bibfield  {title} {\enquote {\bibinfo {title} {{QUANTUM ESPRESSO: a modular and open-source software project for quantum simulations of materials}},}\ }\href {\doibase 10.1088/0953-8984/21/39/395502} {\bibfield  {journal} {\bibinfo  {journal} {J. Phys. Condens. Matter}\ }\textbf {\bibinfo {volume} {21}},\ \bibinfo {pages} {395502} (\bibinfo {year} {2009})}\BibitemShut {NoStop}%
\bibitem [{\citenamefont {Carnimeo}\ \emph {et~al.}(2023)\citenamefont {Carnimeo}, \citenamefont {Affinito}, \citenamefont {Baroni}, \citenamefont {Baseggio}, \citenamefont {Bellentani}, \citenamefont {Bertossa}, \citenamefont {Delugas}, \citenamefont {Ruffino}, \citenamefont {Orlandini}, \citenamefont {Spiga},\ and\ \citenamefont {Giannozzi}}]{Carnimeo2023}%
  \BibitemOpen
  \bibfield  {author} {\bibinfo {author} {\bibfnamefont {I.}~\bibnamefont {Carnimeo}}, \bibinfo {author} {\bibfnamefont {F.}~\bibnamefont {Affinito}}, \bibinfo {author} {\bibfnamefont {S.}~\bibnamefont {Baroni}}, \bibinfo {author} {\bibfnamefont {O.}~\bibnamefont {Baseggio}}, \bibinfo {author} {\bibfnamefont {L.}~\bibnamefont {Bellentani}}, \bibinfo {author} {\bibfnamefont {R.}~\bibnamefont {Bertossa}}, \bibinfo {author} {\bibfnamefont {P.~D.}\ \bibnamefont {Delugas}}, \bibinfo {author} {\bibfnamefont {F.~F.}\ \bibnamefont {Ruffino}}, \bibinfo {author} {\bibfnamefont {S.}~\bibnamefont {Orlandini}}, \bibinfo {author} {\bibfnamefont {F.}~\bibnamefont {Spiga}}, \ and\ \bibinfo {author} {\bibfnamefont {P.}~\bibnamefont {Giannozzi}},\ }\bibfield  {title} {\enquote {\bibinfo {title} {{Quantum ESPRESSO: One Further Step toward the Exascale}},}\ }\href {\doibase 10.1021/acs.jctc.3c00249} {\bibfield  {journal} {\bibinfo  {journal} {J. Chem. Theory Comput.}\ }\textbf {\bibinfo {volume} {19}},\ \bibinfo {pages}
  {6992--7006} (\bibinfo {year} {2023})}\BibitemShut {NoStop}%
\bibitem [{\citenamefont {Matthews}\ \emph {et~al.}(2020)\citenamefont {Matthews}, \citenamefont {Cheng}, \citenamefont {Harding}, \citenamefont {Lipparini}, \citenamefont {Stopkowicz}, \citenamefont {Jagau}, \citenamefont {Szalay}, \citenamefont {Gauss},\ and\ \citenamefont {Stanton}}]{matthews2020coupled}%
  \BibitemOpen
  \bibfield  {author} {\bibinfo {author} {\bibfnamefont {D.~A.}\ \bibnamefont {Matthews}}, \bibinfo {author} {\bibfnamefont {L.}~\bibnamefont {Cheng}}, \bibinfo {author} {\bibfnamefont {M.~E.}\ \bibnamefont {Harding}}, \bibinfo {author} {\bibfnamefont {F.}~\bibnamefont {Lipparini}}, \bibinfo {author} {\bibfnamefont {S.}~\bibnamefont {Stopkowicz}}, \bibinfo {author} {\bibfnamefont {T.-C.}\ \bibnamefont {Jagau}}, \bibinfo {author} {\bibfnamefont {P.~G.}\ \bibnamefont {Szalay}}, \bibinfo {author} {\bibfnamefont {J.}~\bibnamefont {Gauss}}, \ and\ \bibinfo {author} {\bibfnamefont {J.~F.}\ \bibnamefont {Stanton}},\ }\bibfield  {title} {\enquote {\bibinfo {title} {{Coupled-cluster techniques for computational chemistry: The CFOUR program package}},}\ }\href {\doibase 10.1063/5.0004837} {\bibfield  {journal} {\bibinfo  {journal} {J. Chem. Phys.}\ }\textbf {\bibinfo {volume} {152}},\ \bibinfo {pages} {214108} (\bibinfo {year} {2020})}\BibitemShut {NoStop}%
\bibitem [{\citenamefont {Stanton}\ \emph {et~al.}()\citenamefont {Stanton}, \citenamefont {Gauss}, \citenamefont {Cheng}, \citenamefont {Harding}, \citenamefont {Matthews},\ and\ \citenamefont {Szalay}}]{cfourcode}%
  \BibitemOpen
  \bibfield  {author} {\bibinfo {author} {\bibfnamefont {J.~F.}\ \bibnamefont {Stanton}}, \bibinfo {author} {\bibfnamefont {J.}~\bibnamefont {Gauss}}, \bibinfo {author} {\bibfnamefont {L.}~\bibnamefont {Cheng}}, \bibinfo {author} {\bibfnamefont {M.~E.}\ \bibnamefont {Harding}}, \bibinfo {author} {\bibfnamefont {D.~A.}\ \bibnamefont {Matthews}}, \ and\ \bibinfo {author} {\bibfnamefont {P.~G.}\ \bibnamefont {Szalay}},\ }\href@noop {} {\enquote {\bibinfo {title} {{CFOUR, Coupled-Cluster techniques for Computational Chemistry, a quantum-chemical program package}},}\ }\bibinfo {note} {{W}ith contributions from {A}. {A}sthana, {A}.{A}. {A}uer, {R}.{J}. {B}artlett, {U}. {B}enedikt, {C}. {B}erger, {D}.{E}. {B}ernholdt, {S}. {B}laschke, {Y}. {J}. {B}omble, {S}. {B}urger, {O}. {C}hristiansen, {D}. {D}atta, {F}. {E}ngel, {R}. {F}aber, {J}. {G}reiner, {M}. {H}eckert, {O}. {H}eun, {M}. Hilgenberg, {C}. {H}uber, {T}.-{C}. {J}agau, {D}. {J}onsson, {J}. {J}us{\'e}lius, {T}. Kirsch, {M}.-{P}. {K}itsaras, {K}. {K}lein,
  {G}.{M}. {K}opper, {W}.{J}. {L}auderdale, {F}. {L}ipparini, {J}. {L}iu, {T}. {M}etzroth, {L}.{A}. {M}{\"u}ck, {D}.{P}. {O}'{N}eill, {T}. {N}ottoli, {J}. {O}swald, {D}.{R}. {P}rice, {E}. {P}rochnow, {C}. {P}uzzarini, {K}. {R}uud, {F}. {S}chiffmann, {W}. {S}chwalbach, {C}. {S}immons, {S}. {S}topkowicz, {A}. {T}ajti, {T.} Uhlirova, {J}. {V}{\'a}zquez, {F}. {W}ang, {J}.{D}. {W}atts, {P.} Yerg{\"u}n. {C}. {Z}hang, {X}. {Z}heng, and the integral packages {MOLECULE} ({J}. {A}lml{\"o}f and {P}.{R}. {T}aylor), {PROPS} ({P}.{R}. {T}aylor), {ABACUS} ({T}. {H}elgaker, {H}.{J}. {A}a. {J}ensen, {P}. {J}{\o}rgensen, and {J}. {O}lsen), and {ECP} routines by {A}. {V}. {M}itin and {C}. van {W}{\"u}llen. {F}or the current version, see \url{https://www.cfour.de}.}\BibitemShut {Stop}%
\bibitem [{\citenamefont {Lagard{\`e}re}\ \emph {et~al.}(2018)\citenamefont {Lagard{\`e}re}, \citenamefont {Jolly}, \citenamefont {Lipparini}, \citenamefont {Aviat}, \citenamefont {Stamm}, \citenamefont {Jing}, \citenamefont {Harger}, \citenamefont {Torabifard}, \citenamefont {Cisneros}, \citenamefont {Schnieders}, \citenamefont {Gresh}, \citenamefont {Maday}, \citenamefont {Ren}, \citenamefont {Ponder},\ and\ \citenamefont {Piquemal}}]{thp1}%
  \BibitemOpen
  \bibfield  {author} {\bibinfo {author} {\bibfnamefont {L.}~\bibnamefont {Lagard{\`e}re}}, \bibinfo {author} {\bibfnamefont {L.~H.}\ \bibnamefont {Jolly}}, \bibinfo {author} {\bibfnamefont {F.}~\bibnamefont {Lipparini}}, \bibinfo {author} {\bibfnamefont {F.}~\bibnamefont {Aviat}}, \bibinfo {author} {\bibfnamefont {B.}~\bibnamefont {Stamm}}, \bibinfo {author} {\bibfnamefont {Z.~F.}\ \bibnamefont {Jing}}, \bibinfo {author} {\bibfnamefont {M.}~\bibnamefont {Harger}}, \bibinfo {author} {\bibfnamefont {H.}~\bibnamefont {Torabifard}}, \bibinfo {author} {\bibfnamefont {G.~A.}\ \bibnamefont {Cisneros}}, \bibinfo {author} {\bibfnamefont {M.~J.}\ \bibnamefont {Schnieders}}, \bibinfo {author} {\bibfnamefont {N.}~\bibnamefont {Gresh}}, \bibinfo {author} {\bibfnamefont {Y.}~\bibnamefont {Maday}}, \bibinfo {author} {\bibfnamefont {P.~Y.}\ \bibnamefont {Ren}}, \bibinfo {author} {\bibfnamefont {J.~W.}\ \bibnamefont {Ponder}}, \ and\ \bibinfo {author} {\bibfnamefont {J.~P.}\ \bibnamefont {Piquemal}},\ }\bibfield  {title}
  {\enquote {\bibinfo {title} {{Tinker-HP: A massively parallel molecular dynamics package for multiscale simulations of large complex systems with advanced point dipole polarizable force fields}},}\ }\href {\doibase 10.1039/C7SC04531J} {\bibfield  {journal} {\bibinfo  {journal} {Chem. Sci.}\ }\textbf {\bibinfo {volume} {9}},\ \bibinfo {pages} {956--972} (\bibinfo {year} {2018})}\BibitemShut {NoStop}%
\bibitem [{\citenamefont {Adjoua}\ \emph {et~al.}(2021)\citenamefont {Adjoua}, \citenamefont {Lagard{\`e}re}, \citenamefont {Jolly}, \citenamefont {Durocher}, \citenamefont {Very}, \citenamefont {Dupays}, \citenamefont {Wang}, \citenamefont {Inizan}, \citenamefont {C{\'e}lerse}, \citenamefont {Ren}, \citenamefont {Ponder},\ and\ \citenamefont {Piquemal}}]{thp2}%
  \BibitemOpen
  \bibfield  {author} {\bibinfo {author} {\bibfnamefont {O.}~\bibnamefont {Adjoua}}, \bibinfo {author} {\bibfnamefont {L.}~\bibnamefont {Lagard{\`e}re}}, \bibinfo {author} {\bibfnamefont {L.-H.}\ \bibnamefont {Jolly}}, \bibinfo {author} {\bibfnamefont {A.}~\bibnamefont {Durocher}}, \bibinfo {author} {\bibfnamefont {T.}~\bibnamefont {Very}}, \bibinfo {author} {\bibfnamefont {I.}~\bibnamefont {Dupays}}, \bibinfo {author} {\bibfnamefont {Z.}~\bibnamefont {Wang}}, \bibinfo {author} {\bibfnamefont {T.~J.}\ \bibnamefont {Inizan}}, \bibinfo {author} {\bibfnamefont {F.}~\bibnamefont {C{\'e}lerse}}, \bibinfo {author} {\bibfnamefont {P.}~\bibnamefont {Ren}}, \bibinfo {author} {\bibfnamefont {J.~W.}\ \bibnamefont {Ponder}}, \ and\ \bibinfo {author} {\bibfnamefont {J.-P.}\ \bibnamefont {Piquemal}},\ }\bibfield  {title} {\enquote {\bibinfo {title} {{Tinker-HP: Accelerating Molecular Dynamics Simulations of Large Complex Systems with Advanced Point Dipole Polarizable Force Fields Using GPUs and Multi-GPU Systems}},}\
  }\href {\doibase 10.1021/acs.jctc.0c01164} {\bibfield  {journal} {\bibinfo  {journal} {J. Chem. Theory Comput.}\ }\textbf {\bibinfo {volume} {17}},\ \bibinfo {pages} {2034--2053} (\bibinfo {year} {2021})}\BibitemShut {NoStop}%
\bibitem [{\citenamefont {Eastman}\ \emph {et~al.}(2012)\citenamefont {Eastman}, \citenamefont {Friedrichs}, \citenamefont {Chodera}, \citenamefont {Radmer}, \citenamefont {Bruns}, \citenamefont {Ku}, \citenamefont {Beauchamp}, \citenamefont {Lane}, \citenamefont {Wang}, \citenamefont {Shukla}, \citenamefont {Tye}, \citenamefont {Houston}, \citenamefont {Stich}, \citenamefont {Klein}, \citenamefont {Shirts},\ and\ \citenamefont {Pande}}]{eastman2013openmm}%
  \BibitemOpen
  \bibfield  {author} {\bibinfo {author} {\bibfnamefont {P.}~\bibnamefont {Eastman}}, \bibinfo {author} {\bibfnamefont {M.~S.}\ \bibnamefont {Friedrichs}}, \bibinfo {author} {\bibfnamefont {J.~D.}\ \bibnamefont {Chodera}}, \bibinfo {author} {\bibfnamefont {R.~J.}\ \bibnamefont {Radmer}}, \bibinfo {author} {\bibfnamefont {C.~M.}\ \bibnamefont {Bruns}}, \bibinfo {author} {\bibfnamefont {J.~P.}\ \bibnamefont {Ku}}, \bibinfo {author} {\bibfnamefont {K.~A.}\ \bibnamefont {Beauchamp}}, \bibinfo {author} {\bibfnamefont {T.~J.}\ \bibnamefont {Lane}}, \bibinfo {author} {\bibfnamefont {L.-P.}\ \bibnamefont {Wang}}, \bibinfo {author} {\bibfnamefont {D.}~\bibnamefont {Shukla}}, \bibinfo {author} {\bibfnamefont {T.}~\bibnamefont {Tye}}, \bibinfo {author} {\bibfnamefont {M.}~\bibnamefont {Houston}}, \bibinfo {author} {\bibfnamefont {T.}~\bibnamefont {Stich}}, \bibinfo {author} {\bibfnamefont {C.}~\bibnamefont {Klein}}, \bibinfo {author} {\bibfnamefont {M.~R.}\ \bibnamefont {Shirts}}, \ and\ \bibinfo {author} {\bibfnamefont
  {V.~S.}\ \bibnamefont {Pande}},\ }\bibfield  {title} {\enquote {\bibinfo {title} {{OpenMM 4: A Reusable, Extensible, Hardware Independent Library for High Performance Molecular Simulation}},}\ }\href {\doibase 10.1021/ct300857j} {\bibfield  {journal} {\bibinfo  {journal} {J. Chem. Theory Comput.}\ }\textbf {\bibinfo {volume} {9}},\ \bibinfo {pages} {461–469} (\bibinfo {year} {2012})}\BibitemShut {NoStop}%
\bibitem [{\citenamefont {Eastman}\ \emph {et~al.}(2017)\citenamefont {Eastman}, \citenamefont {Swails}, \citenamefont {Chodera}, \citenamefont {McGibbon}, \citenamefont {Zhao}, \citenamefont {Beauchamp}, \citenamefont {Wang}, \citenamefont {Simmonett}, \citenamefont {Harrigan}, \citenamefont {Stern}, \citenamefont {Wiewiora}, \citenamefont {Brooks},\ and\ \citenamefont {Pande}}]{eastman2017openmm}%
  \BibitemOpen
  \bibfield  {author} {\bibinfo {author} {\bibfnamefont {P.}~\bibnamefont {Eastman}}, \bibinfo {author} {\bibfnamefont {J.}~\bibnamefont {Swails}}, \bibinfo {author} {\bibfnamefont {J.~D.}\ \bibnamefont {Chodera}}, \bibinfo {author} {\bibfnamefont {R.~T.}\ \bibnamefont {McGibbon}}, \bibinfo {author} {\bibfnamefont {Y.}~\bibnamefont {Zhao}}, \bibinfo {author} {\bibfnamefont {K.~A.}\ \bibnamefont {Beauchamp}}, \bibinfo {author} {\bibfnamefont {L.-P.}\ \bibnamefont {Wang}}, \bibinfo {author} {\bibfnamefont {A.~C.}\ \bibnamefont {Simmonett}}, \bibinfo {author} {\bibfnamefont {M.~P.}\ \bibnamefont {Harrigan}}, \bibinfo {author} {\bibfnamefont {C.~D.}\ \bibnamefont {Stern}}, \bibinfo {author} {\bibfnamefont {R.~P.}\ \bibnamefont {Wiewiora}}, \bibinfo {author} {\bibfnamefont {B.~R.}\ \bibnamefont {Brooks}}, \ and\ \bibinfo {author} {\bibfnamefont {V.~S.}\ \bibnamefont {Pande}},\ }\bibfield  {title} {\enquote {\bibinfo {title} {{OpenMM 7: Rapid development of high performance algorithms for molecular dynamics}},}\
  }\href {\doibase 10.1371/journal.pcbi.1005659} {\bibfield  {journal} {\bibinfo  {journal} {PLoS Comput. Biol.}\ }\textbf {\bibinfo {volume} {13}},\ \bibinfo {pages} {e1005659} (\bibinfo {year} {2017})}\BibitemShut {NoStop}%
\bibitem [{\citenamefont {Eastman}\ \emph {et~al.}(2023)\citenamefont {Eastman}, \citenamefont {Galvelis}, \citenamefont {Peláez}, \citenamefont {Abreu}, \citenamefont {Farr}, \citenamefont {Gallicchio}, \citenamefont {Gorenko}, \citenamefont {Henry}, \citenamefont {Hu}, \citenamefont {Huang}, \citenamefont {Kr\"{a}mer}, \citenamefont {Michel}, \citenamefont {Mitchell}, \citenamefont {Pande}, \citenamefont {Rodrigues}, \citenamefont {Rodriguez-Guerra}, \citenamefont {Simmonett}, \citenamefont {Singh}, \citenamefont {Swails}, \citenamefont {Turner}, \citenamefont {Wang}, \citenamefont {Zhang}, \citenamefont {Chodera}, \citenamefont {De~Fabritiis},\ and\ \citenamefont {Markland}}]{eastman2023openmm}%
  \BibitemOpen
  \bibfield  {author} {\bibinfo {author} {\bibfnamefont {P.}~\bibnamefont {Eastman}}, \bibinfo {author} {\bibfnamefont {R.}~\bibnamefont {Galvelis}}, \bibinfo {author} {\bibfnamefont {R.~P.}\ \bibnamefont {Peláez}}, \bibinfo {author} {\bibfnamefont {C.~R.~A.}\ \bibnamefont {Abreu}}, \bibinfo {author} {\bibfnamefont {S.~E.}\ \bibnamefont {Farr}}, \bibinfo {author} {\bibfnamefont {E.}~\bibnamefont {Gallicchio}}, \bibinfo {author} {\bibfnamefont {A.}~\bibnamefont {Gorenko}}, \bibinfo {author} {\bibfnamefont {M.~M.}\ \bibnamefont {Henry}}, \bibinfo {author} {\bibfnamefont {F.}~\bibnamefont {Hu}}, \bibinfo {author} {\bibfnamefont {J.}~\bibnamefont {Huang}}, \bibinfo {author} {\bibfnamefont {A.}~\bibnamefont {Kr\"{a}mer}}, \bibinfo {author} {\bibfnamefont {J.}~\bibnamefont {Michel}}, \bibinfo {author} {\bibfnamefont {J.~A.}\ \bibnamefont {Mitchell}}, \bibinfo {author} {\bibfnamefont {V.~S.}\ \bibnamefont {Pande}}, \bibinfo {author} {\bibfnamefont {J.~P.}\ \bibnamefont {Rodrigues}}, \bibinfo {author} {\bibfnamefont
  {J.}~\bibnamefont {Rodriguez-Guerra}}, \bibinfo {author} {\bibfnamefont {A.~C.}\ \bibnamefont {Simmonett}}, \bibinfo {author} {\bibfnamefont {S.}~\bibnamefont {Singh}}, \bibinfo {author} {\bibfnamefont {J.}~\bibnamefont {Swails}}, \bibinfo {author} {\bibfnamefont {P.}~\bibnamefont {Turner}}, \bibinfo {author} {\bibfnamefont {Y.}~\bibnamefont {Wang}}, \bibinfo {author} {\bibfnamefont {I.}~\bibnamefont {Zhang}}, \bibinfo {author} {\bibfnamefont {J.~D.}\ \bibnamefont {Chodera}}, \bibinfo {author} {\bibfnamefont {G.}~\bibnamefont {De~Fabritiis}}, \ and\ \bibinfo {author} {\bibfnamefont {T.~E.}\ \bibnamefont {Markland}},\ }\bibfield  {title} {\enquote {\bibinfo {title} {Openmm 8: Molecular dynamics simulation with machine learning potentials},}\ }\href {\doibase 10.1021/acs.jpcb.3c06662} {\bibfield  {journal} {\bibinfo  {journal} {J. Phys. Chem. B}\ }\textbf {\bibinfo {volume} {128}},\ \bibinfo {pages} {109–116} (\bibinfo {year} {2023})}\BibitemShut {NoStop}%
\bibitem [{\citenamefont {Krack}\ and\ \citenamefont {Parrinello}(2004)}]{krack2004quickstep}%
  \BibitemOpen
  \bibfield  {author} {\bibinfo {author} {\bibfnamefont {M.}~\bibnamefont {Krack}}\ and\ \bibinfo {author} {\bibfnamefont {M.}~\bibnamefont {Parrinello}},\ }\enquote {\bibinfo {title} {{Quickstep: Make the Atoms Dance}},}\ in\ \href@noop {} {\emph {\bibinfo {booktitle} {High performance computing in chemistry}}},\ Vol.~\bibinfo {volume} {25},\ \bibinfo {editor} {edited by\ \bibinfo {editor} {\bibfnamefont {J.}~\bibnamefont {Grotendorst}}}\ (\bibinfo  {publisher} {NIC-Directors},\ \bibinfo {year} {2004})\ pp.\ \bibinfo {pages} {29--51}\BibitemShut {NoStop}%
\bibitem [{\citenamefont {VandeVondele}\ \emph {et~al.}(2005)\citenamefont {VandeVondele}, \citenamefont {Krack}, \citenamefont {Mohamed}, \citenamefont {Parrinello}, \citenamefont {Chassaing},\ and\ \citenamefont {Hutter}}]{vandevondele2005quickstep}%
  \BibitemOpen
  \bibfield  {author} {\bibinfo {author} {\bibfnamefont {J.}~\bibnamefont {VandeVondele}}, \bibinfo {author} {\bibfnamefont {M.}~\bibnamefont {Krack}}, \bibinfo {author} {\bibfnamefont {F.}~\bibnamefont {Mohamed}}, \bibinfo {author} {\bibfnamefont {M.}~\bibnamefont {Parrinello}}, \bibinfo {author} {\bibfnamefont {T.}~\bibnamefont {Chassaing}}, \ and\ \bibinfo {author} {\bibfnamefont {J.}~\bibnamefont {Hutter}},\ }\bibfield  {title} {\enquote {\bibinfo {title} {{Quickstep: Fast and accurate density functional calculations using a mixed Gaussian and plane waves approach}},}\ }\href {\doibase 10.1016/j.cpc.2004.12.014} {\bibfield  {journal} {\bibinfo  {journal} {Comput. Phys. Commun.}\ }\textbf {\bibinfo {volume} {167}},\ \bibinfo {pages} {103--128} (\bibinfo {year} {2005})}\BibitemShut {NoStop}%
\bibitem [{\citenamefont {Schade}\ \emph {et~al.}(2023)\citenamefont {Schade}, \citenamefont {Kenter}, \citenamefont {Elgabarty}, \citenamefont {Lass}, \citenamefont {K{\"u}hne},\ and\ \citenamefont {Plessl}}]{schade2023breaking}%
  \BibitemOpen
  \bibfield  {author} {\bibinfo {author} {\bibfnamefont {R.}~\bibnamefont {Schade}}, \bibinfo {author} {\bibfnamefont {T.}~\bibnamefont {Kenter}}, \bibinfo {author} {\bibfnamefont {H.}~\bibnamefont {Elgabarty}}, \bibinfo {author} {\bibfnamefont {M.}~\bibnamefont {Lass}}, \bibinfo {author} {\bibfnamefont {T.~D.}\ \bibnamefont {K{\"u}hne}}, \ and\ \bibinfo {author} {\bibfnamefont {C.}~\bibnamefont {Plessl}},\ }\bibfield  {title} {\enquote {\bibinfo {title} {{Breaking the exascale barrier for the electronic structure problem in ab-initio molecular dynamics}},}\ }\href {\doibase 10.1177/109434202311776} {\bibfield  {journal} {\bibinfo  {journal} {Int. J. High Perform. Comput. Appl.}\ }\textbf {\bibinfo {volume} {37}},\ \bibinfo {pages} {530--538} (\bibinfo {year} {2023})}\BibitemShut {NoStop}%
\bibitem [{\citenamefont {Das}\ \emph {et~al.}(2023)\citenamefont {Das}, \citenamefont {Kanungo}, \citenamefont {Subramanian}, \citenamefont {Panigrahi}, \citenamefont {Motamarri}, \citenamefont {Rogers}, \citenamefont {Zimmerman},\ and\ \citenamefont {Gavini}}]{das2023GordonBell}%
  \BibitemOpen
  \bibfield  {author} {\bibinfo {author} {\bibfnamefont {S.}~\bibnamefont {Das}}, \bibinfo {author} {\bibfnamefont {B.}~\bibnamefont {Kanungo}}, \bibinfo {author} {\bibfnamefont {V.}~\bibnamefont {Subramanian}}, \bibinfo {author} {\bibfnamefont {G.}~\bibnamefont {Panigrahi}}, \bibinfo {author} {\bibfnamefont {P.}~\bibnamefont {Motamarri}}, \bibinfo {author} {\bibfnamefont {D.}~\bibnamefont {Rogers}}, \bibinfo {author} {\bibfnamefont {P.}~\bibnamefont {Zimmerman}}, \ and\ \bibinfo {author} {\bibfnamefont {V.}~\bibnamefont {Gavini}},\ }\bibfield  {title} {\enquote {\bibinfo {title} {{Large-Scale Materials Modeling at Quantum Accuracy: Ab Initio Simulations of Quasicrystals and Interacting Extended Defects in Metallic Alloys}},}\ }in\ \href {\doibase 10.1145/3581784.3627037} {\emph {\bibinfo {booktitle} {Proceedings of the International Conference for High Performance Computing, Networking, Storage and Analysis}}},\ \bibinfo {series and number} {SC ’23}\ (\bibinfo  {publisher} {ACM},\ \bibinfo {year} {2023})\
  pp.\ \bibinfo {pages} {1--12}\BibitemShut {NoStop}%
\bibitem [{\citenamefont {Giannozzi}\ \emph {et~al.}(2020)\citenamefont {Giannozzi}, \citenamefont {Baseggio}, \citenamefont {Bonfà}, \citenamefont {Brunato}, \citenamefont {Car}, \citenamefont {Carnimeo}, \citenamefont {Cavazzoni}, \citenamefont {de~Gironcoli}, \citenamefont {Delugas}, \citenamefont {Ferrari~Ruffino}, \citenamefont {Ferretti}, \citenamefont {Marzari}, \citenamefont {Timrov}, \citenamefont {Urru},\ and\ \citenamefont {Baroni}}]{Giannozzi2020}%
  \BibitemOpen
  \bibfield  {author} {\bibinfo {author} {\bibfnamefont {P.}~\bibnamefont {Giannozzi}}, \bibinfo {author} {\bibfnamefont {O.}~\bibnamefont {Baseggio}}, \bibinfo {author} {\bibfnamefont {P.}~\bibnamefont {Bonfà}}, \bibinfo {author} {\bibfnamefont {D.}~\bibnamefont {Brunato}}, \bibinfo {author} {\bibfnamefont {R.}~\bibnamefont {Car}}, \bibinfo {author} {\bibfnamefont {I.}~\bibnamefont {Carnimeo}}, \bibinfo {author} {\bibfnamefont {C.}~\bibnamefont {Cavazzoni}}, \bibinfo {author} {\bibfnamefont {S.}~\bibnamefont {de~Gironcoli}}, \bibinfo {author} {\bibfnamefont {P.}~\bibnamefont {Delugas}}, \bibinfo {author} {\bibfnamefont {F.}~\bibnamefont {Ferrari~Ruffino}}, \bibinfo {author} {\bibfnamefont {A.}~\bibnamefont {Ferretti}}, \bibinfo {author} {\bibfnamefont {N.}~\bibnamefont {Marzari}}, \bibinfo {author} {\bibfnamefont {I.}~\bibnamefont {Timrov}}, \bibinfo {author} {\bibfnamefont {A.}~\bibnamefont {Urru}}, \ and\ \bibinfo {author} {\bibfnamefont {S.}~\bibnamefont {Baroni}},\ }\bibfield  {title} {\enquote
  {\bibinfo {title} {{Quantum ESPRESSO toward the exascale}},}\ }\href {\doibase 10.1063/5.0005082} {\bibfield  {journal} {\bibinfo  {journal} {J. Chem. Phys.}\ }\textbf {\bibinfo {volume} {152}},\ \bibinfo {pages} {154105} (\bibinfo {year} {2020})}\BibitemShut {NoStop}%
\bibitem [{\citenamefont {Kirsch}\ \emph {et~al.}(2022)\citenamefont {Kirsch}, \citenamefont {Olsen}, \citenamefont {Bolnykh}, \citenamefont {Meloni}, \citenamefont {Ippoliti}, \citenamefont {Rothlisberger}, \citenamefont {Cascella},\ and\ \citenamefont {Gauss}}]{cfour}%
  \BibitemOpen
  \bibfield  {author} {\bibinfo {author} {\bibfnamefont {T.}~\bibnamefont {Kirsch}}, \bibinfo {author} {\bibfnamefont {J.~M.~H.}\ \bibnamefont {Olsen}}, \bibinfo {author} {\bibfnamefont {V.}~\bibnamefont {Bolnykh}}, \bibinfo {author} {\bibfnamefont {S.}~\bibnamefont {Meloni}}, \bibinfo {author} {\bibfnamefont {E.}~\bibnamefont {Ippoliti}}, \bibinfo {author} {\bibfnamefont {U.}~\bibnamefont {Rothlisberger}}, \bibinfo {author} {\bibfnamefont {M.}~\bibnamefont {Cascella}}, \ and\ \bibinfo {author} {\bibfnamefont {J.}~\bibnamefont {Gauss}},\ }\bibfield  {title} {\enquote {\bibinfo {title} {{Wavefunction-Based Electrostatic-Embedding QM/MM Using CFOUR through MiMiC}},}\ }\href {\doibase 10.1021/acs.jctc.1c00878} {\bibfield  {journal} {\bibinfo  {journal} {J. Chem. Theory Comput.}\ }\textbf {\bibinfo {volume} {18}},\ \bibinfo {pages} {13--24} (\bibinfo {year} {2022})}\BibitemShut {NoStop}%
\bibitem [{\citenamefont {Lamoureux}\ and\ \citenamefont {Roux}(2003)}]{lamoureux2003modeling}%
  \BibitemOpen
  \bibfield  {author} {\bibinfo {author} {\bibfnamefont {G.}~\bibnamefont {Lamoureux}}\ and\ \bibinfo {author} {\bibfnamefont {B.}~\bibnamefont {Roux}},\ }\bibfield  {title} {\enquote {\bibinfo {title} {{Modeling induced polarization with classical Drude oscillators: Theory and molecular dynamics simulation algorithm}},}\ }\href {\doibase 10.1063/1.1589749} {\bibfield  {journal} {\bibinfo  {journal} {J. Chem. Phys.}\ }\textbf {\bibinfo {volume} {119}},\ \bibinfo {pages} {3025--3039} (\bibinfo {year} {2003})}\BibitemShut {NoStop}%
\bibitem [{\citenamefont {Chiariello}\ \emph {et~al.}(2021)\citenamefont {Chiariello}, \citenamefont {Alfonso-Prieto}, \citenamefont {Ippoliti}, \citenamefont {Fahlke},\ and\ \citenamefont {Carloni}}]{clc2021}%
  \BibitemOpen
  \bibfield  {author} {\bibinfo {author} {\bibfnamefont {M.~G.}\ \bibnamefont {Chiariello}}, \bibinfo {author} {\bibfnamefont {M.}~\bibnamefont {Alfonso-Prieto}}, \bibinfo {author} {\bibfnamefont {E.}~\bibnamefont {Ippoliti}}, \bibinfo {author} {\bibfnamefont {C.}~\bibnamefont {Fahlke}}, \ and\ \bibinfo {author} {\bibfnamefont {P.}~\bibnamefont {Carloni}},\ }\bibfield  {title} {\enquote {\bibinfo {title} {{Mechanisms Underlying Proton Release in CLC-type F\textsuperscript{--}/H\textsuperscript{+} Antiporters}},}\ }\href {\doibase 10.1021/acs.jpclett.1c00361} {\bibfield  {journal} {\bibinfo  {journal} {J. Phys. Chem. Lett.}\ }\textbf {\bibinfo {volume} {12}},\ \bibinfo {pages} {4415--4420} (\bibinfo {year} {2021})}\BibitemShut {NoStop}%
\bibitem [{\citenamefont {Schackert}\ \emph {et~al.}(2023)\citenamefont {Schackert}, \citenamefont {Biedermann}, \citenamefont {Abdolvand}, \citenamefont {Minniberger}, \citenamefont {Song}, \citenamefont {Plested}, \citenamefont {Carloni},\ and\ \citenamefont {Sun}}]{glutamate2023}%
  \BibitemOpen
  \bibfield  {author} {\bibinfo {author} {\bibfnamefont {F.~K.}\ \bibnamefont {Schackert}}, \bibinfo {author} {\bibfnamefont {J.}~\bibnamefont {Biedermann}}, \bibinfo {author} {\bibfnamefont {S.}~\bibnamefont {Abdolvand}}, \bibinfo {author} {\bibfnamefont {S.}~\bibnamefont {Minniberger}}, \bibinfo {author} {\bibfnamefont {C.}~\bibnamefont {Song}}, \bibinfo {author} {\bibfnamefont {A.~J.~R.}\ \bibnamefont {Plested}}, \bibinfo {author} {\bibfnamefont {P.}~\bibnamefont {Carloni}}, \ and\ \bibinfo {author} {\bibfnamefont {H.}~\bibnamefont {Sun}},\ }\bibfield  {title} {\enquote {\bibinfo {title} {{Mechanism of Calcium Permeation in a Glutamate Receptor Ion Channel}},}\ }\href {\doibase 10.1021/acs.jcim.2c01494} {\bibfield  {journal} {\bibinfo  {journal} {J. Chem. Inf. Model.}\ }\textbf {\bibinfo {volume} {63}},\ \bibinfo {pages} {1293--1300} (\bibinfo {year} {2023})}\BibitemShut {NoStop}%
\bibitem [{\citenamefont {Chiariello}\ \emph {et~al.}(2020)\citenamefont {Chiariello}, \citenamefont {Bolnykh}, \citenamefont {Ippoliti}, \citenamefont {Meloni}, \citenamefont {Olsen}, \citenamefont {Beck}, \citenamefont {Rothlisberger}, \citenamefont {Fahlke},\ and\ \citenamefont {Carloni}}]{clc2020}%
  \BibitemOpen
  \bibfield  {author} {\bibinfo {author} {\bibfnamefont {M.~G.}\ \bibnamefont {Chiariello}}, \bibinfo {author} {\bibfnamefont {V.}~\bibnamefont {Bolnykh}}, \bibinfo {author} {\bibfnamefont {E.}~\bibnamefont {Ippoliti}}, \bibinfo {author} {\bibfnamefont {S.}~\bibnamefont {Meloni}}, \bibinfo {author} {\bibfnamefont {J.~M.~H.}\ \bibnamefont {Olsen}}, \bibinfo {author} {\bibfnamefont {T.}~\bibnamefont {Beck}}, \bibinfo {author} {\bibfnamefont {U.}~\bibnamefont {Rothlisberger}}, \bibinfo {author} {\bibfnamefont {C.}~\bibnamefont {Fahlke}}, \ and\ \bibinfo {author} {\bibfnamefont {P.}~\bibnamefont {Carloni}},\ }\bibfield  {title} {\enquote {\bibinfo {title} {{Molecular Basis of CLC Antiporter Inhibition by Fluoride}},}\ }\href {\doibase 10.1021/jacs.9b13588} {\bibfield  {journal} {\bibinfo  {journal} {J. Am. Chem. Soc.}\ }\textbf {\bibinfo {volume} {142}},\ \bibinfo {pages} {7254--7258} (\bibinfo {year} {2020})}\BibitemShut {NoStop}%
\bibitem [{\citenamefont {{The PLUMED consortium}}(2019)}]{PLUMED}%
  \BibitemOpen
  \bibfield  {author} {\bibinfo {author} {\bibnamefont {{The PLUMED consortium}}},\ }\bibfield  {title} {\enquote {\bibinfo {title} {{Promoting transparency and reproducibility in enhanced molecular simulations}},}\ }\href {\doibase 10.1038/s41592-019-0506-8} {\bibfield  {journal} {\bibinfo  {journal} {Nat. Methods}\ }\textbf {\bibinfo {volume} {16}},\ \bibinfo {pages} {670--673} (\bibinfo {year} {2019})}\BibitemShut {NoStop}%
\bibitem [{\citenamefont {Asgharpour}\ \emph {et~al.}(2022)\citenamefont {Asgharpour}, \citenamefont {Chi}, \citenamefont {Spehr}, \citenamefont {Carloni},\ and\ \citenamefont {Alfonso-Prieto}}]{clc2024}%
  \BibitemOpen
  \bibfield  {author} {\bibinfo {author} {\bibfnamefont {S.}~\bibnamefont {Asgharpour}}, \bibinfo {author} {\bibfnamefont {L.~A.}\ \bibnamefont {Chi}}, \bibinfo {author} {\bibfnamefont {M.}~\bibnamefont {Spehr}}, \bibinfo {author} {\bibfnamefont {P.}~\bibnamefont {Carloni}}, \ and\ \bibinfo {author} {\bibfnamefont {M.}~\bibnamefont {Alfonso-Prieto}},\ }\enquote {\bibinfo {title} {{Fluoride Transport and Inhibition Across CLC Transporters}},}\ in\ \href {\doibase 10.1007/164_2022_593} {\emph {\bibinfo {booktitle} {Handb. Exp. Pharmacol.}}}\ (\bibinfo  {publisher} {Springer International Publishing},\ \bibinfo {year} {2022})\ pp.\ \bibinfo {pages} {81--100}\BibitemShut {NoStop}%
\bibitem [{\citenamefont {Dmitrieva}\ \emph {et~al.}(2023)\citenamefont {Dmitrieva}, \citenamefont {Alleva}, \citenamefont {Alfonso~Prieto}, \citenamefont {Carloni},\ and\ \citenamefont {Fahlke}}]{2023dgot}%
  \BibitemOpen
  \bibfield  {author} {\bibinfo {author} {\bibfnamefont {N.}~\bibnamefont {Dmitrieva}}, \bibinfo {author} {\bibfnamefont {C.}~\bibnamefont {Alleva}}, \bibinfo {author} {\bibfnamefont {M.}~\bibnamefont {Alfonso~Prieto}}, \bibinfo {author} {\bibfnamefont {P.}~\bibnamefont {Carloni}}, \ and\ \bibinfo {author} {\bibfnamefont {C.~M.}\ \bibnamefont {Fahlke}},\ }\bibfield  {title} {\enquote {\bibinfo {title} {{Exploring the transport cycle of DgoT, a bacterial homolog of human vesicular glutamate transporters}},}\ }\href {\doibase 10.1016/j.bpj.2022.11.1363} {\bibfield  {journal} {\bibinfo  {journal} {Biophys. J.}\ }\textbf {\bibinfo {volume} {122}} (\bibinfo {year} {2023}),\ 10.1016/j.bpj.2022.11.1363}\BibitemShut {NoStop}%
\bibitem [{\citenamefont {Dmitrieva}\ \emph {et~al.}(2024)\citenamefont {Dmitrieva}, \citenamefont {Gholami}, \citenamefont {Alleva}, \citenamefont {Carloni}, \citenamefont {Alfonso-Prieto},\ and\ \citenamefont {Fahlke}}]{slc17_2024}%
  \BibitemOpen
  \bibfield  {author} {\bibinfo {author} {\bibfnamefont {N.}~\bibnamefont {Dmitrieva}}, \bibinfo {author} {\bibfnamefont {S.}~\bibnamefont {Gholami}}, \bibinfo {author} {\bibfnamefont {C.}~\bibnamefont {Alleva}}, \bibinfo {author} {\bibfnamefont {P.}~\bibnamefont {Carloni}}, \bibinfo {author} {\bibfnamefont {M.}~\bibnamefont {Alfonso-Prieto}}, \ and\ \bibinfo {author} {\bibfnamefont {C.}~\bibnamefont {Fahlke}},\ }\bibfield  {title} {\enquote {\bibinfo {title} {{Transport mechanism of DgoT, a bacterial homolog of SLC17 organic anion transporters}},}\ }\href {\doibase 10.1101/2024.02.07.579339} {\bibfield  {journal} {\bibinfo  {journal} {bioRxiv}\ } (\bibinfo {year} {2024}),\ 10.1101/2024.02.07.579339}\BibitemShut {NoStop}%
\bibitem [{\citenamefont {Ahmad}\ \emph {et~al.}(2022)\citenamefont {Ahmad}, \citenamefont {Rizzi}, \citenamefont {Capelli}, \citenamefont {Mandelli}, \citenamefont {Lyu},\ and\ \citenamefont {Carloni}}]{ligand_binding}%
  \BibitemOpen
  \bibfield  {author} {\bibinfo {author} {\bibfnamefont {K.}~\bibnamefont {Ahmad}}, \bibinfo {author} {\bibfnamefont {A.}~\bibnamefont {Rizzi}}, \bibinfo {author} {\bibfnamefont {R.}~\bibnamefont {Capelli}}, \bibinfo {author} {\bibfnamefont {D.}~\bibnamefont {Mandelli}}, \bibinfo {author} {\bibfnamefont {W.}~\bibnamefont {Lyu}}, \ and\ \bibinfo {author} {\bibfnamefont {P.}~\bibnamefont {Carloni}},\ }\bibfield  {title} {\enquote {\bibinfo {title} {{Enhanced-Sampling Simulations for the Estimation of Ligand Binding Kinetics: Current Status and Perspective}},}\ }\href {\doibase 10.3389/fmolb.2022.899805} {\bibfield  {journal} {\bibinfo  {journal} {Front. Mol. Biosci.}\ }\textbf {\bibinfo {volume} {9}},\ \bibinfo {pages} {899805} (\bibinfo {year} {2022})}\BibitemShut {NoStop}%
\bibitem [{\citenamefont {Capelli}\ \emph {et~al.}(2020)\citenamefont {Capelli}, \citenamefont {Lyu}, \citenamefont {Bolnykh}, \citenamefont {Meloni}, \citenamefont {Olsen}, \citenamefont {Rothlisberger}, \citenamefont {Parrinello},\ and\ \citenamefont {Carloni}}]{koff2020}%
  \BibitemOpen
  \bibfield  {author} {\bibinfo {author} {\bibfnamefont {R.}~\bibnamefont {Capelli}}, \bibinfo {author} {\bibfnamefont {W.}~\bibnamefont {Lyu}}, \bibinfo {author} {\bibfnamefont {V.}~\bibnamefont {Bolnykh}}, \bibinfo {author} {\bibfnamefont {S.}~\bibnamefont {Meloni}}, \bibinfo {author} {\bibfnamefont {J.~M.~H.}\ \bibnamefont {Olsen}}, \bibinfo {author} {\bibfnamefont {U.}~\bibnamefont {Rothlisberger}}, \bibinfo {author} {\bibfnamefont {M.}~\bibnamefont {Parrinello}}, \ and\ \bibinfo {author} {\bibfnamefont {P.}~\bibnamefont {Carloni}},\ }\bibfield  {title} {\enquote {\bibinfo {title} {{Accuracy of Molecular Simulation-Based Predictions of \textit{k}\textsubscript{off} Values: A Metadynamics Study}},}\ }\href {\doibase 10.1021/acs.jpclett.0c00999} {\bibfield  {journal} {\bibinfo  {journal} {J. Phys. Chem. Lett.}\ }\textbf {\bibinfo {volume} {11}},\ \bibinfo {pages} {6373--6381} (\bibinfo {year} {2020})}\BibitemShut {NoStop}%
\bibitem [{\citenamefont {Xu}\ \emph {et~al.}(2004)\citenamefont {Xu}, \citenamefont {Zhao}, \citenamefont {Xu}, \citenamefont {Peng}, \citenamefont {Huang}, \citenamefont {Arnold},\ and\ \citenamefont {Ding}}]{idh1_ref}%
  \BibitemOpen
  \bibfield  {author} {\bibinfo {author} {\bibfnamefont {X.}~\bibnamefont {Xu}}, \bibinfo {author} {\bibfnamefont {J.}~\bibnamefont {Zhao}}, \bibinfo {author} {\bibfnamefont {Z.}~\bibnamefont {Xu}}, \bibinfo {author} {\bibfnamefont {B.}~\bibnamefont {Peng}}, \bibinfo {author} {\bibfnamefont {Q.}~\bibnamefont {Huang}}, \bibinfo {author} {\bibfnamefont {E.}~\bibnamefont {Arnold}}, \ and\ \bibinfo {author} {\bibfnamefont {J.}~\bibnamefont {Ding}},\ }\bibfield  {title} {\enquote {\bibinfo {title} {{Structures of Human Cytosolic NADP-dependent Isocitrate Dehydrogenase Reveal a Novel Self-regulatory Mechanism of Activity}},}\ }\href {\doibase 10.1074/jbc.m404298200} {\bibfield  {journal} {\bibinfo  {journal} {J. Biol. Chem.}\ }\textbf {\bibinfo {volume} {279}},\ \bibinfo {pages} {33946--33957} (\bibinfo {year} {2004})}\BibitemShut {NoStop}%
\bibitem [{\citenamefont {Remmel}(2022)}]{Remmel_CEN}%
  \BibitemOpen
  \bibfield  {author} {\bibinfo {author} {\bibfnamefont {A.}~\bibnamefont {Remmel}},\ }\bibfield  {title} {\enquote {\bibinfo {title} {{What exascale computing could mean for chemistry}},}\ }\href {https://cen.acs.org/physical-chemistry/computational-chemistry/exascale-computing-mean-chemistry/100/i31} {\bibfield  {journal} {\bibinfo  {journal} {C\&EN Global Enterp.}\ }\textbf {\bibinfo {volume} {100}} (\bibinfo {year} {2022})}\BibitemShut {NoStop}%
\bibitem [{\citenamefont {Bolnykh}\ \emph {et~al.}(2021)\citenamefont {Bolnykh}, \citenamefont {Rossetti}, \citenamefont {Rothlisberger},\ and\ \citenamefont {Carloni}}]{Slava_WIREs_2021}%
  \BibitemOpen
  \bibfield  {author} {\bibinfo {author} {\bibfnamefont {V.}~\bibnamefont {Bolnykh}}, \bibinfo {author} {\bibfnamefont {G.}~\bibnamefont {Rossetti}}, \bibinfo {author} {\bibfnamefont {U.}~\bibnamefont {Rothlisberger}}, \ and\ \bibinfo {author} {\bibfnamefont {P.}~\bibnamefont {Carloni}},\ }\bibfield  {title} {\enquote {\bibinfo {title} {{Expanding the boundaries of ligand--target modeling by exascale calculations}},}\ }\href {\doibase 10.1002/wcms.1535} {\bibfield  {journal} {\bibinfo  {journal} {Wiley Interdiscip. Rev. Comput. Mol. Sci.}\ }\textbf {\bibinfo {volume} {11}},\ \bibinfo {pages} {e1535} (\bibinfo {year} {2021})}\BibitemShut {NoStop}%
\bibitem [{\citenamefont {Martin}\ \emph {et~al.}(2022)\citenamefont {Martin}, \citenamefont {Sheynkman}, \citenamefont {Lightstone}, \citenamefont {Nussinov},\ and\ \citenamefont {Cheng}}]{Martin_2022}%
  \BibitemOpen
  \bibfield  {author} {\bibinfo {author} {\bibfnamefont {W.}~\bibnamefont {Martin}}, \bibinfo {author} {\bibfnamefont {G.}~\bibnamefont {Sheynkman}}, \bibinfo {author} {\bibfnamefont {F.~C.}\ \bibnamefont {Lightstone}}, \bibinfo {author} {\bibfnamefont {R.}~\bibnamefont {Nussinov}}, \ and\ \bibinfo {author} {\bibfnamefont {F.}~\bibnamefont {Cheng}},\ }\bibfield  {title} {\enquote {\bibinfo {title} {{Interpretable artificial intelligence and exascale molecular dynamics simulations to reveal kinetics: Applications to Alzheimer's disease}},}\ }\href {\doibase 10.1016/j.sbi.2021.09.001} {\bibfield  {journal} {\bibinfo  {journal} {Curr. Opin. Struct. Biol.}\ }\textbf {\bibinfo {volume} {72}},\ \bibinfo {pages} {103--113} (\bibinfo {year} {2022})}\BibitemShut {NoStop}%
\bibitem [{\citenamefont {Beck}, \citenamefont {Carloni},\ and\ \citenamefont {Asthagiri}(2024)}]{Beck_2024}%
  \BibitemOpen
  \bibfield  {author} {\bibinfo {author} {\bibfnamefont {T.~L.}\ \bibnamefont {Beck}}, \bibinfo {author} {\bibfnamefont {P.}~\bibnamefont {Carloni}}, \ and\ \bibinfo {author} {\bibfnamefont {D.~N.}\ \bibnamefont {Asthagiri}},\ }\bibfield  {title} {\enquote {\bibinfo {title} {{All-Atom Biomolecular Simulation in the Exascale Era}},}\ }\href {\doibase 10.1021/acs.jctc.3c01276} {\bibfield  {journal} {\bibinfo  {journal} {J. Chem. Theory Comput.}\ }\textbf {\bibinfo {volume} {20}},\ \bibinfo {pages} {1777--1782} (\bibinfo {year} {2024})}\BibitemShut {NoStop}%
\bibitem [{\citenamefont {Carpenter}\ \emph {et~al.}(2022)\citenamefont {Carpenter}, \citenamefont {Utz}, \citenamefont {Narasimhamurthy},\ and\ \citenamefont {Suarez}}]{carpenter_2022_6090425}%
  \BibitemOpen
  \bibfield  {author} {\bibinfo {author} {\bibfnamefont {P.}~\bibnamefont {Carpenter}}, \bibinfo {author} {\bibfnamefont {U.-H.}\ \bibnamefont {Utz}}, \bibinfo {author} {\bibfnamefont {S.}~\bibnamefont {Narasimhamurthy}}, \ and\ \bibinfo {author} {\bibfnamefont {E.}~\bibnamefont {Suarez}},\ }\bibfield  {title} {\enquote {\bibinfo {title} {{Heterogeneous High Performance Computing}},}\ }\href {\doibase 10.5281/zenodo.6090425} {\bibfield  {journal} {\bibinfo  {journal} {{Zenodo}}\ } (\bibinfo {year} {2022}),\ 10.5281/zenodo.6090425}\BibitemShut {NoStop}%
\bibitem [{\citenamefont {Suarez}\ \emph {et~al.}(2022)\citenamefont {Suarez}, \citenamefont {Eicker}, \citenamefont {Moschny}, \citenamefont {Pickartz}, \citenamefont {Clauss}, \citenamefont {Plugaru}, \citenamefont {Herten}, \citenamefont {Michielsen},\ and\ \citenamefont {Lippert}}]{suarez_2022_6508394}%
  \BibitemOpen
  \bibfield  {author} {\bibinfo {author} {\bibfnamefont {E.}~\bibnamefont {Suarez}}, \bibinfo {author} {\bibfnamefont {N.}~\bibnamefont {Eicker}}, \bibinfo {author} {\bibfnamefont {T.}~\bibnamefont {Moschny}}, \bibinfo {author} {\bibfnamefont {S.}~\bibnamefont {Pickartz}}, \bibinfo {author} {\bibfnamefont {C.}~\bibnamefont {Clauss}}, \bibinfo {author} {\bibfnamefont {V.}~\bibnamefont {Plugaru}}, \bibinfo {author} {\bibfnamefont {A.}~\bibnamefont {Herten}}, \bibinfo {author} {\bibfnamefont {K.}~\bibnamefont {Michielsen}}, \ and\ \bibinfo {author} {\bibfnamefont {T.}~\bibnamefont {Lippert}},\ }\bibfield  {title} {\enquote {\bibinfo {title} {{Modular Supercomputing Architecture}},}\ }\href {\doibase 10.5281/zenodo.6508394} {\bibfield  {journal} {\bibinfo  {journal} {{Zenodo}}\ } (\bibinfo {year} {2022}),\ 10.5281/zenodo.6508394}\BibitemShut {NoStop}%
\bibitem [{\citenamefont {Carrasco-Busturia}\ \emph {et~al.}(2024)\citenamefont {Carrasco-Busturia}, \citenamefont {Ippoliti}, \citenamefont {Meloni}, \citenamefont {Rothlisberger},\ and\ \citenamefont {Olsen}}]{carrasco2024multiscale}%
  \BibitemOpen
  \bibfield  {author} {\bibinfo {author} {\bibfnamefont {D.}~\bibnamefont {Carrasco-Busturia}}, \bibinfo {author} {\bibfnamefont {E.}~\bibnamefont {Ippoliti}}, \bibinfo {author} {\bibfnamefont {S.}~\bibnamefont {Meloni}}, \bibinfo {author} {\bibfnamefont {U.}~\bibnamefont {Rothlisberger}}, \ and\ \bibinfo {author} {\bibfnamefont {J.~M.~H.}\ \bibnamefont {Olsen}},\ }\bibfield  {title} {\enquote {\bibinfo {title} {Multiscale biomolecular simulations in the exascale era},}\ }\href {\doibase 10.1016/j.sbi.2024.102821} {\bibfield  {journal} {\bibinfo  {journal} {Curr. Opin. Struct. Biol.}\ }\textbf {\bibinfo {volume} {86}},\ \bibinfo {pages} {102821} (\bibinfo {year} {2024})}\BibitemShut {NoStop}%
\bibitem [{jup(2024)}]{jupiter_technical}%
  \BibitemOpen
  \href@noop {} {\enquote {\bibinfo {title} {{JUPITER Technical Overview}},}\ }\bibinfo {howpublished} {\url{https://www.fz-juelich.de/en/ias/jsc/jupiter/tech}} (\bibinfo {year} {2024}),\ \bibinfo {note} {date accessed: 2024-03-27}\BibitemShut {NoStop}%
\bibitem [{\citenamefont {Unke}\ \emph {et~al.}(2021)\citenamefont {Unke}, \citenamefont {Chmiela}, \citenamefont {Sauceda}, \citenamefont {Gastegger}, \citenamefont {Poltavsky}, \citenamefont {Sch\"{u}tt}, \citenamefont {Tkatchenko},\ and\ \citenamefont {M\"{u}ller}}]{Unke2021}%
  \BibitemOpen
  \bibfield  {author} {\bibinfo {author} {\bibfnamefont {O.~T.}\ \bibnamefont {Unke}}, \bibinfo {author} {\bibfnamefont {S.}~\bibnamefont {Chmiela}}, \bibinfo {author} {\bibfnamefont {H.~E.}\ \bibnamefont {Sauceda}}, \bibinfo {author} {\bibfnamefont {M.}~\bibnamefont {Gastegger}}, \bibinfo {author} {\bibfnamefont {I.}~\bibnamefont {Poltavsky}}, \bibinfo {author} {\bibfnamefont {K.~T.}\ \bibnamefont {Sch\"{u}tt}}, \bibinfo {author} {\bibfnamefont {A.}~\bibnamefont {Tkatchenko}}, \ and\ \bibinfo {author} {\bibfnamefont {K.-R.}\ \bibnamefont {M\"{u}ller}},\ }\bibfield  {title} {\enquote {\bibinfo {title} {{Machine Learning Force Fields}},}\ }\href {\doibase 10.1021/acs.chemrev.0c01111} {\bibfield  {journal} {\bibinfo  {journal} {Chem. Rev.}\ }\textbf {\bibinfo {volume} {121}},\ \bibinfo {pages} {10142--10186} (\bibinfo {year} {2021})}\BibitemShut {NoStop}%
\bibitem [{\citenamefont {Kocer}, \citenamefont {Ko},\ and\ \citenamefont {Behler}(2022)}]{Kocer2022}%
  \BibitemOpen
  \bibfield  {author} {\bibinfo {author} {\bibfnamefont {E.}~\bibnamefont {Kocer}}, \bibinfo {author} {\bibfnamefont {T.~W.}\ \bibnamefont {Ko}}, \ and\ \bibinfo {author} {\bibfnamefont {J.}~\bibnamefont {Behler}},\ }\bibfield  {title} {\enquote {\bibinfo {title} {{Neural Network Potentials: A Concise Overview of Methods}},}\ }\href {\doibase 10.1146/annurev-physchem-082720-034254} {\bibfield  {journal} {\bibinfo  {journal} {Annu. Rev. Phys. Chem.}\ }\textbf {\bibinfo {volume} {73}},\ \bibinfo {pages} {163--186} (\bibinfo {year} {2022})}\BibitemShut {NoStop}%
\bibitem [{\citenamefont {Mato}\ \emph {et~al.}(2021)\citenamefont {Mato}, \citenamefont {Duster}, \citenamefont {Guidez},\ and\ \citenamefont {Lin}}]{Mato2021}%
  \BibitemOpen
  \bibfield  {author} {\bibinfo {author} {\bibfnamefont {J.}~\bibnamefont {Mato}}, \bibinfo {author} {\bibfnamefont {A.~W.}\ \bibnamefont {Duster}}, \bibinfo {author} {\bibfnamefont {E.~B.}\ \bibnamefont {Guidez}}, \ and\ \bibinfo {author} {\bibfnamefont {H.}~\bibnamefont {Lin}},\ }\bibfield  {title} {\enquote {\bibinfo {title} {{Adaptive-Partitioning Multilayer Dynamics Simulations: 1. On-the-Fly Switch between Two Quantum Levels of Theory}},}\ }\href {\doibase 10.1021/acs.jctc.1c00556} {\bibfield  {journal} {\bibinfo  {journal} {J. Chem. Theory Comput.}\ }\textbf {\bibinfo {volume} {17}},\ \bibinfo {pages} {5456--5465} (\bibinfo {year} {2021})}\BibitemShut {NoStop}%
\end{thebibliography}%

\end{document}